\documentclass{vldb}

\vldbTitle{You Say `What', I Hear `Where' and `Why'---(Mis-)Interpreting \SQL{} to Derive Fine-Grained Provenance}
\vldbAuthors{Tobias M\"uller, Benjamin Dietrich, Torsten Grust}
\vldbDOI{https://doi.org/10.14778/3236187.3236204}
\vldbVolume{11}
\vldbNumber{11}
\vldbYear{2018}

\pdfoutput=1
\newif\ifdraft\draftfalse
\ifdraft
  \usepackage{showlabels}
\else
  \usepackage[keeplastbox]{flushend}
\fi

\newif\ifrendermap\rendermaptrue
\ifdraft
\else
  \rendermaptrue
\fi

\usepackage[utf8]{inputenc}

\usepackage[T1]{fontenc}

\usepackage{etoolbox}

\usepackage{setspace}

\usepackage{lmodern}

\usepackage{cite}

\usepackage{mathdots}
\newcommand{\ndots}{\relax\ensuremath{\mathinner{\ldotp\!\ldotp\!\ldotp}}}
\newcommand{\ncdots}{\relax\ensuremath{\mathinner{\cdotp\!\cdotp\!\cdotp}}}

\usepackage{colonequals}

\usepackage{mathtools}

\usepackage{mathpartir-array}

\usepackage{textcomp}

\usepackage{nicefrac}

\usepackage{stmaryrd}

\usepackage[clock]{ifsym}

\newsavebox{\naturalN}
\sbox\naturalN{$\mathbb{N}$}
\usepackage{bbold}

\usepackage{graphicx}

\usepackage{quoting}

\usepackage{booktabs}
\usepackage{dcolumn}
\newcolumntype{.}{D{.}{.}{-1}}
\usepackage[table]{xcolor}
\usepackage{multirow}
\usepackage{bigdelim}

\colorlet{bodycolor}{black}

\usepackage{wrapfig}
\makeatletter
\newcommand{\leftright}{%
  \ifnum `l=\WF@place left\else right\fi}
\makeatother

\usepackage[flushleft,alwaysadjust]{paralist}

\usepackage{stfloats}

\usepackage{nonfloat}
\usepackage{multicol}

\usepackage{amsmath}
\usepackage{amssymb}

\newenvironment{goal}[1]{%
  \noindent\textbf{Goal}~(#1).\itshape}{}
\newcounter{definition}
\newenvironment{definition}[1]{%
  \refstepcounter{definition}%
  \noindent\textbf{Definition~\thedefinition}~(#1).\itshape}{}

\usepackage[font=bf]{caption}
\DeclareCaptionLabelFormat{contfig}{Figure~#2 (continued)}
\usepackage[labelformat=simple,font+=normalsize]{subcaption}
\usepackage{url}

\usepackage{tikz}
\usetikzlibrary{calc}
\usetikzlibrary{shadows}
\usetikzlibrary{arrows.meta}
\usetikzlibrary{fit}
\usetikzlibrary{patterns}
\usetikzlibrary{decorations.shapes}
\usetikzlibrary{shapes.symbols}
\usetikzlibrary{fpu}
\makeatletter
\global\let\tikz@ensure@dollar@catcode=\relax
\makeatother

\pgfdeclarelayer{background}
\pgfdeclarelayer{annotation}
\pgfdeclarelayer{foreground}
\pgfsetlayers{background,annotation,main,foreground}

\usepackage{pgfplots}

\pgfplotsset{colormap={height}{
  color( 0)=(black!10)
  color(90)=(black!90)}}

\usepackage{bigstrut}
\usepackage{array}
\newcommand{\col}[1]{\ensuremath{\sql{#1}}}
\newcolumntype{H}{>{\columncolor{black}\color{white}}c}
\newcolumntype{P}{>{\columncolor{white}\color{white}}c}
\newcommand{\colhd}[1]{\multicolumn{1}{H}{\strut\col{#1}}}

\newcommand{\keyhd}[1]{\multicolumn{1}{H}{\strut\underline{\smash{\col{#1}}}}}
\newenvironment{littbl}{%
  \renewcommand{\arraystretch}{0.8}
  \renewcommand{\arraycolsep}{1pt}
  \renewcommand{\tabcolsep}{1pt}\ttfamily\ignorespaces}{}
\newcommand{\tabname}[2]{%
  \multicolumn{#1}{@{}l}{%
    \begin{tikzpicture}[baseline]
      \node [inner sep=1pt] (T) {\phantom{#2}};
      \draw [fill=black]
            (T.south west) {[rounded corners=2pt] -- (T.north west)  --
            (T.north east)} -- (T.south east) -- cycle;
      \node [inner sep=1pt,text=white] {#2};
    \end{tikzpicture}}}
\newcommand{\outputname}[2]{%
  \multicolumn{#1}{@{}l}{%
    \begin{tikzpicture}[baseline]
      \node [inner sep=1pt] (T) {\phantom{#2}};
      \draw [draw=black,fill=white]
            (T.south west) {[rounded corners=2pt] -- (T.north west)  --
            (T.north east)} -- (T.south east) -- cycle;
      \node [inner sep=1pt,text=black] {#2};
    \end{tikzpicture}}}

\usepackage{arydshln}

\usepackage{fancyvrb}
\usepackage{fancybox}
\usepackage{listings}
\makeatletter
\def\lst@visiblespace{\textnormal{\textvisiblespace\,}}
\makeatother
\lstMakeShortInline[basicstyle=\fontfamily\ttdefault\selectfont]|
\let\origthelstnumber\thelstnumber
\makeatletter
\newcommand{\suppressnumber}{%
  \lst@AddToHook{OnNewLine}{%
    \let\thelstnumber\relax%
     \advance\c@lstnumber-\@ne\relax}}
\newcommand{\reactivatenumber}{%
  \lst@AddToHook{OnNewLine}{%
   \let\thelstnumber\origthelstnumber%
   \advance\c@lstnumber\@ne\relax}}
\newcommand{\forcenumber}[1]{%
  \lst@AddToHook{OnNewLine}{%
   \let\thelstnumber\origthelstnumber%
   \setcounter{lstnumber}{\numexpr#1-1\relax}}}
\makeatother
\makeatletter
\lst@Key{countblanklines}{true}[t]%
    {\lstKV@SetIf{#1}\lst@ifcountblanklines}
\lst@AddToHook{OnEmptyLine}{%
    \lst@ifnumberblanklines\else%
       \lst@ifcountblanklines\else%
         \advance\c@lstnumber-\@ne\relax%
       \fi%
    \fi}
\makeatother
\newcommand{\SQLbuiltin}{\ttfamily}
\lstdefinelanguage{sql}{
  keywordstyle={[1]\bfseries},
  keywordstyle={[2]\SQLbuiltin},
  countblanklines=false,
  numberblanklines=false,
  morestring=[d]{'},
  morecomment=[l]{--},
  morekeywords={SELECT,FROM,WHERE,DISTINCT,ON,UNION,ALL,GROUP,BY,
    HAVING,SUM,MAX,RANK,FIRST_VALUE,WITH,RECURSIVE,VALUES,AS,AND,ROW,NOT,EXISTS,
    CREATE,OR,REPLACE,FUNCTION,RETURNS,LANGUAGE,SQL},
  morekeywords=[2]{dist,steps,line,generate_series,
    greatest,sqrt,degrees,atan,round,bool_or}
}


\usepackage[capitalise]{cleveref}

\crefname{section}{Section}{Sections}
\crefname{subsection}{Section}{Sections}
\crefname{subsubsection}{Section}{Sections}
\crefname{appendix}{Appendix}{Appendices}
\crefname{equation}{Equation}{Equations}
\crefname{figure}{Figure}{Figures}
\crefname{table}{Table}{Tables}
\crefname{subfigure}{Figure}{Figures}
\crefname{subtable}{Table}{Tables}
\crefformat{item}{\itshape #1.}
\creflabelformat{rule}{(\oldstylenums{#1})}
\crefname{rule}{Rule}{Rules}

\usepackage{pgf}

\usepackage{microtype}

\newcommand{\SQL}   {S\kern-0.07emQ\kern-0.07emL}
\newcommand{\Pg}    {Postgre\SQL}
\newcommand{\TPCH}  {\mbox{TPC-H}}
\newcommand{\Perm}  {\textit{Perm}}
\newcommand{\GProM} {\textit{GProM}}
\newcommand{\tpchQ}[1]{\textit{Q#1}}

\newcommand{\sql}[1]{\textnormal{\ttfamily\textls[-73]{\obeyspaces#1}}\/}

\newcommand{\fn}[1]{\mathit{#1}}
\newcommand{\var}[1]{\sql{\itshape#1}}

\definecolor{prov1}{RGB}{45,162,214}
\definecolor{prov2}{RGB}{255,190,10}
\definecolor{prov3}{RGB}{255,76,79}
\newcommand{\gwherewhy}[2]{%
  \draw[rotate around={-45:#1},thick,draw=#2,fill=#2] #1+(0,1/3*\G) arc [start angle=90,end angle=270,radius=1/3*\G];
  \draw[rotate around={-45:#1},thick,draw=#2,fill=#2!25!white] #1+(0,-1/3*\G) arc [start angle=-90,end angle=90,radius=1/3*\G]}
\newcommand{\gwhy}[2]{\draw[thick,draw=#2,fill=#2!25!white] #1 circle [radius=1/3*\G]}

\makeatletter
\newcommand{\emphbf}[1]{\if\f@series b\textsl{#1}\else\emph{#1}\fi}
\makeatother
\newcommand{\where}{\emphbf{where}}
\newcommand{\Where}{\emphbf{Where}}
\newcommand{\why}  {\emphbf{why}}
\newcommand{\Why}  {\emphbf{Why}}

\tikzset{%
  mouse/.style={circle,inner sep=1pt,fill=black!75,text=white}}

\pgfdeclarepatternformonly{visible}{\pgfqpoint{0pt}{0pt}}{\pgfqpoint{4pt}{4pt}}{\pgfqpoint{3pt}{3pt}}%
{
    \pgfsetlinewidth{0.8pt}
    \pgfpathmoveto{\pgfqpoint{0pt}{0pt}}
    \pgfpathlineto{\pgfqpoint{3.1pt}{3.1pt}}
    \pgfusepath{stroke}
}

\pgfdeclarepatternformonly{stripes}{\pgfpoint{0cm}{0cm}}{\pgfpoint{1.2cm}{1.2cm}}{\pgfpoint{1.2cm}{1.2cm}}
{
  \foreach \i in {0.0, 0.2, 0.4, 0.6, 0.8, 1.0, 1.2}
  {
     \pgfpathmoveto{\pgfpoint{\i cm}{0cm}}
     \pgfpathlineto{\pgfpoint{1.2cm}{1.2cm - \i cm}}
     \pgfpathlineto{\pgfpoint{1.2cm}{1.2cm - \i cm + 0.1cm}}
     \pgfpathlineto{\pgfpoint{\i cm - 0.1cm}{0cm}}
     \pgfpathclose%
     \pgfusepath{fill}
     \pgfpathmoveto{\pgfpoint{0cm}{\i cm}}
     \pgfpathlineto{\pgfpoint{1.2cm - \i cm}{1.2cm}}
     \pgfpathlineto{\pgfpoint{1.2cm - \i cm - 0.1cm}{1.2cm}}
     \pgfpathlineto{\pgfpoint{0cm}{\i cm + 0.1cm}}
     \pgfpathclose%
     \pgfusepath{fill}
   }
}
\colorlet{pixel}{black!40}
\colorlet{nopixel}{black!10}

\newcommand{\msquare}[4]{%
  \begin{tikzpicture}[baseline,x=2.5pt,y=2.5pt]
    \fill[\ifx\empty#1\empty nopixel\else pixel\fi] (0,0) rectangle +(0.9,0.9);
    \fill[\ifx\empty#2\empty nopixel\else pixel\fi] (1,0) rectangle +(0.9,0.9);
    \fill[\ifx\empty#3\empty nopixel\else pixel\fi] (0,1) rectangle +(0.9,0.9);
    \fill[\ifx\empty#4\empty nopixel\else pixel\fi] (1,1) rectangle +(0.9,0.9);
  \end{tikzpicture}}

\newcommand{\pointX}{\ensuremath{\boxtimes}}
\newlength{\G}

\newcounter{rule}
\crefname{rule}{Rule}{Rules}
\creflabelformat{rule}{\textsc{#1}}
\newcommand{\thisisrule}[1]{%
  \renewcommand{\therule}{\textsc{#1}}\refstepcounter{rule}(\textsc{#1})}

\newcommand{\comp}{\Mapsto}
\newcommand{\fresh}{\fn{site}()}

\newcommand{\nvdots}{%
  \ensuremath{\vcenter{\offinterlineskip\hbox{$\cdot$}\vskip-0.35ex\hbox{$\cdot$}\vskip-0.35ex\hbox{$\cdot$}}}}

\newcommand{\rowid}[2] {\ensuremath{{#1}_{#2}}}
\newcommand{\row}{\ensuremath{\rho}}
\newcommand{\Rho}{\ensuremath{\mathrm{P}}}
\newcommand{\ptype}    {\ensuremath{\mathbb{P}}}
\newcommand{\pempty}   {\ensuremath{\varnothing}}
\newcommand{\abs}[1]{#1}
\newcommand{\Y}[1]{\pgfsetfillopacity{0.4}#1\pgfsetfillopacity{1}}
\tikzstyle{circled}=[shape=circle,scale=0.75,solid,draw,font=\normalfont,color=bodycolor,inner sep=0.5pt,line width=0.1ex]
\newcommand{\circled}[1]{\tikz[baseline=(char.base)] {\node[circled] (char) {#1};}}
\newcommand{\location}[1]{\circled{\ensuremath{#1}}}
\newcommand{\logwrite}[1]{\ensuremath{\fn{write}_{\sql{#1}}}}
\newcommand{\logread}[1] {\ensuremath{\fn{read}_{\sql{#1}}}}
\newcommand{\logtable}[1]{\ensuremath{\fn{log}_{\sql{#1}}}}
\tikzstyle{Circled}=[circled,fill=black,text=white]
\newcommand{\Circled}[1]{\tikz[baseline=(char.base)] {\node[Circled] (char) {#1};}}
\newcommand{\toY}{\var{Y}}
\newcommand{\varY}[1]{\var{Y\textsubscript{#1}}}
\newcommand{\partition}{\sql{part}}
\newcommand{\peer}{\sql{rank}}
\newcommand{\ph}[1]{\mathbb{#1}}
\newcommand{\phase}[2]{\smash[t]{\ensuremath{#2^{\ph{#1}}}}}
\newcommand{\twophase}[2]{\langle#1,#2\rangle}
\newcommand{\inphase}[2]{\smash[t]{#2^{\ph{#1}}}}

\DeclareMathOperator*{\setagg}{\bigcup}

\newlength{\circlesize}   %% runtime p1+p2,  p1+p2y
\newlength{\squaresize}   %% runtime norm, p1
\newlength{\barsize}      %% p1 log size
\pgfmathsetmacro{\ci}{1.6}
\pgfmathsetmacro\sq{sqrt(pi * \ci * \ci) / 2}
\tikzstyle{legend}=[rotate=90,right,yshift=0.5pt,font=\tiny]
\tikzstyle{ratio}=[font=\tiny]
\tikzstyle{band}=[line width=0.4pt,densely dotted]
\tikzstyle{trend}=[black!50,-{Computer Modern Rightarrow[width=2pt,length=2pt]},shorten >=1pt]
\tikzstyle{cat}=[black!40,line width=0.4pt,densely dashed,font=\tiny]
\tikzstyle{sketch}=[baseline=0,x=1mm,y=1mm]
\pgfkeys{/pgf/number format/.cd,relative*={1},relative style=fixed}

\newif\ifarXiv\arXivtrue %% false ≡ VLDB camera-ready version

\ifarXiv
\toappear{This paper is an extended version of an article published in the
  Proceedings of the VLDB Endowment (PVLDB, 11(11), August~2018).}
\makeatletter\def\@mkbibcitation{\vskip2cm}\makeatother
\else
\fi

\newif\ifrevision\revisionfalse

\ifrevision
\arXivtrue
\colorlet{changed}{blue!90!black}
\colorlet{bodycolor}{black}
\newenvironment{changed}{%
  \color{changed}\colorlet{bodycolor}{changed}\ignorespaces}{%
  \color{black}\colorlet{bodycolor}{black}\ignorespaces}
\else
\colorlet{changed}{black}
\colorlet{bodycolor}{black}
\newenvironment{changed}{}{}
\fi

\begin{document}

%% ....................................................................
%% VLDB revision
\ifrevision
\pagenumbering{roman}

%% reviewer comments
\newcommand{\reviewersaid}[1]{{\itshape #1}}

\twocolumn[%
  \begin{center}
    \ttlfnt
    You Say `What', I Hear `Where' and `Why' --- \\
    \vskip1ex
    (Mis-)Interpreting \SQL{} to Derive Fine-Grained Provenance \\
    \vskip1.5em\aufnt
    Response to Reviewers (VLDB~2018)
    \vskip1.5em
  \end{center}
]

\smallskip\noindent
We would like to thank all three reviewers for their observations,
suggestions, and pointers to additional literature.  This input
has greatly helped to improve the presentation and contents of the
present paper.  Additional experiments and analyses have been performed
and parts of the material (\emph{e.g.}, the inference rule set and our
discussion of logging) have been significantly extended.  We respond to
all of your comments below.

The suggestions of the reviewers have prompted us to expand and reorganize
aspects of the paper.  An extended version of this work that
contains additional discussion, experiments, and figures in an appendix
is now available
at \emph{arXiv} via
\begin{center}
\url{https://arxiv.org/abs/1805.11517} .
\end{center}
The published version of the paper will point interested readers to
this expanded material.  For your convenience, we have included this
appendix in the present document.

\smallskip\noindent
\begin{changed}
To aid quick reference, we have highlighted sections and paragraphs with changed or
added text in the main paper as well as the appendix.  Further,
eight figures have been added.  Their captions have been color-highlighted,
too.
\end{changed}

\section*{Meta Review}

\begin{itemize}
\item \reviewersaid{In the experimental evaluation, a more
  detailed explanation is needed for the results. Also, a comparison
  against Perm is needed.}

  We have entirely rewritten those parts of the paper that analyse the
  performance of the approach.  A categorization of queries regarding
  their expected costs in Phase~$\ph{1}$ and~$\ph{2}$, a suggestion of
  Reviewer~2, has helped to structure the discussion.  Please see the
  new~\cref{sec:performance-phase1,sec:performance-phase2}.  The new
  scalability experiment (\TPCH{} at scale factor~$10$), discussed
  in~\cref{app:tpch-ten-gb}, adds further aspects of provenance
  derivation over large database instances.

  \cref{sec:perm} contains a direct comparison of our row-level \SQL{}
  interpretation with \Perm.  Here, too, we have now introduced classes
  of queries that greatly help to understand the performance differences
  between the two approaches.  In particular, please
  see~\cref{fig:perm-row-performance} which shows how the advantage of
  interpretation significantly grows with query complexity.  Our
  responses to comments~\emph{1.1} and~\emph{1.6} of Reviewer~2 address
  these points.

\item \reviewersaid{The handling of nested queries should be
  better articulated.}

  Please see our response to comment~\emph{D1} of Reviewer~1.  Query
  shape preservation and a compositional translation enable our approach
  to deliver provenance results for intermediate and nested queries just
  like for the overall top-level query.  Notes about this have been
  added to the paper in~\cref{sec:abs-int}.  Further, the appendix
  gave us the opportunity to show an extended inference rule set that
  now also includes the treatment of (non-)correlated subqueries (see~\cref{app:rules}
  for the added rules and their discussion).

\item \reviewersaid{A more in-depth discussion is needed about the
  overheads for reading and writing logs.}

  We have significantly extended our discussion of log files, their
  exact contents, and their relational implementation (this includes
  indexing and the actual writing/reading realized in terms of \SQL{}
  UDFs).  This new material is found in~\cref{app:log-files}.
  Please also see our responses to comment~\emph{D3} of Reviewer~1 and
  the last comment of Reviewer~3.

  The expected logging effort is tightly coupled with the subject
  query's selectivity, a characteristic that we have now used to
  categorize queries in~\cref{sec:performance}.  As decscribed there,
  logging incurs side effects in Phase~$\ph{1}$.  While this strategy
  aids in preserving the shape of the subject query, it is valid to ask
  how the approach would fare if we abandon side effects.  We have
  experimented with side-effect-free logging, or~\emph{tupling}, and
  found promising performance benefits for some queries. At the same
  time, \emph{tupling} may lead to the duplication of work at query
  runtime, a drawback encountered by \Perm{} and \GProM{} in a very
  similar fashion.  Please see~\cref{sec:performance-phase1} and our
  response to comment~\emph{1.3} of Reviewer~2.

\item \reviewersaid{The differences/similarities between the proposed
  approach and Perm/GProm should be better articulated.}

  Throughout the paper, \Perm{} and \GProM{} are the primary points of
  reference since these system pursue a very similar goal: provenance
  derivation for \SQL{} dialects actually found in practice.
  \cref{sec:from-cell-to-row,sec:perm} detail how \SQL{} interpretation
  and \Perm{}/\GProM{} deviate in their approaches:
  \begin{compactitem}
  \item our \textbf{coverage of \SQL{}} extends beyond \Perm{}'s (\emph{e.g}, when
    it comes to recursion and complex types) and \GProM{}'s (\emph{e.g.}, regarding
    subqueries),
  \item we derive provenance at the \textbf{finer cell-level of granularity} (but can
    gracefully and efficiently scale back to \Perm{}'s row-level, see~\cref{sec:from-cell-to-row}),
  \item we pursue a \textbf{non-invasive \SQL{}-level approach} that, unlike \Perm{}
    and \GProM{}, does not reach inside the RDBMS,
  \item query instrumentation and interpretation aim to \textbf{preserve query
    shape} (as opposed to \Perm{}'s wide-table approach that may duplicate
    computation which has performance implications),
  \item we employ \textbf{dependency sets to represent provenance} (as opposed to
    \Perm{}'s fully normalized redundant representation that may lead to
    a severe size blow-up for nested queries, see~\cref{sec:perm}),
  \item we still may employ \GProM-style \textbf{provenance-aware optimizations}
    although we can formulate these as local and simple \SQL{} (as opposed
    to RDBMS-internal algebraic) rewrites (\cref{sec:perm}) and, lastly,
  \item \SQL{} interpretation derives provenance with \textbf{lower space overhead}
    (also see~\cref{fig:provenance-rep-size}) and \textbf{lower cost} (across \TPCH{},
    interpretation incurs a mean slowdown of~$5.1$ while \Perm{} imposes
    a factor of~$18.9$).
  \end{compactitem}
  Please also see our response to comment~\emph{1.7} of Reviewer~2.
\end{itemize}

\section*{Reviewer~1}

\begin{itemize}
\item \reviewersaid{D1: To my understanding, provenance relates a
query's individual input and output data items helps to not only build
trust in query results, but also support back propagation from the
output to the input so that possible input error or wrong queries can be
identified. So, when a query is nested, it might involve some
intermediate relations. In this case, does the provenance result (e.g.
shown in Figure 16(c) and (d)) cover only cells/rows in the source
tables or paths along the source cells/rows, intermediate cells/rows to
the result?}

This is a very relevant question in the context of this work since an
understanding of the intermediate results (of subqueries) can be vital
to understand the overall logic of and data flow in a complex \SQL{{}}
query.  Since our approach adopts \emph{query shape preservation}
(see~\cref{sec:parametricity}), \emph{a nested subquery will remain a
nested subquery} after instrumentation (Phase~$\ph{1}$) and
interpretation (Phase~$\ph{2}$).  Just like for the top-level query, the
interpretation of each subquery will yield a table of dependency sets that
holds the \where- and \why-provenance for the subquery's results.  The approach
thus delivers provenance for all intermediate results.

To make such intermediate provenance tangible and immediately visible, users may use
\SQL{}'s \sql{WITH} clause (common table expression) and assign an
explicit table name, say~$r$, to any subquery of interest.  After
interpretation, the provenance for the intermediate result will be
available in table~$\inphase{2}{r}$.  See, for example, common table
expression~$\inphase{2}{\sql{max\_scan}}$
in~\cref{fig:visibility-phase2}, whose evaluation will yield the
provenance of the intermediate \emph{max scan} that lies at the heart of
the visibility query example. We have added notes about this important
observation to the discussion of inference~\cref{rule:with}
(\cref{sec:abs-int}) where the required notions are in context.

\item \reviewersaid{D2: It is hard to evaluate the proposed solution.
Is it only a basic solution, or an optimal one? Some discussions are expected.}

The focus of the present work is to be
\begin{compactitem}
\item \textbf{comprehensive} with respect to its coverage of modern
  \SQL{} dialects and its ability to explain the full \where- and
  \why-provenance of complex \SQL{} queries,
\item \textbf{practical}, as it is designed as a non-invasive source-to-source
  transformation, readily implementable on a wide range of existing \SQL{} RDBMSs, and
\item \textbf{efficient}, with its careful design to preserve the shape of
  data and queries that aims to not overwhelm query optimizers and engines.
\end{compactitem}

For most of the benchmarked queries, we (significantly) improve on
\Perm{}---as the major existing proposal in this space---regarding runtime and
scalability (please also see our response to comment~\emph{1.2} of Reviewer~2), but
we make \emph{no claims} regarding optimal performance. In fact, as a
non-invasive source-level transformation approach, we are in the hands
of the underlying database query engine.  There are a number of ways in
which the runtime performance of the approach can be improved and we
discuss
\begin{compactitem}
\item the use of tupling instead of logging in Phase~$\ph{1}$ (a new
  discussion in~\cref{sec:performance-phase1}),
\item provenance-aware rewrites (inspired by \GProM) in~\cref{sec:perm}, and
\item dependency set representations and indexed log files in
 the new appendix (see~\cref{app:dep-set-rep,app:log-files-relational}).
\end{compactitem}

\item \reviewersaid{D3: More discussions and experiments related to log writing and
reading are expected, so that we can understand the detail of provenance in Phase 1.}

Your comment, quite rightly, touches upon the important role that
the log files assume in our work.  In the course of this revision
we have thus added $1\,\nicefrac{1}{2}$ pages of text and three
new figures that motivate and discuss the exact log file contents---also illustrated
using three deliberately simple sample queries---and how these are
used during interpretation.  You will find this
new material in~\cref{app:log-files}.  This also includes a subsection
on the details of one possible relational implementation of log files, writing, and reading (it is
this implementation that was used in all the experiments we report on in this
paper). Please also see our response to the last comment of Reviewer~3.

\item \reviewersaid{D4: Data provenance has been studied in many
literatures. The authors only compare their work with Perm. Some latest
techniques should be considered in experiments. [Literature suggestions.]}

We thank you for pointing out these references that contain a
recent survey of the provenance research landscape~\cite{what-for-form-from}
and two overviews of Glavic'\ \emph{et al.} work on provenance for \SQL{},
covering Perm~\cite{perm-festschrift} and \GProM~\cite{gprom-overview}.  All
references have been added to the present paper.

The Festschrift contribution on \Perm~\cite{perm-festschrift} contains
slightly updated \SQL{} rewriting rules and thus provides an especially
welcome addition to the list of references.  The nature of the rules and
thus Perm's approach to rely on (very) wide tables to represent
provenance information remains, however.  For common query types,
grouping for example (see Rule~(\textbf{R4}) in Figure~9
of~\cite{perm-festschrift}), \Perm{} duplicates computation and represents
provenance in a normalized, redundant fashion.  The present work, instead,
subscribes to data and query shape preservation, aiming to
\begin{inparaenum}[(1)]
\item avoid performance problems faced by \Perm{} and
\item fare without specific---and invasive---provenance-aware query
  optimizations (\emph{instrumentation choices}) as proposed by \GProM.
\end{inparaenum}

We have extended our comparison with the work of Glavic' \emph{et al.} and
placed the discussion in the new~\cref{sec:perm}.  We have further contacted
Boris Glavic to discuss the performance advantage of \GProM{} over \Perm{}.
Please see our response to comment~\emph{1.7} of Reviewer~2:
even if the advances of \GProM{} are considered, \SQL{} interpretation still
implements provenance derivation at lower (space and time) cost.

\end{itemize}

\section*{Reviewer~2}

\begin{itemize}
\item \reviewersaid{1.1) It would be helpful to have a better understanding
on the types of queries that result in higher overheads. For instance,
do these queries have a high selectivity? Do they have complex nested
subqueries? While authors provide some explanations, it is still not
clear to me which types of query have a higher/lower overhead.
Categorizing these queries could help.}

Query categorization has been a helpful suggestion that greatly supported
us in restructuring the discussion in~\cref{sec:performance}.
Such a categorization can now be found in~\cref{fig:query-buckets}.  The
associated discussion is found in the
new~\cref{sec:performance-phase1,sec:performance-phase2} which---just
like the categorization---assess causes of slowdown and
speed-up separately for Phases~$\ph{1}$ and~$\ph{2}$.  We have categorized all 22~queries
of \TPCH{} but the categories are not specific to the benchmark
and apply to \SQL{} queries in general.
Established query properties
like selectivity, just like you pointed out in your comment, indeed are
a determining factor in Phase~$\ph{1}$, for example.  Other
query characteristics, \emph{e.g.}, the degree of ``connectedness'' of input
and output cells (\emph{i.e.}, the dependency set cardinality), are more specific to
the present work.   The presence of nested queries plays a
significant role in our comparison with \Perm{}---please see our response
to your comment~\emph{1.6} below.

\item \reviewersaid{1.2) The scale factor of TPC-H is set to 1, which
generates approximately 1GB of data. This is very small compared to many
real use cases. I wonder what is the performance like for scale factors
of 100 (100GB) or 1000 (1TB). Even better, how does the performance
changes for varying data sizes? Does the overall slowdown increase
linearly?}

A discussion of the scalability of \SQL{} interpretation was indeed missing
from the original submission.  We have taken this revision opportunity
to repeat all experiments and are now reporting the log sizes, the contribution
of the individual phases, as well as the overall slowdown for the
\TPCH{} benchmark scale factors $\mathit{sf} = 1$ (in~\cref{fig:performance})
and $\mathit{sf} = 10$ (in~\cref{fig:performance-ten-gb} in the
appendix). The new~\cref{app:tpch-ten-gb} discusses our observations
regarding scalability.  In a nutshell, we indeed find that log sizes and
Phase~$\ph{1}$+$\ph{2}$ run times scale linearly with the growing \TPCH{}
data volume.  In effect, the \emph{slowdown remains nearly constant}.
We think this convincingly documents the value of query shape preservation.

\smallskip\noindent
To the best of our knowledge, the present work now is the first to report
on results for \where- and \why-provenance derivation for the entire
\TPCH{} benchmark at scale factors~$1$ and~$10$.  The most recent
(still unpublished, see below) work on \Perm{} and \GProM{} reports measurements for
only half of the queries.

(Please see \emph{``Heuristic and Cost-Based Optimization for
Diverse Provenance Tasks''} by Niu, Kapoor, Glavic, Gawlick, Liu, Krishnaswamy,
and Radhakrishnan. IEEE~TKDE, to appear in~2018, also included as reference~\cite{gprom-opt}.)

\item \reviewersaid{1.3) Not having the approach attached to the query engine makes it less invasive and more applicable to other
RDBMSs, but at the same time some queries  may be poorly
optimized. Authors describe some
example of this on page 9 (under "Logging overhead in Phase 1"), but I
think the trade-offs of this decision (if any) must be better discussed
and emphasized in the paper. For instance, is there any type of query
that would better benefit from a provenance collection approach that
leverages the query optimizer?}

We do think that the non-invasiveness is one defining characteristic of
our approach---not the least because \Perm{} and \GProM{} have opted to
reach into the internals of the underlying RDBMS.  Our focus on data and
query shape preservation is in fact motivated by our goal to support
the query optimizer in identifying efficient plans for instrumented and
interpreting queries.  We think that the scalability results reported
in~\cref{app:tpch-ten-gb}---also prompted by your comment~\emph{1.2}, so
thank you---underline the value of that goal.  In this sense, the query
optimizer indeed \emph{is} leveraged.

As you point out, in our present discussion shape preservation implies
log writing and thus side effects that can be challenging for the
optimizer during plan generation.  While even the most complex queries
feature only a small number of log~\logwrite{$\Box$} call sites (up to a
maximum of~$4$ but $2.5$ on average in~\TPCH{}, see the
new~\circled{\phantom{2}} annotations
in~\cref{fig:performance,fig:performance-ten-gb}), any side effect
avoided surely helps.  We have thus explored side-effect-free logging in
the new~\cref{sec:performance-phase1}.  Such a modification of the
approach could rely on~\emph{tupling} and indeed shows significant
performance advantages in Phase~$\ph{1}$. However, tupling sacrifices
query shape preservation and may lead to the duplication of
computation---incidentally much like in \Perm{} and
\GProM{}---\emph{e.g.}, when scalar or set-based subqueries are to be
processed.   Please see the associated discussion
in~\cref{sec:performance-phase1}.  We thus consider a conditional use of
side-effect-free logging a very interesting avenue for future work.

\item \reviewersaid{1.4) The overhead in space is presented in terms of
number of cells, but it would be better to see this in terms of bytes to
understand the real impact.}

We agree that a log size in  number of bytes (instead of the number of
table cells) provides a measure that is more immediately understood by
the reader. \cref{fig:performance} has been updated accordingly.
Additionally, we have augmented this log size information by the number
of \logwrite{$\Box$} call sites that populate these logs, now displayed
in~\circled{\phantom{2}} in the same figure.  We observe that, on
average, a \TPCH{} query features $2.5$~constructs that make
provenance-relevant value-based decisions.  These measures have also
been added to~\cref{fig:performance-ten-gb} in~\cref{app:tpch-ten-gb}
in which we report on a \TPCH{} scalability experiment (scale factor~$10$).
You will see that the log sizes grow linearly with the benchmark's scale.

\item \reviewersaid{1.5) If the bit set representation performed better
than the native one, why wasn't the bit set representation used for the
row-level representation? This representation cannot be adapted for the
row-level? This is not clear.}

The bit set representation provides a general implementation of
type~\sql{int\;MULTISET} (see the~\SQL:2003 standard~\cite[\S\,4.10.3]{sql-2003})
and is perfectly applicable to the derivation of both cell-based and
row-based provenance.  In our comparison with \Perm{} we stick to the
array-based implementation of dependency sets to make the point that an
all-native non-invasive implementation of \SQL{} provenance is a match
for---and can (significantly) outperform---\Perm's approach that reaches
quite deeply into the innards of \Pg.

In fact, your valuable comments on \GProM{} have prompted us to move the
discussion of the non-native bit sets into~\cref{app:dep-set-rep}. That
discussion remains interesting and provides performance benefits but
indeed is orthogonal to our approach.  Instead, we now discuss a \GProM-style
``provenance-aware'' optimization that is easily expressible on the
\SQL{} source level (\emph{i.e.}, natively) and thus is much more in
the spirit of the present work.  Also, these language-level
optimizations have a much more profound impact than bitmap-based
dependency sets.  Please see our response to your comment~\emph{1.7} and~\cref{sec:perm}.

\item \reviewersaid{1.6) It wasn't clear to me why Perm performed better
for some queries (e.g.: Q3, Q6, Q10, etc.). Authors mention that "Perm
sees an advantage if small provenance cardinality keeps the
representation overhead in check", but Q19 has a smaller cardinality
than Q3, and the proposed approach still performs better than Perm for
Q19. More detailed explanations for the different queries is needed.
Perhaps a categorization of the types of queries (see 1.1) could help
here as well.}

In this revision we have rewritten our discussion of the head-to-head
comparison of row-level \SQL{} interpretation with \Perm{}.  Please
see~\cref{sec:perm} and~\cref{fig:perm-row-performance}, in particular.
Following your comment, we have sorted the queries of the
\TPCH{} benchmark into categories of increasing complexity, from simple
scans or narrow joins with aggregation, over joins of increasing width
with grouping and aggregation, to complex queries with (non-)correlated
subqueries.  The resulting figure and its discussion show how the advantage of
row-level \SQL{} interpretation over \Perm{} increases with query complexity.  The
presence of subqueries, in particular, strongly amplifies the space
overhead of \Perm's fully normalized provenance representation.
We think that this categorization adds structure to the comparison
with \Perm{} and thus is more useful to the reader.  We thank you for
suggesting it.

\item \reviewersaid{1.7) I'm assuming GProM wasn't used in the
experiments because it does not support PostgreSQL and nested subqueries
yet, but it seems to me that it would perform better than Perm for some
of the queries. Any intuition or idea on how GProM and the proposed
approach would differ in terms of performance?}

The ability to cover a large subset of \SQL{} is one of the main
motivations behind the present work.  As you correctly say, \Perm{}
and its fully fleshed out implementation thus has been the go-to
system for a direct comparison.  \GProM{} has made interesting advances
(\emph{e.g.}, regarding provenance for transactions and \why-\emph{not} provenance)
but, to this day, only supports a quite limited \SQL{} dialect: only
half of the queries of the \TPCH{} benchmark are executable at all,
primarily due to the absence of any support for subqueries, as you
have pointed out.

Most interestingly, however, \GProM{} also proposes to integrate
\emph{provenance-aware optimizations} (or: instrumentation choices) into
the underlying RDBMS. With these---partially heuristic, partially
cost-based---algebraic rewrites,
\GProM{} reaches even deeper into the RDBMS host than \Perm{}.
We have contacted Boris Glavic to ask whether he knows of any direct
performance comparison of \Perm{} and \GProM{}. According to
him, such a study does not exist.  He added that, for some \TPCH{}
queries, experiments with provenance-aware optimizations showed a
speed-up of query evaluation by ``a factor of~$2$ to~$3$.''  Figure~12
in~\cite{gprom-opt} indicates, that \TPCH{} query~\tpchQ{5} is one~\sql{GROUP\;BY}
query that exhibits such an improvement.  For this query, row-level \SQL{} interpretation
shows an improvement of a factor of~$10$ over \Perm{} (see~\cref{fig:perm-row-performance}).
More generally, while \Perm{} incurs a mean slowdown of~$18.9$ across \TPCH{},
\SQL{} interpretation slows down query evaluation by a factor of~$5.1$
(see~\cref{sec:perm}). This indicates that, even with the
provenance-aware optimization of
\GProM{} enabled, \SQL{} interpretation does incur a lower provenance
tax overall.

However, \GProM-style provenance-aware optimizations still remain a
valid and very interesting idea that also applies to our work.  True to
the spirit of a non-invasive approach, we opt for a variant of the idea
that is easily expressible as a source-to-source transformation on the
\SQL{} language level without reaching into the RDBMS.   With the
revision we have added the discussion of a dependency-set-specific local
\SQL{} rewrite that relates to non-correlated subqueries.  For selected
\TPCH{} queries, this can speed up interpretation considerably (by far
more than a factor~$2$ to~$3$).  Please see~\cref{sec:perm}.

\item \reviewersaid{2) Some related work is missing. I missed the following
related work: [Literature suggestions]}

We would like to thank the reviewer for these suggestions of further
related work, of which three (out of four) have now indeed found their way into
the paper.  You will find these listed as references~\cite{uldbs},
\cite{proql}, and~\cite{smoke-lineage}.

\emph{``Querying Data Provenance''}~\cite{proql} discusses beneficial effects when
access support relations are used to materialize provenance relationships
between rows.  This is very similar to what we observe when we read
join partners off (the tabular implementation of) \logtable{JOIN} log
files: these logs pair provenance-related rows and Phase~$\ph{2}$
effectively uses the log like an access support relation (or join index).
Referenced in~\cref{sec:performance}.

There are aspects of the very recent paper \emph{``Smoke: Fine-Grained Lineage at
Interactive Speed''}~\cite{smoke-lineage} that are indeed closely related
to our work.  An immediate connection exists with their use of
\emph{1-to-N rid indexes} (to represent \sql{GROUP\;BY} provenance) which
\emph{directly} compare to the concept and implementation of our
\logtable{GRP} log files.  A discussion of the exact contents of these
files and their close relationships with the \emph{rid indexes}
of~\cite{smoke-lineage} is found in the
new~\cref{app:log-files-relational}.

Smoke's design decision to piggyback on query evaluation to populate
data structures that represent lineage bears close resemblance to our
side-effecting query instrumentation in Phase~$\ph{1}$.  Smoke's use of
indexes to represent provenance relationships is an inspiring idea that
we plan to pursue further in our context.  Populating an index
constitutes an intrinsic RDBMS-supported side effect (on a hash table or
B$^+$tree, say) that will be more efficient to carry out than the explicit invocation
of a table-updating \SQL{} UDF \logwrite{$\Box$}.

A performance-based comparison of Smoke and the present approach appears
difficult, however: first, the experimental report on Smoke only implements
four (all of which are rather simple, see our categorization
in~\cref{fig:perm-row-performance}) of the 22~\TPCH{} queries. Second,
Smoke relies on a purposely built in-memory engine in which lineage
capture---\emph{i.e.}, row-based
\where-provenance---is tightly integrated with physical plan operators
(projection, selection, hash-join, grouping and aggregation).  The
present work is located in a quite different spot of the design space:
derivation of cell-based
\where- and \why-provenance through source-level query rewriting,
non-invasively implemented on top of existing RDBMSs, aimed to capture
very rich \SQL{} dialects.

\end{itemize}

\section*{Reviewer~3}

\begin{itemize}{}
\item \reviewersaid{The approach finally appeared to me to be
interesting. But I had difficulties to apprehend the authors' vision.
The presentation is a little too descriptive w.r.t. my taste. Different
elements are mixed throughout the article: the illustrative example (by
the way it is well chosen and interesting), general considerations on
provenance, kind of provenance sought, approach and implementation. At
least, the tackled problem could be specified in an easy findable way.
Main concepts as "normalized query", "transformed query", "trace" or
"log files", "interpreter" could be defined more formally. Beyond a
question of taste, this could clarify concepts and help the reading.}

We are grateful for this input since readability (\emph{i.e.}, details
like the selection of examples, consistency of notation, illustrative
figures) is very important to us.  Your suggestion to better
identify/define the cornerstones of this work thus is welcome.  We have
added \textbf{Goal} and \textbf{Definitions} call-outs throughout the text
and have used those to highlight and/or define concepts like
\begin{compactitem}
\item cell-level \where-/\why-provenance for \SQL{} (the computation of which
  constitutes the paper's main goal),
\item dependency set,
\item normalized query,
\item instrumented query,
\item interpreted query, and
\item the core syntactic mapping~$\comp$.
\end{compactitem}
The definitions introduce the sections in which the associated details
are then developed.

\item \reviewersaid{The solution is not fully developed. For example,
some inference rules are missing to cover the benchmarck. If space is
missing, a reference to a complete research report is welcome.}

The inference rule set had been fully developed at the time of
submission of the original paper but you are correct with your
observation that the included subset of rules did not cover the \TPCH{}
benchmark queries. The addition of the appendix provided the opportunity
to rectify this: taken together, the rules of~\cref{fig:rules}
and~\cref{fig:rules-extended} (in~\cref{app:rules}), now cover
\begin{compactitem}
\item the running visibility query example of~\cref{fig:visibility-query},
\item all 22~queries of the \TPCH{} benchmark, plus
\item further advanced \SQL{} constructs, like recursive common table
  expressions (\sql{WITH RECURSIVE}).
\end{compactitem}
The rule subset in the main~\cref{fig:rules} still covers the core rules
that convey all essential ideas behind mapping~$\comp$.  The rule extensions
add constructs like \SQL{}'s \sql{CASE}~expression, (\sql{LEFT}) \sql{OUTER JOIN},
and a discussion of the uniform treatment of (non-)correlated subqueries.

\item \reviewersaid{Due to a dispersed description of log files, it is
uneasy to understand how many they are, what they contain, how they are
managed, etc.}

Since log files play a central role in this work, your comment has
prompted us to add an entirely new section in the appendix (see~\cref{app:log-files})
that clarifies
\begin{compactitem}
\item the exact log file contents (illustrated with the help of a family
  of three sample~\SQL{} queries) as written by Phase~$\ph{1}$,
\item how these contents are to be interpreted in Phase~$\ph{2}$, and
\item the details of a purely relational implementation of logging
  together with all needed \SQL{} routines to write to/read from these logs
  (see~\cref{app:log-files-relational}).
\end{compactitem}
The mentioned relational (or: tabular and indexed) logging is exactly
what has been used in the experiments of~\cref{sec:performance}.  This
section now also indicates the required number of log files for each
\TPCH{} query: see annotations like~\circled{2} at the bottom
of~\cref{fig:performance}.  We find that, on average, a \TPCH{} query writes
to~$2.5$ log files only. (Indeed, logging is only required when the original
query uses expressions of non-parametric type to make a value-based
decision, see~\cref{sec:parametricity}.)

Lastly, our treatment of log files
permeates~\cref{sec:from-values-to-dep-sets,sec:abs-int} of the text since
\begin{compactitem}
\item they are written in Phase~$\ph{1}$ and read from in Phase~$\ph{2}$ and as
  such need to be discussed in both phases, and
\item we decided to abstract from their implementation and separate
  their API (the~\logwrite{$\Box$} and~\logread{$\Box$} functions,
  see~\cref{sec:log-writing-reading}) from a concrete implementation.  A
  reader's reconstruction of our work is free to choose any
  im\-ple\-men\-tation---in an main-memory database context, for example, a
  very simple array-based realization is conceivable---and the
  relational encoding of~\cref{app:log-files-relational} is but
  one option.
\end{compactitem}
We hope that this treatment plus the extensive new material of~\cref{app:log-files}
paints a picture sufficiently complete that no details of logging remain in the
dark.
\end{itemize}

\clearpage
\pagenumbering{arabic}
\fi

%% ....................................................................
%% text body

\title{%
  You Say `What',~~I Hear `Where' and `Why' --- \\[1ex]
  (Mis-)Interpreting \SQL{} to Derive Fine-Grained Provenance}

\numberofauthors{1}
\author{%
  \alignauthor
  Tobias M\"uller
  \qquad
  Benjamin Dietrich
  \qquad
  Torsten Grust
  \\[1ex]
  \affaddr{Universit\"at T\"ubingen} \\
  \affaddr{T\"ubingen, Germany}
  \\[1ex]
  \email{[ to.mueller, b.dietrich, torsten.grust ]@uni-tuebingen.de}
}

\maketitle

\begin{abstract}
  \SQL{} declaratively specifies \emph{what} the desired
  output of a query is.  This work shows that a non-standard
  interpretation of the \SQL{} semantics can, instead, disclose
  \emph{where} a piece of the output originated in the input and
  \emph{why} that piece found its way into the result. We derive such
  data provenance for very rich \SQL{} dialects---including recursion,
  windowed aggregates, and user-defined functions---at the fine-grained
  level of individual table cells. The approach is non-invasive and
  implemented as a compositional source-level \SQL{} rewrite: an input \SQL{}
  query is transformed into its own interpreter that wields data
  dependencies instead of regular values. We deliberately design this
  transformation to preserve the shape of both data and query, which
  allows provenance derivation to scale to complex queries without
  overwhelming the underlying database system.
\end{abstract}

\section{Data Provenance Explains \\ Complex \SQL{} Queries}
\label{sec:provenance-explains-SQL}

\textbf{A complex \SQL{} query.}
In a hilly landscape, which marks are visible from your current
location?  That will depend on your position's altitude and the height
of the terrain around you: valleys are obscured by nearby ridges, while
peaks, even if remote, may still be in view.  The two-dimensional sketch
of~\cref{fig:terrain-2d} suggests one answer to the question: first, compute the
running maximum (or: \emph{max scan}) of view angles between our
location~\pointX{} and the ever farther hill tops before us.  Second, a mark is
visible iff its angle is at least as large as the maximum
angle~$\alpha_i$ we have measured so far.  We thus can spot the tree
(its view angle~$\alpha_3$ exceeds the current maximum of~$\alpha_2$)
while marks~$p_1$ and~$p_2$ are obscured.

\begin{figure}
  \newlength{\X}\setlength{\X}{2\G}
  \newcommand{\maxscan}[1]{\alpha_{\mskip-1mu#1}}
  \centering\small
  \begin{tikzpicture}[x=\X,y=\X,font=\scriptsize]
    %% height map
    \fill[color of colormap={200}] ( 1,0) rectangle ++(\X,1*\X);
    \fill[color of colormap={200}] ( 2,0) rectangle ++(\X,1*\X);
    \fill[color of colormap={200}] ( 3,0) rectangle ++(\X,1*\X);
    \fill[color of colormap={300}] ( 4,0) rectangle ++(\X,2*\X);
    \fill[color of colormap={400}] ( 5,0) rectangle ++(\X,3*\X);
    \fill[color of colormap={400}] ( 6,0) rectangle ++(\X,3*\X);
    \fill[color of colormap={400}] ( 7,0) rectangle ++(\X,3*\X);
    \fill[color of colormap={400}] ( 8,0) rectangle ++(\X,3*\X);
    \fill[color of colormap={200}] ( 9,0) rectangle ++(\X,1*\X);
    \fill[color of colormap={300}] (10,0) rectangle ++(\X,2*\X);
    \fill[color of colormap={700}] (11,0) rectangle ++(\X,6*\X);
    \fill[color of colormap={700}] (12,0) rectangle ++(\X,6*\X);
    \fill[color of colormap={700}] (13,0) rectangle ++(\X,6*\X);
    \fill[color of colormap={500}] (14,0) rectangle ++(\X,4*\X);

    %% visible spots
    \fill[pattern=visible,pattern color=white] ( 1,0) rectangle ++(\X,1*\X);
    \fill[pattern=visible,pattern color=white] ( 2,0) rectangle ++(\X,1*\X);
    \fill[pattern=visible,pattern color=white] ( 3,0) rectangle ++(\X,1*\X);
    \fill[pattern=visible,pattern color=white] ( 4,1) rectangle ++(\X,1*\X);
    \fill[pattern=visible,pattern color=white] ( 5,2) rectangle ++(\X,1*\X);
    \fill[pattern=visible,pattern color=white] (11,5) rectangle ++(\X,1*\X);
    % visible tree
    \begin{scope}[line width=0.4pt,xshift=1/2*\X,yshift=1/2*\X]
      \node[cloud, cloud puffs=7, aspect=1.2, inner sep=2pt, draw, name=tree] at (11,6.2) {};
      \draw (tree.south)+(-0.07,-0.03) .. controls +( 0.02,-0.3) ..  +(-0.15,-0.4);
      \draw (tree.south)+( 0.07,-0.03) .. controls +(-0.02,-0.3) ..  +( 0.15,-0.4);
    \end{scope}

    %% height legend
    \draw (15.5,0) -- (15.5,6);
    \foreach \h in {200,300,...,700}
      { \draw (15.5,-1.5+1/100*\h) -- ++(1mm,0) node[font=\tiny,right] {\h\,\textrm{m}};
      }

    \begin{scope}[xshift=1/2*\X,yshift=1/2*\X]
      %% observation point X
      \node[font=\Large] at (0,0) {\pointX};

      %% angle max scan
      \node at ( 1,-1) {$0\textnormal{\textdegree}$};
      \node at ( 2,-1) {$0\textnormal{\textdegree}$};
      \node at ( 3,-1) {$0\textnormal{\textdegree}$};
      \node at ( 4,-1) {$\maxscan{1}$};
      \node at ( 5,-1) {$\maxscan{2}$};
      \node at ( 6,-1) {$\maxscan{2}$};
      \node at ( 7,-1) {$\maxscan{2}$};
      \node at ( 8,-1) {$\maxscan{2}$};
      \node at ( 9,-1) {$\maxscan{2}$};
      \node at (10,-1) {$\maxscan{2}$};
      \node at (11,-1) {$\maxscan{3}$};
      \node at (12,-1) {$\maxscan{3}$};
      \node at (13,-1) {$\maxscan{3}$};
      \node at (14,-1) {$\maxscan{3}$};
    \end{scope}
    \draw[thick,->] (1,-1) -- (15.5,-1) node[pos=0.5,below] {\sql{MAX($\maxscan{}$)} scan};

    %% eye
    \begin{scope}[xshift=-10,yshift=10,scale=0.7,rotate=18]
      \draw[line cap=round] (0,0) -- +(xyz polar cs:angle= 30,radius=1);
      \draw[line cap=round] (0,0) -- +(xyz polar cs:angle=-30,radius=1);
      \draw (0.7,0) circle [y radius=0.38, x radius=0.2];
      \fill (0.8,0) circle [y radius=0.19, x radius=0.1];
    \end{scope}

    %% view angles
    \begin{scope}[line width=0.4pt,gray]
      \draw[dash pattern=on 3pt off 1pt] (0.1,1) -- ( 4,2);
      \draw[dash pattern=on 3pt off 1pt] (0.1,1) -- ( 5,3);
      \draw[dash pattern=on 3pt off 1pt] (0.1,1) -- (11,6);
      \draw (8.5,3) arc [start angle=0, end angle=28.00, radius=3.5*\X];
      \draw (4.5,2) arc [start angle=0, end angle=28.00, radius=1.5*\X];
      \draw (3.5,1) arc [start angle=0, end angle=14.036, radius=3.5*\X];
      \node at (3.1,1.35) {$\maxscan{1}$};
      \node at (4.1,2.3) {$\maxscan{2}$};
      \node at (8,3.8) {$\maxscan{3}$};
    \end{scope}

    %% invisible spots (since α < current max scan angle)
    \begin{scope}[xshift=1/2*\X,yshift=1/2*\X]
      \node at (10,2) {$p_1$}; % since α < α₂
      \node at (14,4) {$p_2$}; % since α < α₃
    \end{scope}

    %% point whose (maximum) angle is used in visibility computation of p1
    %% (=> where and why-dependency)
    %{\setlength{\G}{\X}\gwherewhy{(11.5,5.5)}{prov1};}
  \end{tikzpicture}
  \vskip-3mm
  \caption{Visibility in a two-dimensional hilly landscape:
    spots marked \protect\tikz\protect\fill[pattern=visible] (0,0) rectangle (0.25,0.25);
    are visible from~\pointX{}.  The max scan encounters the view
    angles~$0\textnormal{\textdegree} < \maxscan{1} < \maxscan{2} <
    \maxscan{3}$ from left to right.}
  \label{fig:terrain-2d}
\end{figure}

\begin{figure}[t]
  \begin{narrow}{-2mm}{-2mm}
  \small
  \subcaptionbox{%
    Height map with our location~\pointX{}.\label{fig:terrain-3d}}[0.73\linewidth]{%
    \centering
    \begin{tikzpicture}[x=\G,y=\G]
      %% axes
      \begin{scope}[very thin,font=\tiny]
        \begin{scope}[yshift=-\G]
          \draw (0,0) -- (20,0);
          \foreach \x in {0,5,...,20}
            { \draw (\x,0) -- (\x,-0.5) node[below] {\x};
            }
          \node[font=\scriptsize] at (21,0) {\col{x}};
        \end{scope}
        \begin{scope}[xshift=-\G]
          \draw (0,0) -- (0,20);
          \foreach \y in {0,5,...,20}
            { \draw (0,\y) -- (-0.5,\y) node[left] {\y};
            }
          \node[font=\scriptsize,anchor=base] at (0,21) {\col{y}};
        \end{scope}
      \end{scope}

      %% visibility mark
      \pgfdeclareplotmark{visible}{%
        \pgfpathrectangle{\pgfpoint{-1/2*\G}{-1/2*\G}}{\pgfpoint{\G}{\G}}
        \pgfsetfillpattern{visible}{white}
        \pgfusepath{fill}}
      %% height map
      \ifrendermap
      \begin{axis}[x=\G,y=\G,at={(0,0)},anchor={origin},hide axis]
        \addplot[%
          scatter,
          only marks,
          scatter src=explicit,
          scatter/use mapped color={fill=mapped color},
          mark=square*,mark size=1/2*\G,mark options={draw=white}]
          table[meta=h]
          {visibility-map-21x21-simpler.xyhv};
        \addplot[%
          scatter,
          only marks,
          scatter src=explicit symbolic,
          scatter/classes={true={mark=visible},false={mark=none}}]
          table[meta=visible]
          {visibility-map-21x21-simpler.xyhv};
      \end{axis}
      \fi

      %% legend
      \node[font=\scriptsize,xshift=1/2*\G,anchor=base] at (22,21) {\col{alt}};
      \foreach \h in {0,100,...,900}
        { \fill[color of colormap={\h}] (22,11+1/100*\h)++(0,-1/2*\G) rectangle ++(\G,\G);
        }
      \node[font=\tiny,right] at (23,11) {  0\,\textrm{m}};
      \node[font=\tiny,right] at (23,20) {900\,\textrm{m}};

      \fill[black!30] (22,8)++(0,-1/2*\G) rectangle ++(\G,\G);
      \fill[pattern=visible,pattern color=white] (22,8)++(0,-1/2*\G) rectangle +(\G,\G);
      \node[font=\tiny,right,align=left] at (23,8) {visible \\ from~\pointX};

      %% observation point X (cross)
      \node[font=\tiny] at (17,10) {\pointX};
    \end{tikzpicture}
  }%
  \subcaptionbox{%
    Table~\sql{map}.\label{fig:terrain-map}}[0.25\linewidth]{%
    \centering
    \begin{littbl}
      \begin{tabular}{|c|c|r|@{}c@{}}
        \tabname{3}{\strut{\col{map}}} \\
        \keyhd{x} & \keyhd{y} & \colhd{alt} \\
        \strut
        0        & 0        & 400.0                         \\
        0        & 1        & 400.0                         \\
        0        & 2        & 400.0                         \\
        $\nvdots$ & $\nvdots$ & \multicolumn{1}{c|}{$\nvdots$} \\
        17       & 10       & 200.0                         & \rlap{\,\pointX}  \\
        $\nvdots$ & $\nvdots$ & \multicolumn{1}{c|}{$\nvdots$} \\
        20       & 18       & 0.0                           \\
        20       & 19       & 0.0                           \\
        20       & 20       & 0.0                           \\
        \cline{1-3}
      \end{tabular}
    \end{littbl}
    \vspace*{8mm}
  }
  \end{narrow}
  \vskip-3mm
  \caption{Height map of three-dimensional terrain and its tabular
    encoding.  Again, spots marked~\protect\tikz\protect\fill[pattern=visible] (0,0) rectangle (0.25,0.25);
    are visible from~\pointX{}.}
  \label{fig:terrain}
\end{figure}
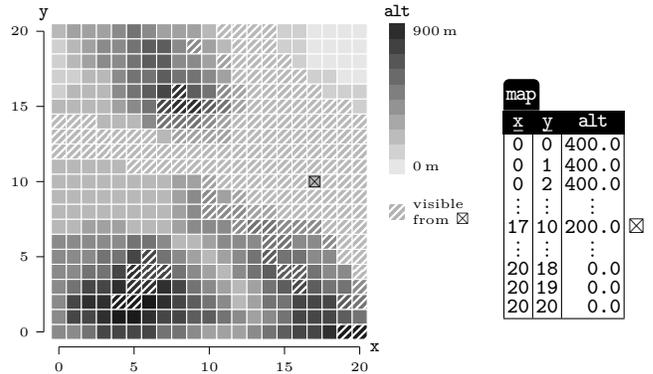

The \emph{max scan} technique does apply in three dimensions, but things
get a bit more complicated. \Cref{fig:terrain-3d} depicts the height map
of a sample terrain in which shades of grey indicate altitude
and~\pointX{} at $(x,y) = (17,10)$ marks our location again.  If we
encode this terrain in a table~$\col{map}$, see~\cref{fig:terrain-map},
we can use the \SQL{} query of~\cref{fig:visibility-query} to compute
the visible spots~(\tikz\fill[pattern=visible] (0,0) rectangle
(0.25,0.25);). The query uses a common table expression
(CTE, \sql{WITH}\,\dots) to structure the computation.  We can spot the
\emph{max scan} in~\crefrange{line:max-scan-start}{line:max-scan-end},
but the interplay of local table definitions, user-defined and builtin
functions, and complex query logic (\emph{e.g.}, the use of window
functions) weaves a tangled web that is hard to see through.  How does
this query work and how does it adapt the two-dimensional \emph{max
scan} idea?

\smallskip\noindent
\textbf{Data provenance} offers answers to these and further
questions~\cite{provenance-in-databases,provenance-past-present-future,uldbs,what-for-form-from}.
Provenance relates a query's individual input and output data items
(table cells, say), sheds light on query internals and bugs, and helps
to build trust in query results---a critical service to data-dependent
science and society~\cite{computational-reproducibility}.
In our present
case, we may hope that provenance helps to understand how the visibility
\emph{max scan} has been tweaked to function in three-dimensional
terrain.

\begin{figure}
  \centering\small
\begin{lstlisting}[language=sql,lineskip=-2pt]
-- Distance between points (x1,y1) and (x2,y2)
CREATE FUNCTION
dist(x1 int, y1 int, x2 int, y2 int) RETURNS float AS
$$
 SELECT sqrt((x2 - x1)^2 + (y2 - y1)^2)
$$ LANGUAGE SQL;

-- Number of steps on the line (x1,y1)-(x2,y2)
CREATE FUNCTION
steps(x1 int, y1 int, x2 int, y2 int) RETURNS int AS
$$
 SELECT greatest(abs(x2 - x1), abs(y2 - y1))
$$ LANGUAGE SQL;

-- Points (x,y) on the line (x1,y1)-(x2,y2)
CREATE FUNCTION
line(x1 int, y1 int, x2 int, y2 int) RETURNS TABLE(x int, y int) AS
$$
 SELECT x1 + round(i * ((x2 - x1) / steps(x1, y1, x2, y2))) AS x,
        y1 + round(i * ((y2 - y1) / steps(x1, y1, x2, y2))) AS y
 FROM   generate_series(0, steps(x1, y1, x2, y2)) AS i
$$ LANGUAGE SQL;

WITH
-- (1) Ray from %\lst@commentstyle\pointX% to (x1,y1) has points (rx,ry)
rays(x1, y1, rx, ry) AS (
 SELECT m.x AS x1, m.y AS y1, l.x AS rx, l.y AS ry
 FROM   map AS m,
        LATERAL line(17, 10, m.x, m.y) AS l(x,y)
 WHERE  m.x IN (0,20) OR m.y IN (0,20)   -- points on the border
),
-- (2) Angle between point (x,y) and %\lst@commentstyle\pointX%
angles(x, y, angle) AS (
 SELECT m.x, m.y,
        degrees(atan((m.alt - 200) /   -- %\lst@commentstyle\pointX% is at altitude 200m
               (dist(m.x, m.y, 17, 10)))) AS angle
 FROM   map AS m
 WHERE  ROW(m.x, m.y) <> ROW(17, 10)
),
-- (3) Line of sight along each ray (uses a max scan) %\label{line:max-scan-start}%
max_scan(x, y, angle, max_angle) AS (
 SELECT r.rx AS x, r.ry AS y, a.angle, MAX(a.angle) OVER (
          PARTITION BY r.x1, r.y1
          ORDER BY dist(17, 10, r.rx, r.ry)) AS max_angle
 FROM   rays AS r, angles AS a
 WHERE  ROW(r.rx, r.ry) = ROW(a.x, a.y)
),                                                    %\label{line:max-scan-end}%
-- (4) Assemble visibility map from all lines of sight
visible(x, y, "visible?") AS (
 SELECT s.x, s.y, bool_or(s.angle >= s.max_angle) AS "visible?"
 FROM   max_scan AS s
 GROUP BY s.x, s.y
)
SELECT v.x, v.y, v."visible?"
FROM   visible AS v;
\end{lstlisting}
  \vskip-3mm
  \caption{\SQL{} query to compute visibility in three-dimen\-sional terrain encoded in
    table~\sql{map}.  A row~$(x,y,\sql{true})$ in result table~\sql{visible}
    indicates that spot~$(x,y)$ is visible from~\pointX{}.}
  \label{fig:visibility-query}
\end{figure}

Clearly, the benefits of data provenance grow with the complexity of the
query logic it is able to explain.  As modern query languages continue
to gain expressive constructs~\cite{sql-2016} and algorithms of
increasing intricacy are cast into relational queries (\emph{e.g.},
graph processing and machine learning
tasks~\cite{grail,spark-sql-ml,mad-lib}), the gap between queries found
in practice and existing approaches for provenance derivation widens
considerably,
however~\cite{general-lineage-tracing,semirings,provenance-in-databases,what-for-form-from}.
The principal languages of study have been the (positive) relational
algebra and its \SQL{} equivalent.  Grouping and aggregation can be
handled by some approaches~\cite{lineage-tracing,perm-rewriting} but are
already considered challenging.  In this light, the derivation of
database provenance for complex queries found ``outside the lab''
appears elusive.

\smallskip\noindent
We set out to bridge this gap and enable the derivation of
\textbf{fine-grained data provenance for a significantly richer family of \SQL{}
queries}.  The admissable query dialect includes
\begin{compactitem}
\item common table expressions including the recursive kind
  (\sql{WITH RECURSIVE}\dots),
\item window functions with arbitrary frame specifications as well as
  grouping and aggregation,
\item scalar or table-valued builtin and user-defined functions,
\item complex types (\emph{e.g.}, row values and arrays), and
\item subqueries without or with dependencies (through~\sql{LATERAL}
  or correlation) to their enclosing query.
\end{compactitem}
We aim for compositionality, \emph{i.e.}, these and further constructs
may be nested arbitrarily as long as \SQL{}'s scoping and typing rules are obeyed.

\smallskip\noindent
The approach is based on a \textbf{non-standard interpretation of the
\SQL{} semantics}.  This new interpretation focuses on the dependencies
between input and output data items---the items' values play a
secondary role only.  The required interpreter is
systematically derived from the original value-based query and
formulated in \SQL{} itself.  As long as we can perform this derivation
for a SQL construct or idiom, the approach is ready to embrace it. While
we work with \Pg{} in what follows, the method may be implemented on top
of any \SQL{}-based RDBMS.  \emph{No engine internals need to be altered.}

\smallskip
\begin{changed}
\begin{goal}{\Where- and \Why-Provenance}
Given a \SQL{} subject query~$q$ and its output table~$t$, for each
cell~$o$ of~$t$ compute
\begin{compactitem}
\item which input table cells were \emph{copied or transformed}
  to determine $o$'s value, and \hfill\textnormal{[\where-provenance]}
\item which input table cells were \emph{inspected to decide} that~$o$
  is present in the output at all. \hfill\textnormal{[\why-provenance]}
\end{compactitem}
\end{goal}
\end{changed}

\smallskip\noindent
This understanding of \where- and \why-provenance largely coincides with
that of earlier
work~\cite{general-lineage-tracing,perm-rewriting,why-and-where}---\cref{sec:related-work}
notes where we deviate. Together, both types of provenance characterize
the exact set of input table cells that were sourced by query~$q$,
providing invaluable information for query explanation and
debugging~\cite{observing-sql-queries-habitat}. Such complete cell-level provenance
provides the most detailed insight into query behavior but comes at a
size and speed cost.  We thus also outline how coarser granularity may
be traded for performance.

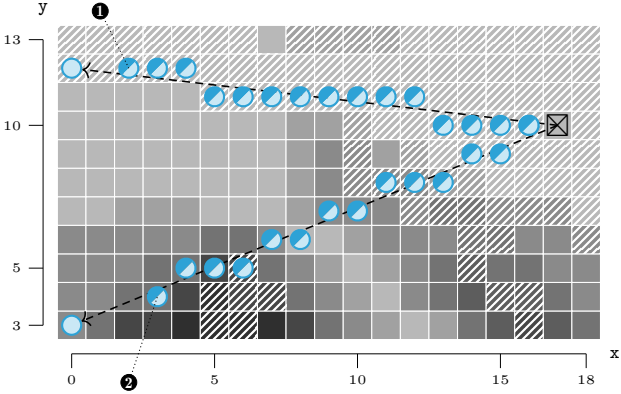
\begin{figure}
  \centering\small
  \setlength{\G}{3.8mm}
  \begin{tikzpicture}[x=\G,y=\G]
    %% axes
    \begin{scope}[very thin,font=\tiny]
      \begin{scope}[yshift=-\G]
        \draw (0,3) -- (18,3);
        \foreach \x in {0,5,10,15,18}
          { \draw (\x,3) -- (\x,2.5) node[below] {\x};
          }
        \node[font=\scriptsize] at (19,3) {\col{x}};
      \end{scope}
      \begin{scope}[xshift=-\G]
        \draw (0,3) -- (0,13);
        \foreach \y in {3,5,10,13}
          { \draw (0,\y) -- (-0.5,\y) node[left] {\y};
          }
        \node[font=\scriptsize,anchor=base] at (0,14) {\col{y}};
      \end{scope}
    \end{scope}

    %% visibility mark
    \pgfdeclareplotmark{visible}{%
      \pgfpathrectangle{\pgfpoint{-1/2*\G}{-1/2*\G}}{\pgfpoint{\G}{\G}}
      \pgfsetfillpattern{visible}{white}
      \pgfusepath{fill}}
    %% height map
    \begin{pgfonlayer}{background}
      \ifrendermap
      \begin{axis}[x=\G,y=\G,at={(0*\G,3*\G)},anchor={origin},hide axis,
        xmin=0,xmax=18,ymin=3,ymax=13]
        \addplot[%
          scatter,
          only marks,
          scatter src=explicit,
          scatter/use mapped color={fill=mapped color},
          mark=square*,mark size=1/2*\G,mark options={draw=white}]
          table[meta=h]
          {visibility-map-21x21-simpler.xyhv};
        \addplot[%
          scatter,
          only marks,
          scatter src=explicit symbolic,
          scatter/classes={true={mark=visible},false={mark=none}}]
          table[meta=visible]
          {visibility-map-21x21-simpler.xyhv};
      \end{axis}
      \fi
    \end{pgfonlayer}

    %% rays
    \begin{scope}[densely dashed,semithick,->,shorten >=4pt]
      \draw (17,10) -- (0,12);
      \draw (17,10) -- (0, 3);
    \end{scope}

    \begin{pgfonlayer}{foreground}
      %% observation point X (cross)
      \node[font=\tiny] at (17,10) {\large\pointX};
      %% fixed sample of clicked points and where/why-provenance for non-OCG PDF viewers
      %% (these are the visible points in file visibility-provenance-21x21-simpler.tex)
      %% (3,4)
      \gwherewhy{(6,5)}{prov1};
      \gwherewhy{(3,4)}{prov1};
      \gwherewhy{(11,8)}{prov1};
      \gwherewhy{(12,8)}{prov1};
      \gwherewhy{(7,6)}{prov1};
      \gwherewhy{(4,5)}{prov1};
      \gwherewhy{(16,10)}{prov1};
      \gwherewhy{(5,5)}{prov1};
      \gwherewhy{(14,9)}{prov1};
      \gwherewhy{(13,8)}{prov1};
      \gwherewhy{(15,9)}{prov1};
      \gwherewhy{(10,7)}{prov1};
      \gwherewhy{(9,7)}{prov1};
      \gwherewhy{(8,6)}{prov1};
      \gwhy{(0,3)}{prov1};
      %% (2,12)
      \gwherewhy{(2,12)}{prov1};
      \gwherewhy{(11,11)}{prov1};
      \gwherewhy{(3,12)}{prov1};
      \gwherewhy{(5,11)}{prov1};
      \gwherewhy{(16,10)}{prov1};
      \gwherewhy{(12,11)}{prov1};
      \gwherewhy{(4,12)}{prov1};
      \gwherewhy{(13,10)}{prov1};
      \gwherewhy{(8,11)}{prov1};
      \gwherewhy{(6,11)}{prov1};
      \gwherewhy{(14,10)}{prov1};
      \gwherewhy{(10,11)}{prov1};
      \gwherewhy{(9,11)}{prov1};
      \gwherewhy{(15,10)}{prov1};
      \gwherewhy{(7,11)}{prov1};
      \gwhy{(0,12)}{prov1};
      %% pointers to (3,4) and (2,12)
      \draw[densely dotted] (1,14) node[Circled] {\textbf{1}} -- (2,12);
      \draw[densely dotted] (2, 1) node[Circled] {\textbf{2}} -- (3, 4);
    \end{pgfonlayer}
  \end{tikzpicture}
  \caption{Excerpt of terrain map after provenance derivation.
    We find that the (non-)visibility of
    spots~\protect\Circled{\textbf{1}} and~\protect\Circled{\textbf{2}} is \where- as well as \why-dependent on
    the points marked~\protect\tikz[baseline=-3pt]\protect\gwherewhy{(0,0)}{prov1};
    and only \why-dependent on the two border points
    marked~\protect\tikz[baseline=-3pt]\protect\gwhy{(0,0)}{prov1};.}
  \label{fig:terrain-provenance}
\end{figure}

\smallskip\noindent
\textbf{Data provenance explains queries.}
Once we perform provenance derivation for the \SQL{} query
of~\cref{fig:visibility-query}, we can understand how the data in input
table~\sql{map} (\cref{fig:terrain-map}) is used to compute visibility.
\Cref{fig:terrain-provenance} highlights those points in the input
terrain that determine the visibility of spots~\Circled{\textbf{1}}
and~\Circled{\textbf{2}}. Since we compute the provenance of the entire
query output, we could have selected any spot and investigated the
provenance of its visibility. Provenance analysis reveals that the
query ``shoots'' rays from~\pointX{} to the points at the
border of the map (see the~\tikz[baseline=-3pt]\draw[<-,densely dashed,semithick] (0,0)--(0.5,0);
in~\cref{fig:terrain-provenance}), effectively leaving us with a two-dimensional problem
that can be tackled via the \emph{max scan} technique
of~\cref{fig:terrain-2d}. We see that the visibility of point~$(x,y)$
only depends on the points on the ray between
\pointX{} and~$(x,y)$, \emph{i.e.}, those points visited by the \emph{max scan} so far.
These and similar findings help to untangle the query and build trust in
its result.

\section{From Values to Dependency Sets}
\label{sec:from-values-to-dep-sets}

Regular query evaluation computes the \emph{value} of an output cell~$o$
through the inspection and transformation of input \emph{values}.  In
this work, instead, we focus on~$o$'s \emph{dependency set}:

\smallskip
\begin{changed}
\begin{definition}{Dependency Set}
Given an output cell~$o$, the \emph{dependency set} of~$o$ is the
(possibly empty) set~$\{i_1, i_2, \dots, i_n\}$ of input table cells
that were copied, transformed, or inspected to compute the value of~$o$.
Values are secondary: $o$ and~$i_1,\dots,i_n$ identify the cells themselves,
not their values. We use~$\ptype$ to denote the type of dependency sets.
\end{definition}
\end{changed}

\smallskip\noindent
It is our main hypothesis that a \emph{non-standard interpretation} of
queries provides a solid foundation to reason about this shift of focus
from values to dependency sets~\cite{provenance-as-dependency}.
We pursue a purely
\SQL{}-based implementation of this shift: from the original value-based
\SQL{} query, we generate its dependency-deriving variant---or
\emph{interpreter}, for short---through query transformation.  Since
this variant manipulates
dependency sets and is oblivious to values, we supply just enough
runtime information to guide the interpreter whenever the original query
made a value-based decision.

\smallskip\noindent
\textbf{Overview.} These considerations shape a two-phase approach.  Let~$q$ denote the original
\SQL{} query:
\begin{itemize}
\item[\textbf{Phase~$\ph{1}$:}] Instrument~$q$ to obtain query~$\inphase{1}{q}$
  that performs the same value-based computation as~$q$ and outputs the same result.  Whenever
  $\inphase{1}{q}$ makes a value-based decision (\emph{e.g.}, let a row pass a predicate or
  locate a row inside a window frame), those values relevant to this decision
  are appended to logs as a side effect of evaluation.
\item[\textbf{Phase~$\ph{2}$:}] Evaluate
  interpreter~$\inphase{2}{q}$ that performs dependency derivation.
  Query~$\inphase{2}{q}$ reads, manipulates, and outputs tables of dependency sets.
  To properly replay the decisions made by~$\inphase{1}{q}$,
  $\inphase{2}{q}$ additionally consults the logs written in Phase~$\ph{1}$.
\end{itemize}

\smallskip\noindent
We shed light on Phases~$\ph{1}$ and~$\ph{2}$ and their interaction in the
upcoming~\cref{sec:phase1,sec:phase2}.  The construction of the instrumented
query~$\inphase{1}{q}$ as well as the
interpreter~$\inphase{2}{q}$---both can be built in tandem---are the
subject of~\cref{sec:abs-int}.  Since the evaluation
of~$\inphase{1}{q}$ incurs logging effort and~$\inphase{2}{q}$ needs
to manipulate sets instead of first normal form (1NF) values,
\cref{sec:performance} discusses the sizes of both logs and
result tables, quantifies the impact on query evaluation time, and discusses
\SQL{} interpretation at the coarser row granularity.
\cref{sec:related-work,sec:conclusions} review related efforts and wrap up.

\subsection{Changing Types, Preserving Shape}
\label{sec:parametricity}

Consider~$q$, a general template for a single-table
\sql{SELECT}-\sql{FROM}-\sql{WHERE} block:
$$
q(e,p,t) = \sql{SELECT}~e(x)~\sql{FROM}~t~\sql{AS}~x~\sql{WHERE}~p(x)\enskip.
$$%
The type of~$q$, namely
$
\forall a,b\colon (a \shortrightarrow b) \times (a \shortrightarrow \sql{bool}) \times \{a\} \shortrightarrow \{b\}
$,
is \emph{parametric}~\cite{parametricity} in the row
types~$a$ and $b$ of the input and output tables.\footnote{We use $a \times
b$ to denote pair (or record) types and write $\{a\}$ for the type of
tables with rows of type~$a$.} Any instantiation of type variables~$a$ and $b$ yields a
workable \emph{filter}-\emph{project} query.  If~$t$ is table~$\col{map}$ of~\cref{fig:terrain-map}
and~$e$ projects on its third column~$\col{alt}$ of type~$\sql{real}$,
then $a \equiv \sql{int}\times\sql{int}\times\sql{real}$,
$b \equiv \sql{real}$, and~$q$ has type
$$
\begin{array}{@{}c@{}}
  (\sql{int}\times\sql{int}\times\sql{real}\shortrightarrow\sql{real}) \times (\sql{int}\times\sql{int}\times\sql{real}\shortrightarrow\sql{bool}) \times {}\\
  \{\sql{int}\times\sql{int}\times\sql{real}\} \shortrightarrow \{\sql{real}\} \enskip.
\end{array}
$$%

\smallskip\noindent
With the shift from values (Phase~$\ph{1}$) to dependency sets
(Phase~$\ph{2}$) we are interested in the particular row type
instantiation in which \emph{all column types are replaced by~$\ptype$},
the type of dependency sets.  If we perform this shift for the former
example, we get $a \equiv \ptype\times\ptype\times\ptype$, $b \equiv
\ptype$, yielding query~$\inphase{2}{q}$ of type
$$
(\ptype\times\ptype\times\ptype\shortrightarrow\ptype) \times (\ptype\times\ptype\times\ptype\shortrightarrow\sql{bool}) \times \{\ptype\times\ptype\times\ptype\} \shortrightarrow \{\ptype\}\enskip,
$$%
over tables of dependency sets. Most importantly, $\inphase{2}{q}$ is indifferent to the
choice of row types~\cite{parametricity}: it continues to implement
the \emph{filter}-\emph{project} semantics.

\smallskip\noindent
This parametricity of queries is central to the approach:
\begin{compactitem}
\item The shift to~$\ptype$ in Phase~$\ph{2}$ not only preserves the shape of the query type
  but, largely, also the \emph{syntactic shape} of the \SQL{} query.  We can thus derive
  an interpreter for a given query via a transformation that
  is compositional (will not break in the face of complex queries) and
  extensible (can embrace new constructs as the \SQL{} language grows).
  The query execution plans of the transformed queries resemble those
  of the originals which reduces the risk of overwhelming the query
  processor, an adverse effect that has been observed by earlier
  work on data provenance for \SQL{}~\cite{gprom}.
\item The value-based and dependency-based queries read and output tables
  of the same width and row count: we also preserve the \emph{shape of
  the data} (albeit not its type).  A one-to-one correspondence between the cells in
  value-based and dependency-carrying tables admits a straightforward
  association of individual data items with their provenance.
\end{compactitem}

\smallskip\noindent
In Phase~$\ph{2}$, note that predicate~$p$ (of type
$\ptype\times\ptype\times\ptype\shortrightarrow\sql{bool}$) exclusively
receives dependency sets as input.
These dependency sets reveal what influenced the predicates's outcome
but do not let us compute the Boolean value of the original~$p$.
We address this in Phase~$\ph{1}$ in which we instrument the original
query such that the outcome of relevant value-based computation is logged.  The
interpreter of Phase~$\ph{2}$ then uses the log to look up $p$'s
Boolean value and to re-enact the original query's behavior.

\subsection{Phase~{\large$\ph{1}$}: Instrumentation}
\label{sec:phase1}

\begin{changed}
\begin{definition}{Instrumented Query, Phase~$\ph{1}$}
Given a subject query~$q$, its \emph{instrumented variant}~$\inphase{1}{q}$
computes the same output table as~$q$.  Whenever~$q$ evaluates an expression
of non-parametric type to make a relevant value-based decision, $\inphase{1}{q}$
logs the outcome of that decision as a side-effect of query evaluation.
\end{definition}
\end{changed}

%% Normalized visibility query
\begin{figure}[t]
  \centering\small
\begin{lstlisting}[language=sql,mathescape=true]
max_scan(x, y, angle, max_angle) AS (
 SELECT $\var{t}$.r_rx AS x, $\var{t}$.r_ry AS y, $\var{t}$.a_angle AS angle,
        MAX($\var{t}$.a_angle) OVER ($\var{w}$) AS max_angle
 FROM   (SELECT r.x1 AS r_x1, r.y1 AS r_y1,
                r.rx AS r_rx, r.ry AS r_ry,
                a.angle AS a_angle
         FROM   rays AS r, angles AS a
         WHERE  r.rx = a.x AND r.ry = a.y) AS $\var{t}$
 WINDOW $\var{w}$ AS
   (PARTITION BY $\var{t}$.r_x1, $\var{t}$.r_y1
    ORDER BY sqrt(($\var{t}$.r_rx - 17)^2 + ($\var{t}$.r_ry - 10)^2))
)
\end{lstlisting}
  \vskip-2mm
  \caption{Common table expression~\sql{max\_scan} of the visibility query
    (\cref{fig:visibility-query}) after normalization.  UDF~\sql{dist} has
    been inlined into the~\sql{ORDER\,BY} clause.}
  \label{fig:visibility-normalized}
\end{figure}

%% Visibility query (Phase 1)
\begin{figure}[t]
  \begin{narrow}{-1mm}{-1mm}
  \centering\small
\begin{lstlisting}[language=sql,mathescape=true]
$\phase{1}{\sql{max\_scan}}$($\row$, x, y, angle, max_angle) AS (
 SELECT  $\logwrite{WIN}$($\location{4}$, $\var{t}$.$\row$, FIRST_VALUE($\var{t}$.$\row$) OVER ($\var{w}$),%\tikz[remember picture] \node (write1) {};% %\label{line:writewin-start}%
                           RANK() OVER ($\var{w}$)) AS $\row$, %\label{line:writewin-end}%
         $\var{t}$.r_rx AS x, $\var{t}$.r_ry AS y, $\var{t}$.a_angle AS angle,    %\tikz[remember picture] \node (added) {};%
         MAX($\var{t}$.a_angle) OVER ($\var{w}$) AS max_angle
 FROM    (SELECT $\logwrite{JOIN2}$($\location{3}$, r.$\row$, a.$\row$) AS $\row$,%\tikz[remember picture] \node (write2) {};% %\label{line:subq-join-start}% %\label{line:writejoin2}%
                 r.x1 AS r_x1, r.y1 AS r_y1,
                 r.rx AS r_rx, r.ry AS r_ry,
                 a.angle AS a_angle
          FROM   $\phase{1}{\sql{rays}}$ AS r, $\phase{1}{\sql{angles}}$ AS a
          WHERE  r.rx = a.x AND r.ry = a.y) AS $\var{t}$        -- $\lst@commentstyle\inphase{1}{p}(\sql{r},\sql{a})$%\label{line:subq-join-end}%
 WINDOW $\var{w}$ AS
   (PARTITION BY $\var{t}$.r_x1, $\var{t}$.r_y1                         -- $\lst@commentstyle\inphase{1}{f}(\!\var{t})$%\label{line:window-start}%
    ORDER BY sqrt(($\var{t}$.r_rx - 17)^2 + ($\var{t}$.r_ry - 10)^2))   -- $\lst@commentstyle\inphase{1}{g}(\!\var{t})$%\label{line:window-end}%
)
\end{lstlisting}%
  \begin{tikzpicture}[remember picture,overlay,line join=round,line cap=round,black!60]
    \draw[<-] (write1) -- +(0.3,0) -- (added);
    \draw[<-] (write2) -- +(1.1,0) -- (added);
    \node[node font=\small,anchor=center,right] at (added) {added};
  \end{tikzpicture}%
  \end{narrow}
  \vskip-7mm
  \caption{Instrumented variant of CTE~\sql{max\_scan} in Phase~$\ph{1}$.}
  \label{fig:visibility-phase1}
\end{figure}

%%   tables 𝟙(map)+𝟚(map)
\begin{figure}
  \centering\small
  \begin{littbl}
    $
    \begin{array}{@{}c@{}c|c|r|>{\quad\color{black!60}}c@{\quad}c|c|c|@{}c@{}}
                           & \tabname{2}{\strut{\phase{1}{\col{map}}}} & \multicolumn{1}{c}{} & & \tabname{2}{\strut{\phase{2}{\col{map}}}} \\
                           & \keyhd{x} & \keyhd{y} & \colhd{alt} & \multicolumn{1}{c}{\color{black!60}\row} & \colhd{x} & \colhd{y} & \colhd{alt} \\
                           \strut
                           & \multicolumn{1}{|c|}{\texttt{0}}  & \texttt{0}  & \texttt{400.0}              & \rowid{m}{(0,0)}   & \multicolumn{1}{|c|}{\{\col{x}_{(0,0)}\}}   & \{\col{y}_{(0,0)}\}   & \{\col{a}_{(0,0)}\} \\
                           & \multicolumn{1}{|c|}{\texttt{0}}  & \texttt{1}  & \texttt{400.0}              & \rowid{m}{(0,1)}   & \multicolumn{1}{|c|}{\{\col{x}_{(0,1)}\}}   & \{\col{y}_{(0,1)}\}   & \{\col{a}_{(0,1)}\} \\
                           & \multicolumn{1}{|c|}{\texttt{0}}  & \texttt{2}  & \texttt{400.0}              & \rowid{m}{(0,2)}   & \multicolumn{1}{|c|}{\{\col{x}_{(0,2)}\}}   & \{\col{y}_{(0,2)}\}   & \{\col{a}_{(0,2)}\} \\
                           & \multicolumn{1}{|c|}{\nvdots}     & \nvdots     & \multicolumn{1}{c|}{\nvdots}& \nvdots            & \multicolumn{1}{|c|}{\nvdots}               & \nvdots               & \nvdots \\
        \llap{\pointX{}\,} & \multicolumn{1}{|c|}{\texttt{17}} & \texttt{10} & \texttt{200.0}              & \rowid{m}{(17,10)} & \multicolumn{1}{|c|}{\{\col{x}_{(17,10)}\}} & \{\col{y}_{(17,10)}\} & \{\col{a}_{(17,10)}\}  & \rlap{\,\pointX} \\
                           & \multicolumn{1}{|c|}{\nvdots}     & \nvdots     & \multicolumn{1}{c|}{\nvdots}& \nvdots            & \multicolumn{1}{|c|}{\nvdots}               & \nvdots               & \nvdots \\
                           & \multicolumn{1}{|c|}{\texttt{20}} & \texttt{18} & \texttt{0.0}                & \rowid{m}{(20,18)} & \multicolumn{1}{|c|}{\{\col{x}_{(20,18)}\}} & \{\col{y}_{(20,18)}\} & \{\col{a}_{(20,18)}\} \\
                           & \multicolumn{1}{|c|}{\texttt{20}} & \texttt{19} & \texttt{0.0}                & \rowid{m}{(20,19)} & \multicolumn{1}{|c|}{\{\col{x}_{(20,19)}\}} & \{\col{y}_{(20,19)}\} & \{\col{a}_{(20,19)}\} \\
                           & \multicolumn{1}{|c|}{\texttt{20}} & \texttt{20} & \texttt{0.0}                & \rowid{m}{(20,20)} & \multicolumn{1}{|c|}{\{\col{x}_{(20,20)}\}} & \{\col{y}_{(20,20)}\} & \{\col{a}_{(20,20)}\} \\
        \cline{2-4}\cline{6-8}
    \end{array}
    $
  \end{littbl}
  \caption{Table~\sql{map} in Phases~$\ph{1}$ and~$\ph{2}$. A row with key $(\col{x},\col{y})$ = $(x,y)$
    is identified by row~ID~$\row = m_{(x,y)}$.}
  \label{fig:terrain-map-phase12}
\end{figure}

\smallskip\noindent
The instrumentation of~$q$ will be \emph{compositional}:
$q$'s overall instrumentation is assembled from the instrumentation of $q$'s subqueries---
the latter transformations do not interfere and may be performed in isolation.
Here, we exploit this to save page space and focus on CTE
fragment~\sql{max\_scan} of the \SQL{} query
in~\cref{fig:visibility-query}.
Input to instrumentation is a normalized form of the original query in
which individual operations (\emph{e.g.}, joins, window functions,
ordering) are placed in separate subqueries.  The normalized
CTE~\sql{max\_scan} is shown in~\cref{fig:visibility-normalized}.
Normalization, discussed in~\cref{sec:abs-int}, helps to devise
compact sets of query transformation rules.

\smallskip\noindent
\Cref{fig:visibility-phase1} shows~$\phase{1}{\sql{max\_scan}}$, the
instrumented form of~\sql{max\_scan}. (For a query, expression, CTE, or
table named~$n$, we use~$n$, $\phase{1}{n}$, and $\phase{2}{n}$ to refer
to the original and its Phase~$\ph{1}/\ph{2}$ variants.) Where the
original query reads from table~$r$, the instrumented version reads from
$\phase{1}{r}$ in which column~$\row$ carries row
identifiers---otherwise, $r$ and~$\phase{1}{r}$ are identical. Indeed,
$r$ and~$\phase{1}{r}$ may denote the very same table if the underlying
RDBMS externalizes row identity in some form (\emph{e.g.}, through virtual
column~\sql{ctid} in \Pg{} or~\sql{rowid} in IBM~Db2 and Oracle).
Table~$\phase{1}{\col{map}}$ is depicted in~\cref{fig:terrain-map-phase12} on the
left.

\smallskip\noindent
When we log the outcome~$v$ of a computation over a row~\sql{r}, we
write the pair~$(\sql{r.$\row$},v)$ to identify the row once we read the log
back. It is the primary aim of instrumentation to insert calls
to side-effecting functions~$\logwrite{$\Box$}$\sql{($\location{\ell}$,\,r.$\row$,\,$v$)} that perform the required log writing.
Parameter~$\location{\ell}$ distinguishes the calls' locations in the
instrumented \SQL{} text such that one log may hold entries written by
multiple call sites.  Phase~$\ph{2}$ (see below) then uses~$\logread{$\Box$}$\sql{($\location{\ell}$,\,\sql{r.$\row$})}
to obtain~$v$ again.
The approach is indifferent to the actual
realization of~$\logwrite{$\Box$}$
and~$\logread{$\Box$}$. \Cref{sec:abs-int}
shows pseudo code and
\ifarXiv
\cref{app:log-files}
\else
the appendix
\fi
proposes a possible \SQL-internal implementation
of logging.

\smallskip\noindent
In the subquery in~\crefrange{line:subq-join-start}{line:subq-join-end}
of~\cref{fig:visibility-phase1}, the result of the join depends on the
evaluations of predicate~$\inphase{1}{p}(\sql{r},\sql{a}) = \sql{r.rx = a.x AND r.ry
= a.y}$. We make the outcomes of~$\inphase{1}{p}$ available to Phase~$\ph{2}$ via
calls to~\sql{$\logwrite{JOIN2}$($\location{3}$,\,r.$\row$,\,a.$\row$)}
in~\cref{line:writejoin2}.  Note that we chose to not log~$\inphase{1}{p}$'s actual
Boolean value but, equivalently, the fact that rows~\sql{r} and~\sql{a}
are join partners---this refinement saves us from logging
the~\sql{false} outcomes of~$\inphase{1}{p}$ and also simplifies Phase~$\ph{2}$.
The invocation of~\sql{$\logwrite{JOIN2}$} performs log writing and
then returns a newly generated row identifier that represents the joined row~$\var{t}$.

%% Visibility query (Phase 𝟚)
\begin{figure}
  \begin{narrow}{-1mm}{-1mm}
  \centering\small
\begin{lstlisting}[language=sql,mathescape=true]
$\phase{2}{\sql{max\_scan}}$($\row$, x, y, angle, max_angle) AS (
 SELECT  $\var{t}$.$\row$ AS $\row$,
         $\abs{\var{t}\sql{.r\_rx}}$ AS rx, $\abs{\var{t}\sql{.r\_ry}}$ AS ry, $\abs{\var{t}\sql{.a\_angle}}$ AS angle,
         $\abs{\setagg \var{t}\sql{.a\_angle OVER (}\var{w}\sql{)} \mathbin{\Y{\cup}} \Y{\setagg \varY{win}\sql{ OVER (}\var{w}\sql{)}}}$ AS max_angle %\label{line:set-aggr}%
 FROM
   (SELECT $\var{join}$.$\row$,
            $\abs{\sql{r.x1} \mathbin{\Y{\cup}} \Y{\varY{join}}}$ AS r_x1, $\abs{\sql{r.y1} \mathbin{\Y{\cup}} \Y{\varY{join}}}$ AS r_y1,
            $\abs{\sql{r.rx} \mathbin{\Y{\cup}} \Y{\varY{join}}}$ AS r_rx, $\abs{\sql{r.ry} \mathbin{\Y{\cup}} \Y{\varY{join}}}$ AS r_ry,
            $\abs{\sql{a.angle} \mathbin{\Y{\cup}} \Y{\varY{join}}}$ AS a_angle
    FROM    $\phase{2}{\sql{rays}}$ AS r, $\phase{2}{\sql{angles}}$ AS a,
            LATERAL $\logread{JOIN2}$($\location{3}$, r.$\row$, a.$\row$) AS $\var{join}$($\row$)$\Y{\sql{,}}$
            $\Y{\sql{LATERAL}}$ $\abs{\Y{\toY\sql{(r.rx} \cup \sql{a.x} \cup \sql{r.ry} \cup \sql{a.y)}}}$ $\Y{\sql{AS}}$ $\Y{\varY{join}}$ -- %\lst@commentstyle\sql{\toY($\inphase{2}{p}(\sql{r},\sql{a})$)}% %\label{line:Y-join}%
    ) AS $\var{t}$,
    LATERAL $\logread{WIN}$($\location{4}$, $\var{t}$.$\row$) AS $\var{win}$($\row$,$\partition$,$\peer$)$\Y{\sql{,}}$
    $\Y{\sql{LATERAL}}$ $\abs{\Y{\toY\sql{(}\var{t}\sql{.r\_x1} \cup \var{t}\sql{.r\_y1}\sql{)} \mathop{\cup}}}$                      $\mskip5mu$-- %\lst@commentstyle\sql{\toY($\inphase{2}{f}(\!\var{t})$)}% %\label{line:Y-win-start}%
            $\abs{\Y{\toY\sql{(}\phase{2}{\sql{dist}}\sql{(}\pempty, \pempty, \var{t}\sql{.r\_rx}, \var{t}\sql{.r\_ry}\sql{))}}}$ $\Y{\sql{AS}}$ $\Y{\varY{win}}$      -- %\lst@commentstyle\sql{\toY($\inphase{2}{g}(\!\var{t})$)}% %\label{line:Y-win-end}%
 WINDOW $\var{w}$ AS (PARTITION BY $\var{win}$.$\partition$ ORDER BY $\var{win}$.$\peer$)
)
\end{lstlisting}
  \end{narrow}
  \vskip-5mm
  \caption{Interpreter for CTE~\sql{max\_scan} in Phase~$\ph{2}$.}
  \label{fig:visibility-phase2}
\end{figure}

In the window-based query enclosing the join, evaluation depends on the
partitioning and ordering criteria that determine the placement of
row~$\var{t}$ inside window~$\var{w}$
(\cref{line:window-start,line:window-end}
in~\cref{fig:visibility-phase1}). Both criteria are functions
of~$\var{t}$, namely $\inphase{1}{f}(\!\var{t}) =$ \sql{(\var{t}.r\_x1,\var{t}.r\_y1)}
and $\inphase{1}{g}(\!\var{t}) =$
\sql{sqrt(($\var{t}$.r\_rx - 17)\^{}2 + ($\var{t}$.r\_ry - 10)\^{}2)}.
Phase~$\ph{2}$ will not be able to evaluate either function once computation has
shifted from column values to dependency sets.  The invocation
of~\sql{$\logwrite{WIN}$}
in~\crefrange{line:writewin-start}{line:writewin-end} thus writes the
required log entries.  Here, again, we do not log the values
of~$\inphase{1}{f}(\!\var{t})$ and~$\inphase{1}{g}(\!\var{t})$ as is,
but equivalently record~\sql{FIRST\_VALUE(\var{t}.$\row$) OVER
(\var{w})} and~\sql{RANK() OVER (\var{w})}: the former represents
$\var{t}$'s~partition in terms of the identifier of that partition's
first row, the latter gives $\var{t}$'s~position inside that partition.
Once both criteria are logged, the
\sql{$\logwrite{WIN}$($\location{4}$,\,\var{t}.$\row$,\,$\dots$)}~call
returns~$\sql{\var{t}.$\row$}$.

\subsection{Phase~{\large$\ph{2}$}: Interpretation}
\label{sec:phase2}

\begin{changed}
\begin{definition}{Interpreter, Phase~$\ph{2}$}
\emph{Interpreter}~$\inphase{2}{q}$ for instrumented
query~$\inphase{1}{q}$ exclusively manipulates dependency sets: if the
evaluation of a subexpression~$\inphase{1}{e}$ of~$\inphase{1}{q}$
depended on the input table cells~$i_1, i_2,\dots,i_n$, its interpreted
counterpart~$\inphase{2}{e}$ in~$\inphase{2}{q}$ evaluates to the dependency
set~$\{i_1, i_2,\dots,i_n\}$.
\label{def:interpreter}
\end{definition}
\end{changed}

\smallskip\noindent
The definition implies that interpreter~$\inphase{2}{q}$ reads and outputs tables
of the same shape (cardinality and width) as instrumented query~$\inphase{1}{q}$:
where Phase~$\ph{1}$ reads
table~$\inphase{1}{r}$, the interpreter reads~$\inphase{2}{r}$
whose cells hold dependency sets (see table~$\inphase{2}{\sql{map}}$
in~\cref{fig:terrain-map-phase12} on the right).  Note that
corresponding rows in~$\inphase{1}{r}$ and~$\inphase{2}{r}$ share their
identifiers~$\row$ to establish a one-to-one correspondence between the
cells of both tables.

\smallskip\noindent
Singleton dependency sets in source table cells indicate that each of
these cells only depends on itself.  In table~$\inphase{2}{\sql{map}}$,
unique identifier~$\col{x}_{(x,y)}$ represents the cell in
column~$\col{x}$ of the row with $\row = m_{(x,y)}$; likewise,
$\col{y}_{(x,y)}$ and~$\col{a}_{(x,y)}$ represent cells in
columns~$\col{y}$ and~$\col{alt}$, respectively.  These cell identifiers
are entirely abstract and never computed with (cf.\ with
the~\emph{colors} of~\cite{provenance-color}).

\smallskip\noindent
The interpreter for CTE~\sql{max\_scan} is shown
in~\cref{fig:visibility-phase2}.
CTE $\inphase{2}{\sql{max\_scan}}$ preserves the syntactic shape of
$\inphase{1}{\sql{max\_scan}}$ in~\cref{fig:visibility-phase1}: a
window-based aggregation consumes the result of the join between
tables~$\inphase{2}{\col{rays}}$ and~$\inphase{2}{\col{angles}}$.
Computation, however, is over dependency sets instead of values.
Rather than committing early to one of many viable relational
set representations~\cite{set-containment-joins,set-valued-attributes,roaring},
$\inphase{2}{\sql{max\_scan}}$ uses the usual operators~$\cup$/$\setagg$ where
these sets are combined/aggregated.

Following the above definition, the non-standard interpretation of
functions~$\inphase{1}{p}$, $\inphase{1}{f}$, $\inphase{1}{g}$ yields
variants~$\inphase{2}{\underline{\phantom{x}}}$
collecting the dependencies for those columns that influence
the functions' evaluation (cf.\ with~\cref{sec:phase1}):
\begin{align*}
  \inphase{2}{p}(\sql{r},\sql{a}) & = \sql{r.rx} \cup \sql{a.x} \cup \sql{r.ry} \cup \sql{a.y} \\
  \inphase{2}{f}(\!\var{t})       & = \sql{\var{t}.r\_x1} \cup \sql{\var{t}.r\_y1} \\
  \inphase{2}{g}(\!\var{t})       & = \sql{\var{t}.r\_rx} \cup \varnothing \cup \sql{\var{t}.r\_ry} \cup \varnothing \enskip.
\end{align*}
As described in~\cref{sec:parametricity}, these functions exclusively manipulate
dependency sets of type~$\ptype$.  The literals~\sql{17}
and~\sql{10} map to~$\varnothing$ in~$\inphase{2}{g}$ since both
do not depend on any input data whatsoever. Set
aggregate~\sql{$\setagg$\,\var{t}.a\_angle OVER (\var{w})}
in~\cref{line:set-aggr} interprets \sql{MAX(\var{t}.a\_angle) OVER
(\var{w})} in~$\inphase{1}{\sql{max\_scan}}$: according to the \SQL{}
semantics, all \sql{\var{t}.a\_angle} values inside current window~$\var{w}$ are
aggregated to evaluate the~\sql{MAX} window
function~\cite[\S\,4.16.3]{sql-2016}
and thus influence the function's result.

\smallskip\noindent
The interpreter uses~\sql{\toY($D$)} to indicate that
dependency set~$D$ contains cells describing \why-provenance instead of
the default \where-provenance.  We construct the \why-dependency
set~\sql{\toY($\inphase{2}{p}(\sql{r},\sql{a})$)} in~\cref{line:Y-join}
to reflect that predicate~$p$ inspects exactly these cells to decide
whether rows~\sql{r} and~\sql{a} are join partners.  (We
use~\sql{LATERAL} to bind this set to~$\varY{join}$ as it is referenced
multiple times later on.)  Likewise, we form~$\varY{win}$ in
\cref{line:Y-win-start,line:Y-win-end} to collect the
cells~$\sql{\toY($\inphase{2}{f}(\!\var{t})$)} \cup
\sql{\toY($\inphase{2}{g}(\!\var{t})$)}$ that are inspected to decide
how window frames are formed.  \Cref{line:set-aggr} then adds these
\why-dependencies to the provenance of the~\sql{MAX} window aggregate.

$\inphase{2}{\sql{max\_scan}}$ reads the logs written in Phase~$\ph{1}$ to
\begin{inparaenum}[(1)]
\item re-enact~$\inphase{1}{p}$'s filtering decisions and
\item to reconstruct the window frames formed by~$\inphase{1}{f}$ and~$\inphase{1}{g}$.
\end{inparaenum}
Iff~\sql{$\logread{JOIN2}$($\location{3}$,r.$\row$,a.$\row$)} returns
a join row identifier, rows~\sql{r} and~\sql{a} have been found to
partner in Phase~$\ph{1}$.
\sql{$\logread{WIN}$($\location{4}$,\var{t}.$\row$)} retrieves partition representative~\sql{\var{win}.part} and
in-partition position~\sql{\var{win}.rank} to enable the~\sql{WINDOW}
clause to place row~$\var{t}$ inside its proper frame.

%% output of phase 2 query (where provenance)
\begin{figure}
  \centering\small
  \begin{littbl}
    $
    \begin{array}{@{}c@{}c|c|c|@{}c@{}}
                                    & \outputname{2}{\strut\col{output}} \\
                                    & \colhd{x} & \colhd{y} & \colhd{visible?} \\
                                      \strut
                                    & \multicolumn{1}{|c|}{\nvdots}              & \nvdots              & \nvdots \\
      \llap{\Circled{\textbf{1}}\,} & \multicolumn{1}{|c|}{\{\col{x}_{(2,12)}\}} & \{\col{y}_{(2,12)}\} & \begin{Bmatrix*}[r]
                                    &                                                                        \col{x}_{(2,12)},  & \col{y}_{(2,12)},  & \col{a}_{(2,12)}, \\
                                    &                                                                        \col{x}_{(3,12)},  & \col{y}_{(3,12)},  & \col{a}_{(3,12)}, \\
                                    &                                                                        \multicolumn{3}{@{}c}{\dots} \\
                                    &                                                                        \col{x}_{(15,10)}, & \col{y}_{(15,10)}, & \col{a}_{(15,10)},\\
                                    &                                                                        \col{x}_{(16,10)}, & \col{y}_{(16,10)}, & \col{a}_{(16,10)}
                                    &                                                                      \end{Bmatrix*} \\
                                    & \multicolumn{1}{|c|}{\nvdots}               & \nvdots             & \nvdots \\
      \llap{\Circled{\textbf{2}}\,} & \multicolumn{1}{|c|}{\{\col{x}_{(3,4)}\}}   & \{\col{y}_{(3,4)}\} & \begin{Bmatrix*}[r]
                                    &                                                                        \col{x}_{(3,4)},   & \col{y}_{(3,4)},   & \col{a}_{(3,4)}, \\
                                    &                                                                        \col{x}_{(4,5)},   & \col{y}_{(4,5)},   & \col{a}_{(4,5)}, \\
                                    &                                                                        \multicolumn{3}{@{}c}{\dots} \\
                                    &                                                                        \col{x}_{(15,9)},  & \col{y}_{(15,9)},  & \col{a}_{(15,9)},\\
                                    &                                                                        \col{x}_{(16,10)}, & \col{y}_{(16,10)}, & \col{a}_{(16,10)}
                                    &                                                                      \end{Bmatrix*} \\
                                    & \multicolumn{1}{|c|}{\nvdots}               & \nvdots             & \nvdots \\
        \cline{2-4}
    \end{array}
    $
  \end{littbl}
  \caption{\Where-provenance of the visibility of spots~\protect\Circled{\textbf{1}}
    and~\protect\Circled{\textbf{2}} (see~\cref{fig:terrain-provenance}) as derived by
    interpretation in Phase~$\ph{2}$.}
  \label{fig:spots12-provenance}
\end{figure}

\smallskip\noindent
\textbf{Output.} Interpretation for the visibility query
of~\cref{fig:visibility-query} yields the dependency set table
of~\cref{fig:spots12-provenance}.  For a spot in the terrain located
at~$(x,y)$, we learn that its coordinates have been copied over from
input table~$\col{map}$ (the cells in column~$\col{x}$ solely depend
on~$\col{x}_{(x,y)}$; likewise for column~$\col{y}$). Spot visibility,
however, depends on the terrain's altitude along the ray from $(x,y)$ to
$\pointX$. Indeed, \Cref{fig:terrain-provenance} simply is a
visualization of the dependency sets found in column~$\col{visible?}$
of table~\sql{output} in~\cref{fig:spots12-provenance}.

\section{Interpreting \SQL{} in \SQL{}}
\label{sec:abs-int}

The query instrumentation of Phase~$\ph{1}$ and the construction of the
interpreter of Phase~$\ph{2}$ are based on a pair of rule-based \SQL{}
source transformations. We first \emph{normalize} the input query to
facilitate transformation rules that do not face large monolithic
\sql{SELECT}~blocks but may focus on a single \SQL{} clause at a time.

\begin{changed}
\begin{definition}{Normalized Query}
All~\sql{SELECT} blocks in the \emph{normalized query} for subject query~$q$
adhere to the syntactic form shown in~\cref{fig:normal-form}.  Normalization
preserves the semantics of~$q$.
\end{definition}
\end{changed}

\smallskip\noindent
Normalization of the input query rests on the following two cornerstones:
%\setdefaultleftmargin{1em}{}{}{}{}{}
\begin{compactdesc}
\item[Explicitness.] Expand the column list implicit in~\sql{SELECT\;*}.  In~\sql{SELECT} clauses, name
  expressions~$e$ explicitly (\sql{$e$\;AS\;$c$}).  In~\sql{FROM}
  clauses, introduce explicit row aliases for tables or subqueries~$q$
  (\sql{$q$\;AS\;$t$}).  In expressions, use qualified column references
  (\sql{$t$.$c$}) only. Expand~\sql{DISTINCT} into~\sql{DISTINCT\;ON}.
  Trade inline window specifications for explicit~\sql{WINDOW} clauses.
  Inline the bodies of non-recursive UDFs (like~\sql{dist}
  of~\cref{fig:visibility-query}).  Remove syntactic sugar to reduce
  query diversity, \emph{e.g.}, supply empty
  \sql{GROUP\;BY} criteria $g \equiv \sql{()}$, or make defaults like~\sql{OFFSET\;0} and~\sql{LIMIT\;ALL} explicit,
  should any of these be missing.
  %% After \sql{GROUP\;BY}, homogenize the treatment of aggregates
  %% and non-aggregated expressions: wrap the latter in calls to the
  %% pseudo-aggregate~\sql{THE($\cdot$)}\footnote{\sql{THE($\{x,x,\dots,x\}$)}
  %% evaluates to $x$ and fails otherwise.}~\cite{Wadler-Peyton-Jones}.

\item[Clause isolation.] Traverse the query syntax tree bottom up.  Inside a~\sql{SELECT}
  block, isolate its \SQL{} clauses by placing each clause inside a separate
  subquery.  This leads to ``onion-style'' uncorrelated nesting in the \sql{FROM}~clause,
  cf.\ the sketch of the resulting normal form in~\cref{fig:normal-form}.
  On completion, transformation rules like~\labelcref{rule:window} or~\labelcref{rule:group} (see~\cref{fig:rules},
  discussed below) may assume that they encounter
  single-table~\sql{FROM\;$q$\;AS\;$t$} clauses only.
  As an example, see~\cref{fig:visibility-normalized} where the~\sql{WINDOW} clause has been isolated
  from the join of~$\col{rays}$ and~$\col{angles}$.
  \label{step:isolation}
\end{compactdesc}
Normalization preserves query semantics as well as data pro\-ve\-nance.
This holds, in particular, for clause isolation: from inner to outer, the onion's
layers adhere to the evaluation order defined for \SQL{}
clauses in a~\sql{SELECT} block~\cite[\S\,7.5ff]{sql-2016}.

%% SQL:2016 standard: Section 7.5ff
%% PgSQL: https://www.postgresql.org/docs/10/static/sql-select.html
%% Oracle: https://docs.oracle.com/database/122/SQLRF/SELECT.htm#SQLRF01702
%% Blog: https://blog.jooq.org/2016/12/09/a-beginners-guide-to-the-true-order-of-sql-operations/

\begin{figure}
  \centering\small
  \newsavebox{\normalform}
  \begin{lrbox}{\normalform}
  \begin{lstlisting}[language=sql,numbers=none,mathescape=true]
SELECT $\cdots$
FROM  (SELECT DISTINCT ON($\cdots$) $\cdots$
       FROM   (SELECT $\cdots$ $\var{AGG}$($\cdots$) OVER($\var{w}$ $\phi$) $\cdots$
               FROM   (SELECT $\cdots$ $\var{AGG}$($\cdots$) $\cdots$
                       FROM   (SELECT $\cdots$
                               FROM   $q$,$\dots$,$q$
                               WHERE  $p$) AS $\var{t}$
                       GROUP BY $g$
                       HAVING $p$) AS $\var{t}$
               WINDOW $\var{w}$ AS ($\cdots$)) AS $\var{t}$
       ORDER BY $o$) AS $\var{t}$
ORDER BY $o$
OFFSET $n$
LIMIT $n$
  \end{lstlisting}
  \end{lrbox}
  \begin{tikzpicture}
    \fill[black!10] (-2.15,-1.0)  rectangle (3.5,1.9);  % DISTINCT ON
    \fill[black!20] (-1.15,-0.7)  rectangle (3.3,1.6);  % WINDOW
    \fill[black!30] (-0.15,-0.4)  rectangle (3.1,1.3);  % GROUP BY
    \fill[black!40] ( 0.87, 0.15) rectangle (2.9,1.0);  % join (inner)
    \node at (0,0) {\usebox{\normalform}};
  \end{tikzpicture}
  \vskip-5mm
  \caption{Syntactic shape of a normalized~\sql{SELECT} block after \SQL{} clause isolation.
    Any but the innermost layer of the ``onion'' may be missing.}
  \label{fig:normal-form}
\end{figure}

\smallskip
\begin{changed}
\begin{definition}{Syntactic Transformation~$\comp$}
Given a normalized subject query~$q$, the syntax-directed mapping
$$
q \comp \twophase{\phase{1}{q}}{\phase{2}{q}}
$$
derives both $q$'s instrumented variant~$\phase{1}{q}$ and
interpreter~$\phase{2}{q}$. Mapping~$\comp$ is collectively defined by
the inferences rules of~\cref{fig:rules} and
\ifarXiv
\cref{fig:rules-extended} (in~\cref{app:rules}).
\else
the appendix.
\fi
\end{definition}
\end{changed}

\smallskip\noindent
The synchronized derivation
allows $\phase{1}{q}$ and $\phase{2}{q}$ to readily share information
about call sites~$\location{\ell}$ when we place a call
to~\sql{$\logwrite{$\Box$}$($\location{\ell}$,$\dots$)}
in~$\phase{1}{q}$ and its associated
\sql{$\logread{$\Box$}$($\location{\ell}$,$\dots$)} call in~$\phase{2}{q}$.
(The inference rules invoke~$\location{\ell} = \fresh$ to obtain arbitrary yet fresh call site
identifiers~$\location{\ell}$.)

\begin{changed}
\cref{fig:rules} displays a representative subset of the complete rule
set. Taken jointly with the additions of
\ifarXiv
\cref{fig:rules-extended} in
\fi
the appendix, the rules cover the rich \SQL{} dialect characterized in
the introduction and can translate the visibility query
of~\cref{fig:visibility-query} as well as the 22~queries of the \TPCH{}
benchmark (see~\cref{sec:performance} below).
\end{changed}
In the rules' antecedents, we use $\vert q_i \comp
\twophase{\cdot}{\cdot}\vert_{i=1,\dots,n}$ to indicate that all
(sub)queries $q_1,\dots,q_n$ are to be transformed.

\smallskip\noindent
Mapping~$\comp$ proceeds bottom-up and first establishes trivial
interpreters for \SQL{}'s syntactic leaf constructs. No logging is
required in these cases. Rule~\labelcref{rule:literal}: A literal~$l$
represents itself: its interpreter thus returns the empty
set~$\abs{\pempty}$ of input data dependencies.
Rule~\labelcref{rule:colref}: In Phase~$\ph{2}$, a column
reference~\sql{$t$.$c$} holds a set of cell identifiers that represents
\sql{$t$.$c$}'s data dependencies (see~\cref{def:interpreter}). The rule thus simply returns this set.
\cref{rule:table} ensures that Phase~$\ph{1}$ operates over regular base
data held in the cells of~$\phase{1}{\fn{table}}$ while Phase~$\ph{2}$
reads (singleton) dependency sets from~$\phase{2}{\fn{table}}$ that
represent these cells (cf.\ \cref{fig:terrain-map-phase12}).

\smallskip\noindent
Non-leaf rules first invoke~$\comp$ on constituent queries and assemble
the results to form composite instrumentations and interpreters.
\cref{rule:builtin-func} manifests that the evaluation of a built-in
\SQL{} operator~$\oplus$ (returning a single scalar, row, or array
value) depends on all of its $n$~arguments~$e_i$.  The interpreter thus
unions the arguments' dependency sets $\abs{\phase{2}{e_i}}$.
\cref{rule:with} invokes~$\comp$ recursively on the
common table expression~$q_i$ but otherwise preserves the syntactic
shape of the input query (\cref{sec:parametricity}).  The rule does,
however, extend the schemata of all CTEs to expose new column~$\row$
whose row identifiers help to relate the results of Phases~$\ph{1}$
and~$\ph{2}$ (again, see~\cref{fig:terrain-map-phase12}).
\begin{changed}
Shape preservation in~\cref{rule:with}, specifically, presents the
opportunity to use \SQL{}'s \sql{WITH} to assign a name, say~$t$, to any
intermediate query result of interest.  After interpretation,
table~$\inphase{2}{t}$ will hold the \where- and \why-provenance of the
intermediate result. The ability to inspect such intermediate provenance
(as computed by common table expression~$\inphase{2}{\sql{max\_scan}}$, for example,
see~\cref{fig:visibility-phase2}) can be instrumental in the analysis
and debugging of very complex queries.
\end{changed}

\smallskip\noindent
\cref{rule:join} infers the instrumentation and interpreter for $m$-fold
joins.  Such joins (or its simpler variants, see
\ifarXiv
\cref{app:rules})
\else
the appendix)
\fi
form the
innermost layer of the onion. All other \SQL{} clauses of the
current~\sql{SELECT} block are placed in enclosing layers.

As discussed in~\cref{sec:phase1}, instrumented query~$\phase{1}{i}$
invokes~\sql{$\logwrite{JOIN$\langle m\rangle$}$} to record which
combinations of rows satisfied join predicate~$p$ and to obtain a new row
identifier~$\row$ that represents the joined row---otherwise, the input query and
$\inphase{1}{i}$ perform the same computation.  Interpreter~$\phase{2}{i}$ re-enacts
the join based on the log and~\sql{$\logread{JOIN$\langle m\rangle$}$}
as described in~\cref{sec:phase2}.  Since, in the input query, the
evaluation of~$p$ determined the inclusion of a joined row with its
columns~$c_1,\dots,c_n$, we
collect~$\abs{\phase{2}{e_i} \mathbin{\cup} \sql{\toY($\phase{2}{p}$)}}$
to form the full \where- and \why-pro\-ve\-nance for
column~$c_i$.

% ldelim bug / relevant for Tobias:
% turn
%   \ldelim\{{4.6}{*}
% into
%   \ldelim\{{4}{*}

\begin{figure}[t]
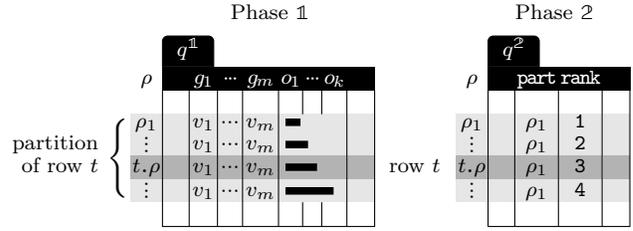

  \newcommand{\rank}[1]{\tikz[baseline=-2pt,inner sep=0mm,outer sep=0mm,x=0.1mm] \draw[line width=1mm] (0,0) -- (#1,0);}%
  \centering\small%
  \begin{tabular}{@{}c@{}c@{}}
    \hskip10mm Phase~$\ph{1}$ & \hskip3mm Phase~$\ph{2}$ \\[1ex]
    \begin{littbl}
      \begin{tabular}{@{\hskip-2mm}rcccc|c|c|c|c|c|c|>{~}l@{}}
                               &                   &                                            & \tabname{2}{~$\inphase{1}{q}$~\strut} \\
                               &                   & $\row$                                     & \colhd{\quad}                              & \colhd{$g_1$}             & \colhd{$\ncdots$}             & \colhd{$g_m$}             & \colhd{$o_1$}                 & \colhd{$\ncdots$}    & \colhd{$o_k$}        & \colhd{\quad} \\
                               &                   & \strut                                     & \multicolumn{1}{|c|}{}                     &                           &                               &                           &                               &                      &                      & \\
                               & \ldelim\{{4.6}{*} & \cellcolor{black!10}$\row_1$               & \multicolumn{1}{|c|}{\cellcolor{black!10}} & \cellcolor{black!10}$v_1$ & \cellcolor{black!10}$\ncdots$ & \cellcolor{black!10}$v_m$ & \multicolumn{3}{l|}{\cellcolor{black!10}\rank{20}}   & \cellcolor{black!10} \\
       \textnormal{partition}  &                   & \cellcolor{black!10}$\nvdots$              & \multicolumn{1}{|c|}{\cellcolor{black!10}} & \cellcolor{black!10}$v_1$ & \cellcolor{black!10}$\ncdots$ & \cellcolor{black!10}$v_m$ & \multicolumn{3}{l|}{\cellcolor{black!10}\rank{30}}   & \cellcolor{black!10} \\
       \textnormal{of row~$t$} &                   & \cellcolor{black!30}\strut\sql{$t$.$\row$} & \multicolumn{1}{|c|}{\cellcolor{black!30}} & \cellcolor{black!30}$v_1$ & \cellcolor{black!30}$\ncdots$ & \cellcolor{black!30}$v_m$ & \multicolumn{3}{l|}{\cellcolor{black!30}\rank{42}}   & \cellcolor{black!30} & \textnormal{row~$t$} \\
                               &                   & \cellcolor{black!10}$\nvdots$              & \multicolumn{1}{|c|}{\cellcolor{black!10}} & \cellcolor{black!10}$v_1$ & \cellcolor{black!10}$\ncdots$ & \cellcolor{black!10}$v_m$ & \multicolumn{3}{l|}{\cellcolor{black!10}\rank{64}}   & \cellcolor{black!10} \\
                               &                   & \strut                                     & \multicolumn{1}{|c|}{}                     &                           &                               &                           &                               &                      &                      & \\
        \cline{4-11}
      \end{tabular}
    \end{littbl}
    &
    \begin{littbl}
      \begin{tabular}{@{\hskip-1mm}ccc|c|c|@{}}
                                                   & \tabname{2}{~$\inphase{2}{q}$~\strut} \\
        $\row$                                     & \colhd{\quad}                              & \colhd{part}             & \colhd{rank}         & \colhd{\quad} \\
        \strut                                     & \multicolumn{1}{|c|}{}                     &                               &                       & \\
        \cellcolor{black!10}$\row_1$               & \multicolumn{1}{|c|}{\cellcolor{black!10}} & \cellcolor{black!10}$\row_1$  & \cellcolor{black!10}1 & \cellcolor{black!10} \\
        \cellcolor{black!10}$\nvdots$              & \multicolumn{1}{|c|}{\cellcolor{black!10}} & \cellcolor{black!10}$\row_1$  & \cellcolor{black!10}2 & \cellcolor{black!10} \\
        \cellcolor{black!30}\strut\sql{$t$.$\row$} & \multicolumn{1}{|c|}{\cellcolor{black!30}} & \cellcolor{black!30}$\row_1$  & \cellcolor{black!30}3 & \cellcolor{black!30} \\
        \cellcolor{black!10}$\nvdots$              & \multicolumn{1}{|c|}{\cellcolor{black!10}} & \cellcolor{black!10}$\row_1$  & \cellcolor{black!10}4 & \cellcolor{black!10} \\
        \strut                                     & \multicolumn{1}{|c|}{}                     &                               &                       & \\
        \cline{2-5}
      \end{tabular}
    \end{littbl}
  \end{tabular}
  \caption{Placement of row~$t$ in a windowed table with clause
    \sql{WINDOW\;$w$\;AS\;(PARTITION\;BY\;$g_1$,$\ndots$,$g_m$\;ORDER\;BY\;$o_1$,$\ndots$,$o_k$)}. All rows
    in $t$'s partition agree on~$\sql{FIRST\_VALUE($t$.$\row$)\;OVER\;($w$)} = \row_1$. In its partition,
    $t$ ranks 3rd (bars\,\protect\rank{24}\,picture the ordering criteria). Pair $(\row_1,\sql{3})$ thus
    exactly pinpoints $t$'s placement in Phase~$\ph{2}$.}
  \label{fig:window-row}
\end{figure}

\smallskip\noindent
\cref{rule:group} instruments \sql{GROUP BY}~queries to collect the row
identifiers of the current group (via the set
aggregate~$\smash{\setagg}\,\{\sql{$t$.$\row$}\}$).
\sql{$\logwrite{GRP}$} logs the resulting row identifier set along with a unique group
identifier~$\row$. When Phase~$\ph{2}$ processes row~$t$, it
invokes~\sql{$\logread{GRP}$($\location{\ell}$,$t$.$\row$)} to retrieve
the identifier~\sql{$\var{group}$.$\row$} of $t$'s group as a stand-in
grouping criterion.  The interpreter thus faithfully re-enacts the
grouping performed in~Phase~$\ph{1}$.  In Phase~$\ph{2}$,
\cref{rule:aggregate-fun} turns a value-based
aggregate~\sql{$\var{AGG}$($\inphase{1}{e}$)} into a set aggregate
$\abs{\smash{\setagg}\,\inphase{2}{e}}$ that collects the dependencies
of all evaluations of its argument~$e$ (this models \SQL{}'s aggregate
semantics~\cite[\S\,4.16.4]{sql-2016}).  To this \where-provenance,
\cref{rule:group} adds the
\why-provenance~$\abs{\smash{\setagg}\,\varY{group}}$ to reflect
\begin{inparaenum}[(1)]
\item that the criteria~$g_i$ jointly determined which group a row belongs to and
\item \begin{changed}% Benjamin
        that \sql{HAVING} predicate~$p$ decided the group's inclusion in the result.
      \end{changed}
\end{inparaenum}

\begin{figure*}
  \begin{narrow}{-5mm}{-5mm}
  \centering\small
  \renewcommand{\arraystretch}{0.9}%
  \begin{tabular}{@{}c@{}}
  \inferrule{\relax}{%
    l \comp \twophase{l}{\abs{\pempty}}}
    \thisisrule{Lit}\label{rule:literal}
  \quad
  \inferrule{\relax}{%
    t\sql{.}c \comp \twophase{t\sql{.}c}{\abs{t\sql{.}c}}}
    \thisisrule{Col}\label{rule:colref}
  \quad
  \inferrule{\relax}{%
    \fn{table} \comp \twophase{\phase{1}{\fn{table}}}
                               {\phase{2}{\fn{table}}}}
    \thisisrule{Table}\label{rule:table}
  \quad
  \inferrule{%
    \oplus \in \{\cdot\mathbin{\sql{+}}\cdot, \cdot\mathbin{\sql{<=}}\cdot, \sql{ROW($\cdot$,$\ndots$,$\cdot$)}, \sql{$\cdot$ IN ($\cdot$,$\ndots$,$\cdot$)}, \ndots\} \and\and
    \bigl\lvert e_i \comp \twophase{\inphase{1}{e_i}}{\inphase{2}{e_i}} \bigr\rvert_{i=1\ndots n}}{%
    \oplus\sql{(}e_1\sql{,}\ndots\sql{,}e_n\sql{)} \comp \twophase{\oplus\sql{(}\inphase{1}{e_1}\sql{,}\ndots\sql{,}\inphase{1}{e_n}\sql{)}}
                                                            {\abs{\inphase{2}{e_1} \cup \cdots \cup \inphase{2}{e_n}}}}
    \thisisrule{Builtin}\label{rule:builtin-func}
  \\[9mm]

  \inferrule{%
    \bigl\lvert q_i \comp \twophase{\inphase{1}{q_i}}{\inphase{2}{q_i}} \bigr\rvert_{i=0\ndots n} \quad
    \inphase{1}{i} =
      \begin{array}{@{}l@{~}l}
      \sql{\textbf{WITH}}~[\sql{\textbf{RECURSIVE}}] & \phase{1}{t_1}\sql{(}\row,c_{11}\sql{,}\ndots\sql{,}c_{1k_1}\sql{)}~\sql{\textbf{AS}}~\sql{(}\inphase{1}{q_1}\sql{),}\ndots\sql{,} \\[1ex]
                                                     & \phase{1}{t_n}\sql{(}\row,c_{n1}\sql{,}\ndots\sql{,}c_{nk_n}\sql{)}~\sql{\textbf{AS}}~\sql{(}\inphase{1}{q_n}\sql{)} \\[1ex]
                                                     & \inphase{1}{q_0}
      \end{array}
    \quad
    \inphase{2}{i} =
      \begin{array}{@{}l@{~}l}
      \sql{\textbf{WITH}}~[\sql{\textbf{RECURSIVE}}] & \phase{2}{t_1}\sql{(}\row,c_{11}\sql{,}\ndots\sql{,}c_{1k_1}\sql{)}~\sql{\textbf{AS}}~\sql{(}\inphase{2}{q_1}\sql{),}\ndots\sql{,} \\[1ex]
                                                     & \phase{2}{t_n}\sql{(}\row,c_{n1}\sql{,}\ndots\sql{,}c_{nk_n}\sql{)}~\sql{\textbf{AS}}~\sql{(}\inphase{2}{q_n}\sql{)} \\[1ex]
                                                     & \inphase{2}{q_0}
      \end{array}
    }{%
    \begin{array}{@{}l@{~}l@{}}
      \sql{\textbf{WITH}}~[\sql{\textbf{RECURSIVE}}] & t_1\sql{(}c_{11}\sql{,}\ndots\sql{,}c_{1k_1}\sql{)}~\sql{\textbf{AS}}~\sql{(}q_1\sql{),}\ndots\sql{,} \\
                                                     & t_n\sql{(}c_{n1}\sql{,}\ndots\sql{,}c_{nk_n}\sql{)}~\sql{\textbf{AS}}~\sql{(}q_n\sql{)} \\
                                                     & q_0
    \end{array}
    \comp
    \twophase{\inphase{1}{i}}{\inphase{2}{i}}}
    \thisisrule{With}\label{rule:with}\\
  \\[2mm]

  \inferrule{%
    m \geqslant 1 \and
    \bigl\lvert e_i \comp \twophase{\inphase{1}{e_i}}{\inphase{2}{e_i}}\bigr\rvert_{i=1\ndots n} \and
    \bigl\lvert q_i \comp \twophase{\inphase{1}{q_i}}{\inphase{2}{q_i}}\bigr\rvert_{i=1\ndots m} \and
    p \comp \twophase{\inphase{1}{p}}{\inphase{2}{p}} \and
    \location{\ell} = \fresh \and\and\and
    \inphase{1}{i} =
      \begin{array}{l@{~}l}
        \sql{\textbf{SELECT}}   & \logwrite{JOIN$\langle m\rangle$}\sql{(}\location{\ell}\sql{,} t_1\sql{.}\row\sql{,}\ndots\sql{,} t_m\sql{.}\row\sql{)}~\sql{\textbf{AS}}~\row\sql{,} \\
                                &  \inphase{1}{e_1}~\sql{\textbf{AS}}~c_1\sql{,} \ndots\sql{,} \inphase{1}{e_n}~\sql{\textbf{AS}}~c_n \\
        \sql{\textbf{FROM}}     & \inphase{1}{q_1}~\sql{\textbf{AS}}~t_1\sql{,} \ndots\sql{,}[\sql{\textbf{LATERAL}}]~\inphase{1}{q_m}~\sql{\textbf{AS}}~t_m \\
        \sql{\textbf{WHERE}}   & \inphase{1}{p}
      \end{array}
    \and
    \inphase{2}{i} =
      \begin{array}{l@{~}l}
        \sql{\textbf{SELECT}}   & \var{join}\sql{.}\row~\sql{\textbf{AS}}~\row\sql{,}~\abs{\inphase{2}{e_1}} \mathbin{\Y{\abs{\cup}}} \Y{\abs{\varY{join}}}~\sql{\textbf{AS}}~c_1\sql{,} \ndots\sql{,} \abs{\inphase{2}{e_n}} \mathbin{\Y{\abs{\cup}}} \Y{\abs{\varY{join}}}~\sql{\textbf{AS}}~c_n \\
        \sql{\textbf{FROM}}     & \inphase{2}{q_1}~\sql{\textbf{AS}}~t_1\sql{,} \ndots\sql{,}[\sql{\textbf{LATERAL}}]~\inphase{2}{q_m}~\sql{\textbf{AS}}~t_m\sql{,}~ \\
                                & \sql{\textbf{LATERAL}}~\logread{JOIN$\langle m\rangle$}\sql{(}\location{\ell}\sql{,} \langle t_1\sql{.}\row\sql{,}\ndots\sql{,} t_m\sql{.}\row\rangle\sql{)}~\sql{\textbf{AS}}~\var{join}\sql{(}\row\sql{)}, \\
                                & \Y{\sql{\textbf{LATERAL}}~\abs{\toY\sql{(}\inphase{2}{p}\sql{)}}~\sql{\textbf{AS}}~\varY{join}}
      \end{array}
    }{%
    \begin{array}{l@{~}l}
      \sql{\textbf{SELECT}}   & e_1~\sql{\textbf{AS}}~c_{1}\sql{,} \ndots\sql{,} e_n~\sql{\textbf{AS}}~c_n \\
      \sql{\textbf{FROM}}     & q_1~\sql{\textbf{AS}}~t_1\sql{,}\ndots\sql{,}[\sql{\textbf{LATERAL}}]~q_m~\sql{\textbf{AS}}~t_m \\
      \sql{\textbf{WHERE}}   & p
    \end{array}
    \comp
    \twophase{\inphase{1}{i}}{\inphase{2}{i}}}
    \thisisrule{Join}\label{rule:join}\\
  \\[2mm]

  \inferrule{%
     q \comp \twophase{\inphase{1}{q}}{\inphase{2}{q}} \and
     p \comp \twophase{\inphase{1}{p}}{\inphase{2}{p}} \and
    \bigl\lvert e_i \comp \twophase{\inphase{1}{e_i}}{\inphase{2}{e_i}}\bigr\rvert_{i=1\ndots n} \and
    \bigl\lvert g_i \comp \twophase{\inphase{1}{g_i}}{\inphase{2}{g_i}}\bigr\rvert_{i=1\ndots m} \and
    \location{\ell} = \fresh \and\and\and
    \inphase{1}{i} =
      \begin{array}{l@{~}l}
        \sql{\textbf{SELECT}}   & \logwrite{GRP}\sql{(}\location{\ell}\sql{,} \smash{\setagg}\,\{t\sql{.}\row\}\sql{)}~\sql{\textbf{AS}}~\row\sql{,} \\
                                & \inphase{1}{e_1}~\sql{\textbf{AS}}~c_1\sql{,} \ndots\sql{,}~\inphase{1}{e_n}~\sql{\textbf{AS}}~c_n \\
        \sql{\textbf{FROM}}     & \inphase{1}{q}~\sql{\textbf{AS}}~t \\
        \sql{\textbf{GROUP BY}} & \inphase{1}{g_1}\sql{,} \ndots\sql{,} \inphase{1}{g_m} \\
        \sql{\textbf{HAVING}}   & \inphase{1}{p}
      \end{array}
    \and
    \inphase{2}{i} =
      \begin{array}{l@{~}l}
        \sql{\textbf{SELECT}}   & \var{group}\sql{.}\row~\sql{\textbf{AS}}~\row\sql{,}~\abs{\inphase{2}{e_1} \Y{\cup \smash[t]{\setagg}\,\varY{group}}}~\sql{\textbf{AS}}~c_1\sql{,} \ndots\sql{,} \abs{\inphase{2}{e_n} \Y{\cup \smash[t]{\setagg}\,\varY{group}}}~\sql{\textbf{AS}}~c_n \\
        \sql{\textbf{FROM}}     & \inphase{2}{q}~\sql{\textbf{AS}}~t\sql{,} \\
                                & \sql{LATERAL}~\logread{GRP}\sql{(}\location{\ell}\sql{,} t\sql{.}\row\sql{)}~\sql{\textbf{AS}}~\var{group}\sql{(}\row\sql{)}, \\
                                & \Y{\sql{LATERAL}~\abs{\toY\sql{(}\inphase{2}{g_1} \cup \cdots \cup \inphase{2}{g_m} \cup \inphase{2}{p}\sql{)}}~\sql{\textbf{AS}}~\varY{group}} \\
        \sql{\textbf{GROUP BY}} & \var{group}\sql{.}\row
      \end{array}
    }{%
    \begin{array}{l@{~}l}
      \sql{\textbf{SELECT}}   & e_1~\sql{\textbf{AS}}~c_1\sql{,} \ndots\sql{,} e_n~\sql{\textbf{AS}}~c_n \\
      \sql{\textbf{FROM}}     & q~\sql{\textbf{AS}}~t \\
      \sql{\textbf{GROUP BY}} & g_1\sql{,} \ndots\sql{,} g_m \\
      \sql{\textbf{HAVING}}   & p
    \end{array}
    \comp
    \twophase{\inphase{1}{i}}{\inphase{2}{i}}}
    \thisisrule{Group}\label{rule:group} \\
  \\[2mm]

  \inferrule{%
    e \comp \twophase{\inphase{1}{e}}{\inphase{2}{e}}}{%
    \sql{\var{AGG}($e$)} \comp \twophase{\sql{\var{AGG}($\inphase{1}{e}$)}}
                                  {\abs{\setagg \inphase{2}{e}}}}
    \thisisrule{Agg}\label{rule:aggregate-fun}
  \quad
  \inferrule{%
    \sql{\var{AGG}($e$)} \comp \twophase{\inphase{1}{a}}{\inphase{2}{a}} \and
     \Y{\varY{} = \abs{\setagg \varY{win}~\sql{\textbf{OVER}}~\sql{(}\var{w}~\phi\sql{)}}}
    }{%
    \sql{\var{AGG}($e$)}~\sql{\textbf{OVER}}~\sql{(}\var{w}~\phi\sql{)}
    \comp
    \twophase{\inphase{1}{a}~\sql{\textbf{OVER}}~\sql{(}\var{w}~\phi\sql{)}}
             {\abs{\inphase{2}{a}~\sql{\textbf{OVER}}~\sql{(}\var{w}~\phi\sql{)}\mathbin{\Y{\cup}} \Y{\varY{}}}}}
    \thisisrule{AggWin}\label{rule:aggregate-window-fun}\\
  \\[3mm]

  \inferrule{%
    q \comp \twophase{\inphase{1}{q}}{\inphase{2}{q}} \and
    \bigl\lvert e_i \comp \twophase{\inphase{1}{e_i}}{\inphase{2}{e_i}}\bigr\rvert_{i=1\ndots n} \and
    \bigl\lvert g_i \comp \twophase{\inphase{1}{g_i}}{\inphase{2}{g_i}}\bigr\rvert_{i=1\ndots m} \and
    \bigl\lvert o_i \comp \twophase{\inphase{1}{o_i}}{\inphase{2}{o_i}}\bigr\rvert_{i=1\ndots k} \and
    \location{\ell} = \fresh \and\and\and
    \inphase{1}{i} =
      \begin{array}{l@{~}l}
        \sql{\textbf{SELECT}}   & \begin{array}[t]{@{}l@{}l@{}}
                                    \logwrite{WIN}\sql{(} & \location{\ell}\sql{,}t\sql{.}\row\sql{,}
                                                            \sql{FIRST\_VALUE(}t\sql{.}\row\sql{)}~\sql{\textbf{OVER}}~\sql{(}\var{w}\sql{),} \\
                                                          & ~~~~~~~~~\sql{RANK()}~\sql{\textbf{OVER}}~\sql{(}\var{w}\sql{))}
                                                            ~\sql{\textbf{AS}}~\row\sql{,}
                                  \end{array} \\
                                & \inphase{1}{e_1}~\sql{\textbf{AS}}~c_1\sql{,} \ndots\sql{,} \inphase{1}{e_n}~\sql{\textbf{AS}}~c_n \\
        \sql{\textbf{FROM}}     & \inphase{1}{q}~\sql{\textbf{AS}}~t \\
        \sql{\textbf{WINDOW}}   & \var{w}~\sql{\textbf{AS}}~
                                    \begin{array}[t]{@{}l@{}l@{~}l@{}l}
                                      \sql{(} & \sql{\textbf{PARTITION BY}} & \inphase{1}{g_1}\sql{,}\ndots\sql{,}\inphase{1}{g_m} & \\
                                              & \sql{\textbf{ORDER BY}}     & \inphase{1}{o_1}~\fn{dir}_1\sql{,}\ndots\sql{,}\inphase{1}{o_k}~\fn{dir}_k & \sql{)}
                                    \end{array}
      \end{array}
    \and
    \inphase{2}{i} =
      \begin{array}{l@{~}l}
        \sql{\textbf{SELECT}}   & t\sql{.}\row~\sql{\textbf{AS}}~\row\sql{,}~\abs{\inphase{2}{e_1}}~\sql{\textbf{AS}}~c_1\sql{,} \ndots\sql{,} \abs{\inphase{2}{e_n}}~\sql{\textbf{AS}}~c_n \\
        \sql{\textbf{FROM}}     & \inphase{2}{q}~\sql{\textbf{AS}}~t\sql{,} \\
                                & \sql{\textbf{LATERAL}}~\logread{WIN}\sql{(}\location{\ell}\sql{,} t\sql{.}\row\sql{)}~\sql{\textbf{AS}}~\var{win}\sql{(\partition,\peer),} \\
                                & \Y{\sql{\textbf{LATERAL}}~\abs{\toY\sql{(}\inphase{2}{g_1} \cup \cdots \cup \inphase{2}{g_m} \cup \inphase{2}{o_1} \cup \cdots \cup \inphase{2}{o_k}\sql{)}}~\sql{\textbf{AS}}~\varY{win}}\\
        \sql{\textbf{WINDOW}}   & \var{w}~\sql{\textbf{AS}}~
                                    \begin{array}[t]{@{}l@{}l@{~}l@{}l}
                                      \sql{(} & \sql{\textbf{PARTITION BY}} & \var{win}\sql{.}\partition & \\
                                              & \sql{\textbf{ORDER BY}}     & \var{win}\sql{.}\peer & \sql{)}
                                    \end{array}
      \end{array}
    }{%
    \begin{array}{l@{~}l}
      \sql{\textbf{SELECT}}   & e_1~\sql{\textbf{AS}}~c_1\sql{,} \ndots\sql{,} e_n~\sql{\textbf{AS}}~c_n \\
      \sql{\textbf{FROM}}     & q~\sql{\textbf{AS}}~t \\
      \sql{\textbf{WINDOW}}   & \var{w}~\sql{\textbf{AS}}~
                                  \begin{array}[t]{@{}l@{}l@{~}l@{}l}
                                    \sql{(} & \sql{\textbf{PARTITION BY}} & g_1\sql{,}\ndots\sql{,}g_m & \\
                                            & \sql{\textbf{ORDER BY}}     & o_1~\fn{dir}_1\sql{,}\ndots\sql{,}o_k~\fn{dir}_k & \sql{)}
                                  \end{array}
    \end{array}
    \comp
    \twophase{\inphase{1}{i}}{\inphase{2}{i}}}
    \thisisrule{Window}\label{rule:window}
  \end{tabular}
  \end{narrow}
  \caption{Excerpt of inference rules $q \comp \twophase{\inphase{1}{i}}{\inphase{2}{i}}$
    that derive instrumented query~$\inphase{1}{i}$ and interpeter~$\inphase{2}{i}$
    from input query $q$.}
  \label{fig:rules}
\end{figure*}

\smallskip\noindent
The rows of a windowed table are partitioned and then ordered before a
window---or: frame---of rows is formed around each input
row~$t$~\cite[\S\,4.15.14]{sql-2016}.  \cref{rule:window} thus injects a
call to~\sql{$\logwrite{WIN}$} that logs the identifier of $t$'s
partition as well as the row's intra-partition position (\cref{fig:window-row}
illustrates).  Later, the interpreter reads the pair back (cf.\
\sql{$\var{win}$.part} and~\sql{$\var{win}$.rank} in $\inphase{2}{i}$)
to correctly place~$t$ among its peers. Since the interpretation of
windowed aggregates preserves their original frame clause~$\phi$
(see~\cref{rule:aggregate-window-fun}), Phase~$\ph{2}$ builds
\where-provenance from exactly those dependency sets found in the frame
around row~$t$.\footnote{The \sql{max\_scan} CTE of~\cref{fig:visibility-query} omits the
default frame clause $\phi\mathrel{\equiv}$ \sql{RANGE BETWEEN UNBOUNDED
PRECEDING AND CURRENT ROW}. \cref{rule:window,rule:aggregate-window-fun}
work for arbitrary frames.}
Again, this coincides with the \SQL{} window
aggregate semantics~\mbox{\cite[\S\,4.16.3]{sql-2016}}. Much like in the
\sql{GROUP BY}~case, the rules add \why-provenance based on the
partitioning and ordering criteria~$g_i$ and~$o_j$, respectively (these
are collected in~$\abs{\varY{win}}$ and then added
in~\cref{rule:aggregate-window-fun}).

\smallskip\noindent
\textbf{Rule characteristics.} The mapping rules for~$\comp$
discussed here exhibit  general properties that are
characteristic for the full rule set:
\begin{compactitem}
\item
  \Why-provenance may be optionally derived in addition to \where-provenance.
  If we omit the \Y{lighter} subexpressions in the definitions of
  the~$\inphase{2}{i}$, interpretation will compute \where-provenance only.
  Since the \why-provenance of an output cell can be substantial (\emph{e.g.},
  in~\cref{rule:aggregate-fun,rule:aggregate-window-fun}, the rows of an
  entire group or window frame contribute their dependency sets), we can
  expect significant time and space savings if we skip the derivation
  of \why-dependencies.
\item During interpretation, provenance sets
  grow monotonically (once found, dependencies are never thrown away).
  This helps to devise a simple and efficient internal representation
  of provenance sets.
\end{compactitem}

%% - BuiltIn, With
%% - Clause isolation allows to consider SQL constructs like joins or window functions seperately:
%%   - Join (simpler cases/no loggin{}g: Map, Select)
%%   - Window/AggWinFun/AggFun
%%   - TblFun

%% - rationale for (selected) rules
%%   - refer to SQL:2016 semantics (if useful/possble)
%%
%%   - Window: representative of a SQL construct with complex semantics: the approach can cope
%%     - may assume single-table FROM: normalization (q represents lower onion layers)
%%     - FIRST_VALUE(t.ρ) OVER (w) identical in partition and unique => identify partition row t is placed in
%%     - RANK() OVER (w) => represent t's position inside the partition: the ⋃ e² OVER (w) will only
%%        aggregate those rows as dependencies that are BEFORE t in the partition
%%        (point to SQL semantics, SQL standard)
%%
%%   - AggFun: result of aggregate depends on all aggregated values ⇒ union all dependencies
%%
%%   - AggWinFun: - result of window aggregate depends on all rows inside window defined by w and ɸ
%%                  ⇒ union all dependencies (see AggFun) inside window
%%                - forming of windows (why-)depends on partition criteria g and ordering criteria o
%%
%%   - TblFun:  tblf(e₁,...,eₙ) yields a set of rows, each of these rows
%%              depend on the same arguments e₁,...,eₙ
%%

\subsection{Log Writing and Reading}
\label{sec:log-writing-reading}

In the absence of concrete values, the interpreters consult logs
(via~$\logread{$\Box$}$ calls) to re-enact relevant value-based
computations performed in Phase~$\ph{1}$.  Pseudo code for
three~$\logread{$\Box$}$ functions and
their~$\logwrite{$\Box$}$ pendants is shown
in~\cref{fig:logwrite-logread}.  These functions invoke lower-level routines
$\var{put}_{\Box}$ and $\var{get}_{\Box}$ that write to and read from
log file~$\var{log}_{\Box}$.
The log files may be realized in various forms,
\emph{e.g.}, in terms of operating system files or indexed relational tables.
\begin{changed}
Below, we discuss the details of logging but abstract from any
particular implementation.
\ifarXiv
\cref{app:log-files-relational}
\else
The appendix
\fi
shows
concrete log contents for a set of sample tables and queries and also
elaborates on a purely relational, indexed encoding of log files.  The \SQL{}-based
implementation described there has been used in the upcoming~\cref{sec:performance}.
\end{changed}

\smallskip\noindent
Lower-level logging routines are:
\begin{compactitem}
\item $\var{put}_{\Box}(\location{\ell},k,e)$: add record $\langle\location{\ell},k,e\rangle$
  to file~$\var{log}_{\Box}$, then return entry~$e$.
\item $\var{get}_{\Box}(\location{\ell},k)$: from~$\var{log}_{\Box}$, return the set of~$e$
  found in records $\langle\location{\ell},k,e\rangle$.  Return~$\varnothing$ if there
  are no matching records.
\item $\var{row}()$: generate and return a new unique row identifier.
\end{compactitem}

\smallskip\noindent
A
record~$\langle\location{\ell},\langle\row_1,\ndots,\row_m\rangle,\row\rangle$
in file~$\var{log}_{\sql{JOIN}}$ indicates that a \sql{FROM} clause at
site~$\location{\ell}$ joined $m$~rows~$\row_1,\dots,\row_m$ to yield a
new row~$\row$. Function~$\logwrite{JOIN$\langle m\rangle$}$ ensures
that this fact is recorded \emph{once} in the log: only if a join
of~$\row_1,\dots,\row_m$ at~$\location{\ell}$ has not been encountered
before (\emph{i.e.},
$\var{get}_{\sql{JOIN}}(\location{\ell},\langle\row_1,\ndots,\row_m\rangle)
= \varnothing$), a log entry with a fresh~$\row$ is made. Phase~$\ph{1}$
may attempt such repeated identical writes to~$\var{log}_{\sql{JOIN}}$
if site~$\location{\ell}$ is located inside a subquery                         %% @#@#@
which the query optimizer decided to evaluate more than once (this may
happen in the \TPCH{} benchmark, for example,
see~\cref{sec:performance}). In such scenarios, $\logwrite{JOIN$\langle
m\rangle$}$ makes sure that its side-effect on~$\var{log}_{\sql{JOIN}}$
is not carried out repeatedly.  This \emph{write once} safeguard also ensures
that~$\logread{JOIN$\langle
m\rangle$}(\location{\ell},\langle\row_1,\ndots,\row_m\rangle)$ will
either yield a set of~0 or~1 row identifiers---recall that the
interpreter~$\inphase{2}{i}$ of~\cref{rule:join} uses this behavior to
properly re-enact the join semantics. Analogous remarks
apply to~$\logwrite{GRP}$/$\logread{GRP}$
and~$\logwrite{WIN}$/$\logread{WIN}$.

\begin{figure}
  \lstnewenvironment{pseudo}{\lstset{mathescape=true,boxpos=t}}{}
  \renewcommand*\thelstnumber{\relax} %% lst option numbers=none yields a LaTeX error
  %% explains {...ϱ...} notation
  \newcommand{\rhoelem}{%
    \begin{tabular}{@{}c@{}}
      match sets $\Rho$ \\ with $\row \in \Rho$
    \end{tabular}}
  \centering\small
  \begin{tabular}{>{\hskip-2mm}l|l@{}}
\begin{pseudo}
%\logwrite{JOIN$\langle m\rangle$}%($\location{\ell}$,$\row_1$,$\ndots$,$\row_m$):
  $\{\row\} \shortleftarrow \var{get}_{\sql{JOIN}}(\location{\ell},\langle\row_1,\!\ndots\!,\row_m\rangle)$
  if $\{\row\} = \varnothing$ then
  $\lfloor$ $\row \shortleftarrow \var{put}_{\sql{JOIN}}(\location{\ell},\langle\row_1,\!\ndots\!,\row_m\rangle,\var{row}())$
  return $\row$
\end{pseudo}
  &
\begin{pseudo}
%\logread{JOIN$\langle m\rangle$}%($\location{\ell}$,$\langle\row_1$,$\ndots$,$\row_m\rangle$):
  $\sql{return}\,\var{get}_{\sql{JOIN}}(\location{\ell},\langle\row_1,\!\ndots\!,\row_m\rangle)$
\end{pseudo}
  \\ \\
\begin{pseudo}
$\logwrite{GRP}$($\location{\ell}$,$\{\row_1,\!\ndots\!,\row_n\}$):
  $\{\row\} \shortleftarrow \var{get}_{\sql{GRP}}(\location{\ell},\{\row_1,\!\ndots\!,\row_n\})$
  if $\{\row\} = \varnothing$ then
  $\lfloor$ $\row \shortleftarrow \var{put}_{\sql{GRP}}(\location{\ell},\{\row_1,\!\ndots\!,\row_n\},\var{row}())$
  return $\row$
\end{pseudo}
  &
\begin{pseudo}
$\logread{GRP}$($\location{\ell}$,$\row$):
  $\sql{return}\,\var{get}_{\sql{GRP}}(\location{\ell},\lst@commentstyle\underbrace{\vrule height 0mm depth 2mm width 0mm\color{black}\{\ndots\!,\row,\!\ndots\}}_{\clap{\textnormal{\scriptsize\rhoelem}}})$
\end{pseudo}
  \\ \\
\begin{pseudo}
$\logwrite{WIN}$($\location{\ell}$,$\row$,$\row_{\fn{part}}$,$\fn{rank}$):
  if $\var{get}_{\sql{WIN}}(\location{\ell},\row) = \varnothing$ then
  $\lfloor$ $\var{put}_{\sql{WIN}}(\location{\ell},\row,\langle\row_{\fn{part}},\fn{rank}\rangle)$
  return $\row$
\end{pseudo}
  &
\begin{pseudo}
$\logread{WIN}$($\location{\ell}$,$\row$):
  $\sql{return}\,\var{get}_{\sql{WIN}}(\location{\ell},\row)$
\end{pseudo}
  \end{tabular}
  \vskip-2mm
  \caption{Pseudo code: \sql{$\logwrite{$\Box$}$}/\sql{$\logread{$\Box$}$} function
    pairs for log writing and reading, $\Box \in \{\sql{JOIN$\langle m\rangle$},
    \sql{GRP},\sql{WIN}\}$.}
  \label{fig:logwrite-logread}
\end{figure}

\section{The Provenance Tax}
\label{sec:performance}

%% - [good]: Q9✔︎ (speedup!)
%% - P1 [bad]: Q17✔︎ (high slowdown despite small log size)
%% - P1+P2Y [bad]: Q11✔︎(slowdowns of 1000×), [good]: Q2✔︎ | Q19✔︎ (slowdown of only about ~2×, small logs!)
%%   (→ relates to prov representation size?)
%% - Pset [disappointing]: Q9 | Q19 , [good]: Q1 | Q10 | Q11 |  | | Q20 | Q22 (speedup of one magnitude)
%%   (→ related to prov cardinality?  But Q10+Q22 show good speedup despite small prov cardinality; maybe
%%    large intermediate prov card?)
%% - row granule [disappointing]: Q13 | Q19 | Q21 (small speedup), [good]: Q1 | Q10 | Q11 | Q20 | (one magnitude faster)

Provenance derivation processes substantially more data than the
value-based computation it explains. First, we trade the value-based query~$q$
for the pair $\twophase{\inphase{1}{q}}{\inphase{2}{q}}$: effectively, the subject query is executed
twice.  Second, we expect that
\begin{compactdesc}
\item[$\inphase{1}{q}$ is costly:] The two queries communicate via log files.
  Log file writes in~$\inphase{1}{q}$ lead to additional data movement
  and incur side effects that may constrain the query optimizer.
\item[$\inphase{2}{q}$ is costly:] Where~$q$ outputs 1NF cell values, $\inphase{2}{q}$
returns entire sets of dependencies.  These dependency sets may be large, \emph{e.g.},
  if $q$ invokes aggregate functions.
\end{compactdesc}
This section aims to quantify how high this ``provenance tax'' indeed
is and how it correlates with general query characteristics.
\begin{changed}
On the way, we demonstrate how variations of the provenance granularity,
dependency set representation, and an awareness of the properties of
set operations can lead to significant runtime improvements.
\end{changed}

The experiments below derive the full \where- and \why-provenance for
all 22~queries of the \TPCH{} benchmark~\cite{tpch}. Here, we set the
benchmark's scale factor to~$1$, \emph{i.e.},
table~\sql{lineitem} holds about $6\,000\,000$ rows.
\begin{changed}
A repetition of the experiments at \TPCH{} scale factor~$10$ shows
how the approach scales with growing database instances: see
\ifarXiv
\cref{app:tpch-ten-gb}
\else
the appendix
\fi
which reports slowdowns and speed-ups nearly identical to those observed in the
discussion below.
\end{changed}

\smallskip\noindent
All queries execute on a \Pg\,9.5 engine
hosted on a Linux (kernel~4.4) machine with two 4-core Intel~Xeon~5570
CPUs, 70\,GB of RAM, and harddisk-based secondary storage.  We report
the average performance of five runs with best and worst execution times ignored.
%%   - (System parameters are: 2×Intel Xeon X5570 @ 2.93GHz, L1 cache: 8MB, total of 8 (= 2×4) cores,
%%      70GB RAM, ext4 filesystem on Seagate Barracuda SATA 7200rpm, 3.0GB/s transfer, PostgreSQL 9.5, Linux kernel 4.4)
Instead of absolute wall clock times we focus on the slowdown---or
speed-up---we observe once we switch from value to dependency set
computation.  In all plots below, a slowdown of $\mathbin{\times} 1$ represents the
evaluation time of the original \TPCH{} queries (no provenance derived).
Queries~\tpchQ{1} to~\tpchQ{22} are displayed across the horizontal axes; the
plots are thus best read column by column.

\smallskip\noindent
\textbf{The (non-)impact of normalization.}
\cref{fig:performance} summarizes the impact of the individual phases of
provenance derivation.  The ``onion-style'' query normalization
(\cref{fig:normal-form}) does not alter the semantics and---on its
own---also appears to preserve query performance (see
the points~\tikz\draw[fill=white] (0,0) rectangle ++(2\squaresize,2\squaresize); cluster
around~$\mathbin{\times} 1$).  We have found RDBMSs to successfully remove
the simple uncorrelated nesting in the~\sql{FROM} clause and generate plans
identical to those for the original \TPCH{} queries.
For~\tpchQ{9}, the explicit onion nesting leads \Pg{} to aggregate first and sort later which
even beats the system's usual planning in which these operations are swapped.
(Out of curiosity, we also fed the 22~original and normalized queries
into HyPer~\cite{hyper} with its advanced query unnesting
procedure~\cite{hyper-unnesting} and found no plan differences at all.)

\smallskip\noindent
\begin{changed}
An analysis of the experiments reveals four major subject query characteristics
that influence the overhead of provenance derivation in Phases~$\ph{1}$ and~$\ph{2}$.
\cref{fig:query-buckets} shows these query categories and how the 22~\TPCH{} queries fit in.
We discuss this categorization below, phase by phase.
\end{changed}

\begin{figure*}
  \centering\small
  %% the plot
  \begin{tikzpicture}[x=0.72mm,y=8mm]
    \begin{scope}[thick]
      %% x axis (TPC-H queries)
      \draw (-22,0) -- (220,0);
      %% y axes
      \draw[->] (-1, 0.05) -- (-1, 3.9);
      \node[anchor=south,align=center] at (-1,3.9) {slowdown};
      \draw[->] (-1,-0.05) -- (-1,-1.7);
      \node[anchor=east,align=right] at (-3,-1.7) {log size \\\relax [kBytes]};
    \end{scope}
    %% queries (TPC-H query Q#/norm ratio/p1 ratio/p1+p2 ratio/p1+p2y ratio/log cells/write sites/show measurement?)
    \foreach \tpch/\n/\p/\pp/\ppy/\logsize/\sites/\measure [count=\x from 0] in {
      %% optimized experiments  (19.05.2018)
      1  /0.98    /4.02    /18.32   /53.88   /255856  /1 /0,
      2  /1.00    /1.27    /1.31    /1.38    /80      /3 /1,
      3  /1.00    /1.96    /2.92    /4.27    /2888    /3 /0,
      4  /1.02    /3.44    /4.12    /8.48    /4544    /2 /1,
      5  /1.06    /1.17    /1.29    /1.50    /752     /2 /0,
      6  /1.02    /2.45    /3.31    /3.66    /4048    /1 /0,
      7  /0.98    /1.11    /1.19    /1.66    /616     /2 /0,
      8  /0.97    /1.20    /2.02    /2.51    /416     /3 /0,
      9  /0.78    /2.46    /4.42    /8.25    /32608   /2 /1,
      10 /0.99    /3.30    /6.43    /17.77   /10824   /3 /0,
      11 /0.98    /16.69   /17.51   /1039.12 /3296    /3 /1,
      12 /1.01    /2.05    /2.64    /3.54    /5376    /4 /0,
      13 /0.99    /13.39   /15.81   /22.63   /175496  /3 /0,
      14 /0.99    /3.45    /5.06    /5.68    /6576    /2 /0,
      15 /1.00    /1.57    /2.71    /6.01    /9784    /2 /0,
      16 /0.98    /4.93    /8.32    /17.06   /10248   /3 /0,
      17 /0.99    /99.12   /99.21   /99.86   /296     /2 /1,
      18 /0.98    /1.18    /1.19    /1.22    /64      /3 /1,
      19 /1.00    /1.11    /1.22    /1.24    /8       /1 /1,
      20 /0.98    /2.08    /2.11    /42.49   /720     /4 /0,
      21 /1.00    /28.20   /28.23   /28.42   /13120   /4 /0,
      22 /1.01    /5.44    /5.74    /11.46   /2184    /3 /0
    } {
      %% measurements: query runtime ratios (+ log size)
      \begin{scope}[yshift=5mm]
        %% normalized query
        \pgfmathsetmacro\y{log10(\n)}
        \draw[black,fill=white] (2+10*\x,\y)+(-\squaresize,-\squaresize) rectangle ++(\squaresize,\squaresize);
        \ifnum\x=0
          \node[legend] at (2+10*\x,\y) {normalized};
        \fi
        %% p1
        \pgfmathsetmacro\y{log10(\p)}
        \draw[fill=black] (4+10*\x,\y)+(-\squaresize,-\squaresize) rectangle ++(\squaresize,\squaresize);
        \ifnum\x=0
          \node[legend] at (4+10*\x,\y) {$\ph{1}$};
        \fi
        %% p1+p2
        \pgfmathsetmacro\y{log10(\pp)}
        \draw[black,fill=white] (6+10*\x,\y) circle (\circlesize);
        \ifnum\x=0
          \node[legend] at (6+10*\x,\y) {$\ph{1}\mathord{+}\ph{2}$};
        \fi
        %% p1+p2y
        \pgfmathsetmacro\y{log10(\ppy)}
        \draw[fill=black] (8+10*\x,\y) circle (1.5pt);
        \ifnum\x=0
          \node[legend] at (8+10*\x,\y) {$\ph{1}\mathord{+}\ph{2}$\toY};
        \fi
        %% selected measurements
        \ifnum\measure=1
          \pgfkeys{/pgf/number format/.cd,std,precision=0}
          \node[legend,anchor=east,black,yshift=-1pt] at (8+10*\x,\y) {\pgfmathprintnumber{\ppy}};
        \fi
      \end{scope}
      %% measurements: p1 log size
      \begin{scope}
        \pgfkeys{/pgf/fpu=true,/pgf/fpu/output format=fixed} %% activate high-precision arithmetics in TikZ
        \pgfmathsetmacro\y{-(log10(\logsize) - 1) / 5}
        \pgfkeys{/pgf/fpu=false}
        \draw[black,fill=black] (5+10*\x,-0.02)+(-\barsize,0) rectangle ++(\barsize,\y);
        \node[legend,black,left=-2pt] at (3+10*\x,0) {$\pgfmathprintnumber{\logsize}$};
        %% # of write sites
        \node[legend,black] at (5.2+10*\x,\y-0.45) {\circled{\sites}};
      \end{scope}
      %% query columns
      \begin{pgfonlayer}{background}
        \pgfmathsetmacro\even{15*(1-mod(int(\tpch),2))}
        \draw[draw=none,fill=black!\even] (0+10*\x,-1.5) rectangle (10+10*\x,3.7);
      \end{pgfonlayer}
      %% TPC-H query number
      \node[anchor=north,font=\scriptsize] at (5+10*\x, -1.5) {\tpchQ{\tpch}};
    };
    %% bands (slowdown)
    \begin{pgfonlayer}{background}
      \begin{scope}[yshift=5mm,band]
        \foreach \slowdown in {1,10,100,1000} {
          \pgfmathsetmacro\y{log10(\slowdown)}
          \draw (-3,\y) -- (220,\y);
          \node[anchor=east] at (-3,\y) {$\mathop{\times} \pgfmathprintnumber{\slowdown}$};
        };
      \end{scope}
      %% bands (log size)
      \begin{scope}[band]
        \foreach \logsize in {1000,1000000} {
          \pgfkeys{/pgf/fpu=true,/pgf/fpu/output format=fixed} %% activate high-precision arithmetics in TikZ
          \pgfmathsetmacro\y{-(log10(\logsize) - 1) / 5}
          \pgfkeys{/pgf/fpu=false}
          \draw (-3,\y) -- (220,\y);
          \node[anchor=east] at (-3,\y) {$\pgfmathprintnumber{\logsize}$};
        };
      \end{scope}
    \end{pgfonlayer}
  \end{tikzpicture}
  \vskip-2mm
  \caption{Normalization
    (\protect\tikz\protect\draw[fill=white] (2.5,2.5) rectangle ++(2\squaresize,2\squaresize);),
    Phase~$\ph{1}$ (\protect\tikz\protect\draw[fill=black] (2.5,2.5) rectangle ++(2\squaresize,2\squaresize);),
    Phases~$\ph{1}$+$\ph{2}$
    (\protect\tikz\protect\draw[fill=white] (0,0) circle (\circlesize);
    without/\protect\tikz\protect\draw[fill=black] (0,0) circle (\circlesize); with \why-provenance)
    relative to value-based \TPCH{}.}
  \label{fig:performance}
\end{figure*}

\begin{figure}
  \centering\small
  \begin{tabular}{@{}c@{\quad}lc@{\ }c@{}}
    \toprule
                   &                               & \textbf{Yes}             & \textbf{No}           \\
                   & \textbf{Query Characteristic} & \textbf{(High Overhead)} & \textbf{(Low Overhead)} \\
    \cmidrule(lr){1-2}\cmidrule(l){3-4}
    \multirow{2}*[-2mm]{$\ph{1}$}
    & \begin{tabular}{@{}l@{}}
        non-selective? \\ \scriptsize (high \# of log writes)
      \end{tabular}
    & \scriptsize\tpchQ{1},\tpchQ{13}
    & \begin{tabular}{@{}c@{}}
        \scriptsize\tpchQ{2},\tpchQ{5},\tpchQ{7},\\
        \scriptsize\tpchQ{17},\tpchQ{18},\tpchQ{19}
      \end{tabular}
    \\
    \cmidrule(r){2-4}
    & \begin{tabular}{@{}l@{}}
          correlation? \\ \scriptsize (repeated side effects)
      \end{tabular}
    & \begin{tabular}{@{}c@{}}
        \scriptsize\tpchQ{4},\tpchQ{11},\tpchQ{17},\\
        \scriptsize\tpchQ{21},\tpchQ{22}
      \end{tabular}
    & \begin{tabular}{@{}c@{}}
        \scriptsize\tpchQ{1},\tpchQ{5},\tpchQ{6},\\
        \scriptsize\tpchQ{7},\tpchQ{12},\tpchQ{18}
      \end{tabular}
    \\
    \cmidrule(lr){1-4}
    \multirow{2}*[-2mm]{$\ph{2}$}
    & \begin{tabular}{@{}l@{}}
         high dependency \\ set cardinality?
      \end{tabular}
    & \begin{tabular}{@{}c@{}}
          \scriptsize\tpchQ{1},\tpchQ{9},\tpchQ{11},\tpchQ{13},   \\
          \scriptsize\tpchQ{15},\tpchQ{16},\tpchQ{20},\tpchQ{22}
       \end{tabular}
    & \begin{tabular}{@{}c@{}}
        \scriptsize\tpchQ{2},\tpchQ{3},\tpchQ{10}, \\
        \scriptsize\tpchQ{17},\tpchQ{19}
      \end{tabular}
    \\
    \cmidrule(r){2-4}
    & \begin{tabular}{@{}l@{}}
         expensive \\ \why-provenance?
      \end{tabular}
    & \scriptsize\tpchQ{1},\tpchQ{11},\tpchQ{20}
    & \begin{tabular}{@{}c@{}}
        \scriptsize\tpchQ{5},\tpchQ{6},\tpchQ{8},\tpchQ{14}, \\
        \scriptsize\tpchQ{18},\tpchQ{19},\tpchQ{21}
       \end{tabular}
    \\
    \bottomrule
  \end{tabular}
  \vskip-2mm
  \caption{%
    \color{changed}
    Query characteristics that influence provenance overhead in
    Phases~$\ph{1}$ and~$\ph{2}$ (all \TPCH{} queries categorized).}
  \label{fig:query-buckets}
\end{figure}

\begin{changed}
\subsection{Phase~{\large $\ph{1}$}: Impact of Logging}
\label{sec:performance-phase1}

Relative to the original
\TPCH{} queries, we observe a geometric mean slowdown of~$3.4$ in
Phase~$\ph{1}$ (see~\tikz[baseline=-1pt]\draw[fill] (0,0)
rectangle ++(2\squaresize,2\squaresize); in~\cref{fig:performance}). The
gaps
\begin{tikzpicture}[sketch]
  \draw[fill=white] (0,0) rectangle ++(2\squaresize,2\squaresize);
  \draw[fill] (1.5,1.5) rectangle ++(2\squaresize,2\squaresize);
\end{tikzpicture}
are a measure of the logging effort that the instrumented queries
invest.
The log sizes and call site counts in~\circled{\phantom{2}} at the
bottom of~\cref{fig:performance} show that, on average, a \TPCH{} query
contains~$2.5$~\logwrite{$\Box$} call sites that log just below~24\,MB
of data if we use the tabular representation of logs described
in
\ifarXiv
\cref{app:log-files-relational}.
\else
the appendix.
\fi

\smallskip\noindent
\textbf{Selectivity and logging overhead.} Selective filters and joins
reduce data \emph{as well as} log volume (recall the placement
of~\logwrite{JOIN$\langle m\rangle$} in~\cref{rule:join}).
Queries~\tpchQ{2},
\tpchQ{18}, \tpchQ{19} show this most clearly and only induce negligible
overhead in Phase~$\ph{1}$.  The opposite holds for~\tpchQ{1} (whose
non-selective predicates let almost all~$6\,000\,000$ rows
of~\sql{lineitem} pass) and \tpchQ{13} (in which a left outer join
requires the logging of the identifiers of both qualifying and
non-qualifying pairs of rows).  Both queries log substantial data volumes
and exhibit large Phase~$\ph{1}$ overhead.

\smallskip\noindent
\textbf{Correlation and side-effecting log writes.}
\logwrite{$\Box$} call sites located in a subquery that the RDBMS fails
to decorrelate and unnest (a problem that already occurs with the
original \TPCH{} query~\cite{tpch-pain-points}) trigger the functions'
guard against writing identical log entries repeatedly,
see~\cref{sec:log-writing-reading}.  In our implementation, this
increases the cost of~\logwrite{$\Box$} to about~$0.14\,\text{ms}$ per
call. Queries~\tpchQ{4}, \tpchQ{11}, \tpchQ{17}, \tpchQ{21}, and
\tpchQ{22} contain such correlated subqueries in their~\sql{WHERE} and~\sql{HAVING}
clauses and show the Phase~$\ph{1}$ cost of avoiding these unwanted side
effects.

\smallskip\noindent
\textbf{Logging without side effects?} \emph{Tupling}~\cite{tupling}
suggests a functionally pure implementation alternative in which a row is extended
by \emph{an extra column} that holds its associated log entry: instead of
issuing a side-effecting log write,
\logwrite{JOIN$\langle 2\rangle$} constructs and returns a
pair $(\row,\langle\row_1,\row_2\rangle)$ to indicate that rows~$\row_1$
and~$\row_2$ were joined to form a new row~$\row$, for example.  For~\tpchQ{17} and~\tpchQ{21},
this shows a promising runtime improvement of factor~$76$ and~$28$ in Phase~$\ph{1}$, respectively.
However, tupling complicates the treatment of constructs like scalar or~\sql{IN} subqueries
which, effectively, now need to be executed twice (once yielding the original, once the extended rows).
Queries like~\tpchQ{18} thus are penalized.  Tupling bears a promising performance advantage
in Phase~$\ph{1}$ but we would
\begin{compactenum}[(1)]
\item lose fully compositional query transformation (the use of tupling would
  be conditioned on the absence of the mentioned query constructs) and
\item sacrifice query shape preservation (\cref{sec:parametricity}) and ultimately
  face the same problems as \Perm{} and \GProM{}~(\cref{sec:perm}).
\end{compactenum}
We consider the conditional use of tupling an interesting item of future work.
\end{changed}

\begin{changed}
\subsection{\hskip-2mm Phase~{\large $\ph{2}$}: Computing with Dependency Sets}
\label{sec:performance-phase2}

We derive provenance through the composition of Phases~$\ph{1}$ and~$\ph{2}$.
Measurement~\tikz\draw[fill=white] (0,0) circle (\circlesize);
in~\cref{fig:performance} thus reflects the \emph{overall slowdown if both
phases are executed in sequence}.
We find a mean slowdown of factor~$4.6$ (visualized by the
\begin{tikzpicture}[sketch]
  \draw[fill=white] (0,-0.5) rectangle ++(2\squaresize,2\squaresize);
  \draw[fill=white] (2,2) circle (\circlesize);
\end{tikzpicture} gaps) compared to value-based query
evaluation.

\smallskip\noindent
\textbf{Dependency set cardinality.}
Where a value-based query manipulates an 1NF cell value~$v$, its
interpreter will construct $v$'s---possibly large---dependency set: if
we consider the entire \TPCH{} benchmark and form the mean, we find that
each output data cell depends on about $10\,000$~input cells.
When a single cell holds an aggregate of a group (or window) of rows, its dependency set
cardinality directly reflects the group (or window) size, see~\cref{rule:aggregate-fun}.
Foremost, this affects~\tpchQ{1} and its eight aggregates, one of
which (column~\sql{sum\_charge}) yields a \where-dependency set of about~$4\,500\,000$ elements
per output cell.  As an aggregation-heavy OLAP benchmark, \TPCH{} generally constitutes
a challenging workload in this respect (see~\cref{fig:query-buckets}).

\smallskip\noindent
\textbf{Expensive \why-provenance.}
Recall that we can selectively enable the derivation of \why-provenance
in Phase~$\ph{2}$. If we do, we experience larger overall overheads as
marked by the points~\tikz\draw[fill] (0,0) circle (\circlesize);
in~\cref{fig:performance}, with a mean overall slowdown of factor~$9.0$.
While the logs encode the outcome of a predicate~$p$, this does not suffice
to derive \why-provenance: we now also need to interpret~$p$ (\emph{i.e.}, evaluate~$\inphase{2}{p}$)
to learn which input items influenced $p$'s value.  For~\tpchQ{11}, in particular,
this requires the interpretation of a complex \sql{HAVING~$p$} clause where~$p$ contains
a three-way join and aggregation.  \tpchQ{1} now additionally derives
how aggregates depend on grouping critera (see subexpressions~$g_i$ in~\cref{rule:group}),
doubling~\sql{sum\_charge}'s dependency cardinality to~$9\,000\,000$ input cells per output cell.

\smallskip\noindent
\textbf{Dependency set representation.}
Given these substantial dependency cardinalities, it is expected that
Phase~$\ph{2}$ can benefit from efficient set representations.
\ifarXiv
\cref{app:dep-set-rep}
\else
The appendix
\fi
indeed makes this observation if we replace the
\Pg{}-native set encoding based on type~\sql{int[]} by bit sets~\cite{roaring}.
\end{changed}

\begin{figure*}[t]
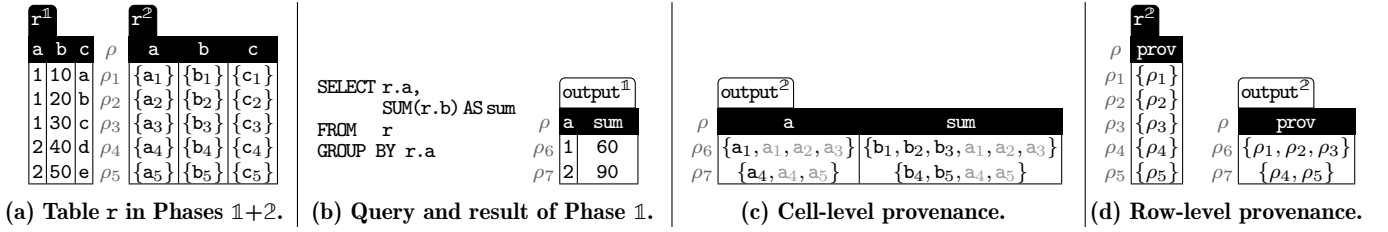

  \begin{narrow}{-2mm}{-2mm}
  \centering\small
  \subcaptionbox{Table~$\sql{r}$ in Phases~$\ph{1}$+$\ph{2}$.\label{fig:prov-row-input}}{%
    \centering
    \begin{littbl}
      \setlength\minrowclearance{1pt}%
      \begin{tabular}{c|c|l|>{\,\color{black!60}}c<{\,}c|c|c|}
        \tabname{3}{$\inphase{1}{\sql{r}}$\strut}       &           & \tabname{3}{$\inphase{2}{\sql{r}}$\strut}                                 \\
        \colhd{a}               & \colhd{b} & \colhd{c} &  $\row$   & \colhd{a}               & \colhd{b}    & \colhd{c}                        \\
        \multicolumn{1}{|c|}{1} & 10        & a         &  $\row_1$ & \multicolumn{1}{|c|}{$\{\sql{a}_1\}$} & $\{\sql{b}_1\}$ & $\{\sql{c}_1\}$ \\
        \multicolumn{1}{|c|}{1} & 20        & b         &  $\row_2$ & \multicolumn{1}{|c|}{$\{\sql{a}_2\}$} & $\{\sql{b}_2\}$ & $\{\sql{c}_2\}$ \\
        \multicolumn{1}{|c|}{1} & 30        & c         &  $\row_3$ & \multicolumn{1}{|c|}{$\{\sql{a}_3\}$} & $\{\sql{b}_3\}$ & $\{\sql{c}_3\}$ \\
        \multicolumn{1}{|c|}{2} & 40        & d         &  $\row_4$ & \multicolumn{1}{|c|}{$\{\sql{a}_4\}$} & $\{\sql{b}_4\}$ & $\{\sql{c}_4\}$ \\
        \multicolumn{1}{|c|}{2} & 50        & e         &  $\row_5$ & \multicolumn{1}{|c|}{$\{\sql{a}_5\}$} & $\{\sql{b}_5\}$ & $\{\sql{c}_5\}$ \\
        \cline{1-3}\cline{5-7}
      \end{tabular}
    \end{littbl}}
  \hfill\vrule\hfill
  \subcaptionbox{Query and result of Phase~$\ph{1}$.\label{fig:prov-row-query}}{%
    \centering
    \begin{tabular}[t]{c@{}c}
      \renewcommand{\arraystretch}{0.9}%
      $
      \begin{array}[t]{@{}l@{~}l@{}}
        \sql{SELECT}   & \sql{r.a,}              \\
                       & \sql{SUM(r.b)\,AS\,sum} \\
        \sql{FROM}     & \sql{r}                 \\
        \multicolumn{2}{@{}l}{\sql{GROUP BY r.a}}
      \end{array}
      $
      &
      \begin{littbl}
      \setlength\minrowclearance{1pt}%
        \begin{tabular}[t]{>{\color{black!60}}c@{\,}c|c|}
                                & \outputname{2}{$\inphase{1}{\sql{output}}$\strut}    \\
          $\row$                & \colhd{a}               & \colhd{sum} \\
          $\row_6\vphantom{\{}$ & \multicolumn{1}{|c|}{1} & 60  \\
          $\row_7\vphantom{\{}$ & \multicolumn{1}{|c|}{2} & 90  \\
          \cline{2-3}
        \end{tabular}
      \end{littbl}
    \end{tabular}}
  \vrule
  \subcaptionbox{Cell-level provenance.\label{fig:prov-row-cells}}{%
    \centering
    \begin{littbl}
      \setlength\minrowclearance{1pt}%
      \begin{tabular}{@{}>{\color{black!60}}c@{\,}c|c|@{}}
        & \outputname{2}{$\inphase{2}{\sql{output}}$\strut} \\
        $\row$ & \colhd{a} & \colhd{sum} \\
        $\row_6$ & \multicolumn{1}{|c|}{$\{\sql{a}_1,\Y{\sql{a}_1},\Y{\sql{a}_2},\Y{\sql{a}_3}\}$} & $\{\sql{b}_1,\sql{b}_2,\sql{b}_3,\Y{\sql{a}_1},\Y{\sql{a}_2},\Y{\sql{a}_3}\}$ \\
        $\row_7$ & \multicolumn{1}{|c|}{$\{\sql{a}_4,\Y{\sql{a}_4},\Y{\sql{a}_5}\}$} & $\{\sql{b}_4,\sql{b}_5,\Y{\sql{a}_4},\Y{\sql{a}_5}\}$ \\
        \cline{2-3}
      \end{tabular}
    \end{littbl}
  }
  \vrule
  \subcaptionbox{Row-level provenance.\label{fig:prov-row-row}}{%
    \centering
    \begin{littbl}
      \setlength\minrowclearance{1pt}%
      \begin{tabular}[b]{@{}>{\color{black!60}}c@{\,}c|@{}}
        & \tabname{1}{$\inphase{2}{\sql{r}}$\strut} \\
        $\row$   & \colhd{prov}        \\
        $\row_1$ & \multicolumn{1}{|c|}{$\{\row_1\}$} \\
        $\row_2$ & \multicolumn{1}{|c|}{$\{\row_2\}$} \\
        $\row_3$ & \multicolumn{1}{|c|}{$\{\row_3\}$} \\
        $\row_4$ & \multicolumn{1}{|c|}{$\{\row_4\}$} \\
        $\row_5$ & \multicolumn{1}{|c|}{$\{\row_5\}$} \\
        \cline{2-2}
      \end{tabular}
    \end{littbl}
    \hskip-3pt
    \begin{littbl}
      \setlength\minrowclearance{1pt}%
      \begin{tabular}[b]{>{\color{black!60}}c@{\,}c|}
        & \outputname{1}{$\inphase{2}{\sql{output}}$\strut} \\
        $\row$   & \colhd{prov}        \\
        $\row_6$ & \multicolumn{1}{|c|}{$\{\row_1,\row_2,\row_3\}$} \\
        $\row_7$ & \multicolumn{1}{|c|}{$\{\row_4,\row_5\}$} \\
        \cline{2-2}
      \end{tabular}
    \end{littbl}
  }
  \end{narrow}
  \vskip-2mm
  \caption{Provenance derivation at cell and row granularities for a simple~\sql{GROUP\,BY} query.}
  \label{fig:prov-row}
\end{figure*}

\smallskip\noindent
\textbf{Beneficial effects of logging.}
Logging incurs overhead in Phase~$\ph{1}$, but Phase~$\ph{2}$ can benefit from
the effort. To exemplify, in the
original~\tpchQ{19}, a join between~\sql{lineitem} and~\sql{part}
accounts for $98\%$ of the execution time.  In the
interpreted~\tpchQ{19}, table $\var{log}_{\sql{JOIN2}}$ acts much like a
join index or access support relation~\cite{join-indices,access-support-rels} from
which the row identifiers of the join partners are read off directly.
As a result, interpretation is about $10\mathbin{\times}$ faster than value-based evaluation.
The situation is similar for~\tpchQ{21}, where $\var{log}_{\sql{JOIN4}}$
assumes the role of a join index for an expensive four-way join.
Additionally, the interpreter saves the evaluation effort for two complex
\sql{$[$NOT$]$ EXISTS($\cdots$)} subqueries: the identifier of the
row that constitutes the existential quantifier's \why-provenance is simply
read off the log tables.
\begin{changed}
Access support relations that materialize provenance relationships
between rows have shown very similar beneficial effects in~\cite{proql}.
\end{changed}

\subsection{Switching From Cell to Row Granularity}
\label{sec:from-cell-to-row}

The present approach derives provenance at the granularity of
\emph{individual table cells}: each output cell is assigned the set of
input cells that influenced its value. We obtain highly detailed insight
into input-output data dependencies but surely pay a price in terms of
interpreter overhead and size of the resulting provenance.  It turns out
that this level of granularity is not firmly baked into the method.  We
can straightforwardly adapt it to operate at the less detailed
\emph{row} level which suffices for many uses and also is the
granularity provided by the majority of existing
work~\cite{lineage-tracing,semirings,glavic-phd,queries-explain-work,diff-data-flow}.
Below, we contrast both granularity levels, sketch how row-level
interpretation can be realized, and assess the resulting performance
advantage.

\smallskip\noindent
For the cell granularity case, consider 5-row input table~\sql{r} whose
Phase~$\ph{1}$ and~$\ph{2}$ variants are shown
in~\cref{fig:prov-row-input}. In~$\phase{2}{\sql{r}}$, each cell is
assigned a singleton dependency set (cf.\
\cref{fig:terrain-map-phase12}). If we use the \sql{GROUP\,BY} query
of~\cref{fig:prov-row-query} as the subject query, Phase~$\ph{1}$ yields
the~$\inphase{1}{\sql{output}}$ table shown in the same figure. Phase~$\ph{2}$
preserves the shape of the output but returns a table whose cells hold
dependency sets (\cref{fig:prov-row-cells}). Cell identifier shades
indicate the provenance kind (\where, \Y{\why}): to arrive at the
aggregate value~\sql{90} of row~$\row_7$, the query had to sum the input
cells $\sql{b}_4$, $\sql{b}_5$ (holding~\sql{40}, \sql{50}) and decide
group membership based on cells $\Y{\sql{a}_4}$, $\Y{\sql{a}_5}$ (both
holding~\sql{2}).

\begin{figure}
  \begin{narrow}{-3mm}{-3mm}
  \centering\small
   %% the plot
  \begin{tikzpicture}[x=0.33mm,y=5.8mm]
    \begin{scope}[thick]
      %% x axis (TPC-H queries)
      \draw (0,-1.3) -- (220,-1.3);
      %% y axes
      \draw[->] (-1, -1.3) -- (-1, 4.2);
      \node[anchor=south,align=center] at (-1,4.2) {slowdown/speed-up};
    \end{scope}
    %% queries (TPC-H query Q#/p2y ratio/p2y-tuple ratio/show measurement?)
    \foreach \tpch/\ppy/\ppytuple/\measure [count=\x from 0] in {
      %% new experiments (19.05.2018)
      1  /49.86   /2.26   /1,  % TPC-H result size: 4
      2  /0.11    /0.04   /0,  % TPC-H result size: 100
      3  /2.31    /0.96   /0,  % TPC-H result size: 10
      4  /5.03    /3.25   /0,  % TPC-H result size: 5
      5  /0.34    /0.15   /0,  % TPC-H result size: 5
      6  /1.21    /1.31   /0,  % TPC-H result size: 1
      7  /0.55    /0.11   /1,  % TPC-H result size: 4
      8  /1.31    /0.89   /1,  % TPC-H result size: 2
      9  /5.78    /1.79   /1,  % TPC-H result size: 175
      10 /14.47   /1.66   /1,  % TPC-H result size: 20
      11 /1022.43 /140.72 /1,  % TPC-H result size: 1048
      12 /1.49    /0.33   /1,  % TPC-H result size: 2
      13 /9.24    /2.64   /0,  % TPC-H result size: 42
      14 /2.23    /0.93   /0,  % TPC-H result size: 1
      15 /4.44    /1.88   /0,  % TPC-H result size: 1
      16 /12.13   /1.61   /1,  % TPC-H result size: 18314
      17 /0.74    /0.53   /0,  % TPC-H result size: 1
      18 /0.03    /0.01   /0,  % TPC-H result size: 57
      19 /0.13    /0.11   /0,  % TPC-H result size: 1
      20 /40.42   /3.87   /1,  % TPC-H result size: 186
      21 /0.22    /0.15   /0,  % TPC-H result size: 411
      22 /6.02    /1.62   /1   % TPC-H result size: 7
    } {
      %% measurements: query runtime ratios
      \begin{scope}[yshift=5mm]
        %% p1+p2y
        \pgfmathsetmacro\y{log10(\ppy)}
        \draw[fill=black] (2+10*\x,\y) circle (1.5pt);
        \ifnum\x=3
          \node[legend] at (2+10*\x,\y) {$\ph{2}$\toY~(cell)};
        \fi
        %% p1+p2y-tuple
        \pgfmathsetmacro\z{log10(\ppytuple)}
        \draw[draw=black,fill=white] (8+10*\x,\z) circle (1.5pt);
        \ifnum\x=3
          \node[legend] at (8+10*\x,\z) {$\ph{2}$\toY~(row)};
        \fi
        %% trend line
        \begin{pgfonlayer}{annotation}
          \draw[trend] (2+10*\x,\y) -- (8+10*\x,\z);
        \end{pgfonlayer}
        %% selected measurements
        \ifnum\measure=1
          \ifdim\ppy pt<1pt\pgfkeys{/pgf/number format/precision=2}
            \else\pgfkeys{/pgf/number format/precision=1}\fi
          \node[legend,anchor=east,black] at (2+10*\x,\y) {\pgfmathprintnumber{\ppy}};
          \ifdim\ppytuple pt<1pt\pgfkeys{/pgf/number format/precision=2}
            \else\pgfkeys{/pgf/number format/precision=1}\fi
          \node[legend,anchor=east,black] at (8+10*\x,\z) {\pgfmathprintnumber{\ppytuple}};
        \fi
      \end{scope}
      %% query columns
      \begin{pgfonlayer}{background}
        \pgfmathsetmacro\even{15*(1-mod(int(\tpch),2))}
        \draw[draw=none,fill=black!\even] (0+10*\x,-1.3) rectangle (10+10*\x,4);
      \end{pgfonlayer}
      %% TPC-H query number
      \node[anchor=east,rotate=90,font=\scriptsize] at (5+10*\x,-1.3) {\tpchQ{\tpch}};
    };
    %% bands (slowdown)
    \begin{pgfonlayer}{background}
      \begin{scope}[yshift=5mm,band]
        \pgfkeys{/pgf/number format/.cd,fixed,precision=2}
        \foreach \slowdown in {0.1,1,10,100,1000} {
          \pgfmathsetmacro\y{log10(\slowdown)}
          \draw (-3,\y) -- (220,\y);
          \node[anchor=east] at (-3,\y) {$\mathop{\times} \pgfmathprintnumber{\slowdown}$};
        };
      \end{scope}
    \end{pgfonlayer}
  \end{tikzpicture}
  \end{narrow}
  \vskip-3mm
  \caption{Deriving cell-level
    (\protect\tikz\protect\draw[fill=black] (0,0) circle (\circlesize);) vs.\ row-level provenance
    (\protect\tikz\protect\draw[fill=white] (0,0) circle (\circlesize);).}
  \label{fig:full-row}
\end{figure}

If we switch to row granularity, Phase~$\ph{1}$ remains unchanged.
Phase~$\ph{2}$ entirely abstracts from the input's columns and thus
assigns one singleton identifier set per row, see the modified
two-column version of~$\inphase{2}{\sql{r}}$
in~\cref{fig:prov-row-row}. A simplified interpreter (discussed below)
tracks row dependencies and finally emits the~$\inphase{2}{\sql{output}}$ table
in~\cref{fig:prov-row-row}. We learn that aggregate value~\sql{60} of
output row~$\row_6$ depends on input rows $\row_1, \row_2, \row_3$,
\emph{i.e.}, exactly those rows of table~\sql{r} that constitute the group in
which column~$\sql{a} = \sql{1}$.

\smallskip\noindent
The rules of
\ifarXiv
\cref{fig:rules,fig:rules-extended}
\else
\cref{fig:rules}
\fi
adapt to row granularity in a
systematic fashion. As mentioned, the definitions of the instrumented
queries~$\inphase{1}{i}$ remain as is.  Where the cell-level
interpreters~$\inphase{2}{i}$ track dependencies column by column, the
new row-level interpreters collect dependencies held in the
single~$\sql{prov}$ column.  In~\cref{rule:join}, $\inphase{2}{i}$ is now
defined as
$$
\inphase{2}{i}=
  \begin{array}{@{\,}l@{~}l@{}}
    \sql{\textbf{SELECT}} & \var{join}\sql{.}\row\sql{,} \abs{t_1\sql{.prov} \mathbin{\cup} \cdots \mathbin{\cup} t_m\sql{.prov}}~\sql{\textbf{AS}}~\sql{prov}\\
    \sql{\textbf{FROM}}   & \inphase{2}{q_1}~\sql{\textbf{AS}}~t_1\sql{,} \ndots\sql{,}[\sql{\textbf{LATERAL}}]~\inphase{2}{q_m}~\sql{\textbf{AS}}~t_m\sql{,} \\
                                & \sql{\textbf{LATERAL}}~\logread{JOIN$\langle m\rangle$}\sql{(}\location{\ell}\sql{,} \langle t_1\sql{.}\row\sql{,}\ndots\sql{,} t_m\sql{.}\row\rangle\sql{)}~\sql{\textbf{AS}}~\var{join}\sql{(}\row\sql{)}
  \end{array}.
$$

%% focus on selected row(s) → predicate in Phase 2, focus on selected column: adapt ColRef rule

%% SPEEDUP / SWITCH TO COARSER ROW GRANULARITY
%%
%% - collect where/why-dependencies of all cells in of a row in a single provenance set
%%   ⇒ row granularity as proposed/implemented by other systems: CWW00, GKT07, Gla10, CAA14, CLMR16 [see Pdgree DFG proposal]
%%   - invariably yields two-column result: ϱ | c₁&...&cₙ
%%   - see fig:prov-row
%% - sample adaptation of a rule JOIN (only i² shown):

%% fig:full-row
%% - expect ●↘○ [decreasing runtime with row granularity]
%% - no savings for Q17: Q17 only returns a single row anyway
%%   - NB: TPC-H result size not large in general -- bigger impact expected for queries with large results

\smallskip\noindent
At row granularity level, we process narrow two-column tables
(columns~$\row$, \sql{prov}) regardless of the width of the input and
output tables.  Also, compared to the cell-level variant, the interpreter
evaluates fewer $\cup/\bigcup$~operations that build smaller dependency sets:
in \TPCH{}, one output row has about~$2\,500$ dependencies on input rows
(mean across the benchmark).  \cref{fig:full-row} documents how
the interpretation overhead drops by an order of magnitude once we switch from cell-
(\tikz\draw[fill=black] (0,0.5) circle (\circlesize);) to
row-level (\tikz\draw[fill=white] (0,0.5) circle (\circlesize);) dependencies.

\subsection{A Comparison with Perm and GProM}
\label{sec:perm}

The computation of row-level dependencies also paves the way for a
direct comparison with
\Perm~\cite{glavic-phd,perm-rewriting,perm-subqueries,perm-festschrift}, a long-running
research effort that makes a genuine attempt at provenance derivation
for \SQL{}. In \Perm, input queries are translated into a multiset
algebra, rewritten and augmented for provenance computation, and then
translated back into \SQL{} for execution on \Pg{}. Unlike the present
work, \Perm{} opts for an invasive approach and adds code that sits
between the query rewriter and planner of~\Pg~8.3.  To any output
row~$o$, \Perm{} attaches all columns of those input rows that
influence~$o$'s computation (influence contribution
semantics~\cite{perm-rewriting})---if~$o$ has~$n$
influencing rows, $o$ is repeated~$n$ times in the result.  For
table~$\sql{r}$ and the~\sql{GROUP\,BY} query
of Figures~\labelcref{fig:prov-row-input}~and~\subref{fig:prov-row-query},
\Perm{} thus emits the table of~\cref{fig:prov-size-perm}.
Row~$(\sql{a},\sql{sum}) =
(\sql{1},\sql{60})$, for example, is contained three times as it depends on all
\begin{wrapfigure}[14]{i}[0mm]{4.35cm}
  \colorlet{overhead}{black!15}
  \centering\small
  \begin{littbl}
    \setlength\minrowclearance{1pt}%
    \begin{tabular}{r@{~}c|c!{\vrule\hskip0.8pt\vrule}>{\columncolor{overhead}}c|>{\columncolor{overhead}}c|>{\columncolor{overhead}}c|@{}}
        & \outputname{3}{\sql{output}\strut} \\
        & \colhd{a}                                   & \multicolumn{1}{H!{\vrule\hskip0.8pt\vrule}}{sum} & \colhd{r.a} & \colhd{r.b} & \colhd{r.c} \\
        \ldelim[{1.7}{*}[\scriptsize result]  & \multicolumn{1}{|c|}{1}                     & 60                                                & 1           & 10          & a \\
        & \multicolumn{1}{|c|}{2}                     & 90                                                & 2           & 40          & d \\
        & \multicolumn{1}{|c|}{\cellcolor{overhead}1} & \cellcolor{overhead}60                            & 1           & 20          & b \\
        & \multicolumn{1}{|c|}{\cellcolor{overhead}1} & \cellcolor{overhead}60                            & 1           & 30          & c \\
        \raisebox{2pt}{\ldelim[{-2.7}{*}[\scriptsize provenance]} & \multicolumn{1}{|c|}{\cellcolor{overhead}2} & \cellcolor{overhead}90                            & 2           & 50          & e \\
        \cline{2-6}
        \multicolumn{3}{c}{} & \multicolumn{3}{c}{$\underbracket[0.6pt]{\mskip80mu}_{\textnormal{provenance}}$}
    \end{tabular}
  \end{littbl}
  \vskip-2mm
  \caption{\Perm's fully normalized representation of result and provenance
    (\protect\tikz[baseline=2pt]\protect\fill[overhead] (0,0) rectangle (0.6,0.3);) for
    the \sql{GROUP\,BY}~query of~\cref{fig:prov-row-query}.}
  \label{fig:prov-size-perm}
\end{wrapfigure}
input rows with~\mbox{$\sql{a} = \sql{1}$}. Recall that row-level \SQL{}
interpretation represents the same provenance information in the
$\inphase{2}{\sql{output}}$ table of~\cref{fig:prov-row-row}.
\begin{changed}
\mbox{In practice,} the resulting redundancy can be significant,
as~\cref{fig:provenance-rep-size} illustrates: across the
\TPCH{} benchmark que\-ries, \Perm{}'s normalized representation of pro\-ve\-nance
consistently requires more space than dependency sets (we measured
a mean factor of~$19$).
\end{changed}

\begin{figure*}[t]
  \begin{minipage}[b]{0.36\linewidth}
  \centering\small
  \begin{tikzpicture}[x=0.36mm,y=3.8mm]
    \begin{scope}[thick]
      %% x axis (TPC-H queries)
      \draw (0,0) -- (119,0);
      %% y axes
      \draw[->] (-1, 0.05) -- (-1, 7.0);
      \node[anchor=south west,align=left] at (-21,6.9) {provenance \\ representation size};
    \end{scope}
    %% queries (TPC-H query Q#/p1+p2y ratio 1GB/p1+p2y ratio 10GB
    \foreach \tpch/\ppy/\perm/\measure [count=\x from 0] in {
      %% provenance sizes: p2Y (#row IDs) vs. Perm (#table cells)
      3  /   .0085/     .2402/0,
      19 /   .0224/     .3145/0,
      10 /   .0392/    1.2260/0,
      8  /  1.1649/   16.1072/0,
      7  /  1.7588/   30.7460/0,
      5  /  2.1429/   35.4603/0,
      12 /  6.0987/   86.7210/0,
      6  / 11.4160/  194.1212/0,
      14 / 13.9095/  197.3477/0,
      9  /108.4992/ 1692.7887/0,
      13 /163.3918/ 2914.4453/1,
      1  /591.6591/15383.1326/0
    } {
      %% measurements: query runtime ratios
      \begin{scope}[yshift=10mm]
        %% p1+p2y-tuple
        \pgfmathsetmacro\y{log10(\ppy)}
        \draw[fill=white] (2+10*\x,\y) circle (1.5pt);
        \ifnum\x=3
          \node[legend] at (2+10*\x,\y) {$\ph{1}\mathord{+}\ph{2}$\toY~(row)};
        \fi
        %% Perm
        \pgfmathsetmacro\z{log10(\perm)}
        \draw[fill=black] (8+10*\x,\z)+(-\squaresize,-\squaresize) rectangle ++(\squaresize,\squaresize);
        \ifnum\x=3
          \node[legend] at (8+10*\x,\z) {\Perm};
        \fi
        %% trend line
        \begin{pgfonlayer}{annotation}
          \ifnum\measure=1
            \pgfmathsetmacro\ratio{\perm > \ppy ? \perm / \ppy : \ppy / \perm}
            \pgfkeys{/pgf/number format/.cd,std,precision=0}
            \draw[trend] (2+10*\x,\y) to node[black,ratio,sloped,above=-1pt] {$\pgfmathprintnumber{\ratio}\scalebox{0.8}{$\times$}$} (8+10*\x,\z);
          \else
            \draw[trend] (2+10*\x,\y) -- (8+10*\x,\z);
          \fi
        \end{pgfonlayer}
      \end{scope}
      %% query columns
      \begin{pgfonlayer}{background}
        \pgfmathsetmacro\odd{15*mod(int(\x),2)}
        \draw[draw=none,fill=black!\odd] (0+10*\x,0) rectangle (10+10*\x,6.7);
      \end{pgfonlayer}
      %% TPC-H query number
      \node[anchor=east,rotate=90,font=\scriptsize] at (5+10*\x, 0) {\tpchQ{\tpch}};
    };
    %% bands (provenance representation size)
    \begin{pgfonlayer}{background}
      \begin{scope}[yshift=10mm,band]
        \foreach \psize [count=\pow from 2] in {0.01,0.1,1,10,100,1000,10000} {
          \pgfmathsetmacro\y{log10(\psize)}
          \draw (-3,\y) -- (119,\y);
          \node[anchor=east] at (-3,\y) {$10^{\pow}$};
        };
      \end{scope}
    \end{pgfonlayer}
  \end{tikzpicture}
  \vskip-2mm
  \caption{%
    \color{changed}
    Size of provenance representation:
    dependency sets~(\protect\tikz\protect\draw[fill=white] (0,0) circle (\circlesize);)
    vs.\ \Perm~(\protect\tikz\protect\draw[fill=black] (2.5,2.5) rectangle ++(2\squaresize,2\squaresize);).}
  \label{fig:provenance-rep-size}
  \end{minipage}
  \hfill
  \begin{minipage}[b]{0.6\linewidth}
  \centering\small
   %% the plot
  \begin{tikzpicture}[x=0.37mm,y=7.8mm]
    \begin{scope}[thick]
      %% x axis (TPC-H queries)
      \draw (0,0) -- (220,0);
      %% y axes
      \draw[->] (-1, 0.05) -- (-1, 4.2);
      \node[anchor=south east,align=right] at (5,4.2) {slowdown};
    \end{scope}
    %% queries (TPC-H query Q#/p1+p2y-tuple ratio/Perm ratio/show measurement?/is nested subquery)
    \foreach \tpch/\ppytuple/\perm/\measure/\cat [count=\x from 0] in {
      %% new experiments (19.05.2018)
      21 /28.34   /2.89    /1 /0,
      10 /4.96    /2.03    /1 /0,
      1  /6.29    /3.00    /0 /0,
      14 /4.39    /2.28    /0 /0,
      6  /3.75    /2.26    /0 /0,
      3  /2.92    /1.97    /0 /0,
      4  /6.69    /7.70    /0 /1,
      12 /2.38    /2.88    /0 /1,
      7  /1.23    /1.96    /0 /1,
      19 /1.22    /2.06    /0 /1,
      8  /2.08    /3.94    /0 /2,
      11 /157.41  /1284.93 /1 /2,
      5  /1.32    /12.94   /1 /2,
      18 /1.19    /12.92   /1 /2,
      9  /4.25    /90.84   /1 /2,
      15 /3.44    /133.73  /1 /3,
      13 /16.04   /1324.15 /1 /3,
      16 /6.55    /1116.87 /1 /3,
      22 /7.06    /7199.48 /1 /3,
      2  /1.31    /-999    /0 /3,       %% DNF
      20 /5.95    /-999    /0 /3,       %% DNF
      17 /99.65   /-999    /0 /3        %% DNF
    } {
      %% measurements: query runtime ratios
      \begin{scope}[yshift=5mm]
        %% p1+p2y-tuple
        \pgfmathsetmacro\y{log10(\ppytuple)}
        \draw[draw=black,fill=white] (2+10*\x,\y) circle (1.5pt);
        \ifnum\x=2
          \node[legend] at (2+10*\x,\y) {$\ph{1}\mathord{+}\ph{2}$\toY~(row)};
        \fi
        %% Perm
        \ifnum-999=\perm
          \typeout{Perm DNF}
        \else
          \pgfmathsetmacro\z{log10(\perm)}
          \ifnum\tpch=22
            \draw[draw=none,fill=black,fill opacity=0.3] (8+10*\x,\z)+(-\squaresize,-\squaresize) rectangle ++(\squaresize,\squaresize);
          \else
            \draw[fill=black] (8+10*\x,\z)+(-\squaresize,-\squaresize) rectangle ++(\squaresize,\squaresize);
          \fi
          \ifnum\x=2
            \node[legend] at (8+10*\x,\z) {\Perm};
          \fi
          %% trend line
          \begin{pgfonlayer}{annotation}
            \ifnum\measure=1
              \pgfmathsetmacro\ratio{\perm > \ppytuple ? \perm / \ppytuple : \ppytuple / \perm}
              \pgfkeys{/pgf/number format/.cd,std,precision=0}
              \draw[trend] (2+10*\x,\y) to node[black,ratio,sloped,above=-3pt] {$\pgfmathprintnumber{\ratio}\scalebox{0.8}{$\times$}$} (8+10*\x,\z);
            \else
              \draw[trend] (2+10*\x,\y) -- (8+10*\x,\z);
            \fi
          \end{pgfonlayer}
        \fi
      \end{scope}
      %% query columns
      \begin{pgfonlayer}{background}
        \pgfmathsetmacro\even{10*\cat}
        \ifnum-999=\perm
          \draw[draw=none,pattern=stripes,pattern color=black!\even] (0+10*\x,0) rectangle (10+10*\x,4);
        \else
          \draw[draw=none,fill=black!\even] (0+10*\x,0) rectangle (10+10*\x,4);
        \fi
      \end{pgfonlayer}
      %% TPC-H query number (highlight those without nested subqueries / prov semantics comparable to perm)
      \iftrue %% \ifnum\subq=1
        \node[anchor=east,rotate=90,font=\scriptsize] at (5+10*\x, 0) {\tpchQ{\tpch}};
      \else
        \node[anchor=east,rotate=90,font=\scriptsize] at (5+10*\x, 0) {\textbf{\tpchQ{\tpch}}};
      \fi
    };
    %% bands (slowdown)
    \begin{pgfonlayer}{background}
      \begin{scope}[yshift=5mm,band]
        \foreach \slowdown in {1,10,100,1000} {
          \pgfmathsetmacro\y{log10(\slowdown)}
          \draw (-3,\y) -- (220,\y);
          \node[anchor=east] at (-3,\y) {$\mathop{\times} \pgfmathprintnumber{\slowdown}$};
        };
      \end{scope}
    \end{pgfonlayer}
    %% categories
    \begin{pgfonlayer}{foreground}
      \begin{scope}[cat]
        \draw ( 60, 0.0) -- ( 60, 4.7);
        \draw (100, 0.0) -- (100, 4.2);
        \draw (150, 0.0) -- (150, 4.7);
        \node[black!75,align=center] at ( 30,4.5) {\circled{A} scans/joins \\ (+ \sql{G\;BY}) + agg};
        \node[black!75,align=center] at (105,4.5) {\circled{B} increasingly wider joins \\ + \sql{G\;BY} + agg + subqueries};
        \node[black!75,align=center] at (185,4.5) {\circled{C} complex queries \\  + nested subqueries};
      \end{scope}
    \end{pgfonlayer}
  \end{tikzpicture}
  \vskip-2mm
  \caption{Head-to-head: interpretation at row granularity
    (\protect\tikz\protect\draw[fill=white] (0,0) circle (\circlesize);) and
    \Perm{} (\protect\tikz\protect\draw[fill=black] (2.5,2.5) rectangle ++(2\squaresize,2\squaresize);).}
  \label{fig:perm-row-performance}
  \end{minipage}
\end{figure*}

\smallskip\noindent
\textbf{Row-level interpretation vs.\ \Perm.}
These space considerations and our earlier observations about query characteristics
(\cref{sec:performance-phase1,sec:performance-phase2}) are also reflected in~\cref{fig:perm-row-performance}.
In this head-to-head
slowdown comparison of row-level \SQL{} interpretation~(\tikz\draw[fill=white] (0,0.5) circle (\circlesize);\,,
including Phases~$\ph{1}$ and~$\ph{2}$) and
\Perm{}~(\tikz\draw[fill=black] (1.5,1.5) rectangle ++(2\squaresize,2\squaresize);),
a trend
\begin{tikzpicture}[sketch]
  \draw[trend] (0,-0.25) -- (2.5,1.5);
  \draw[fill=white] (0,-0.25) circle (\circlesize);
  \draw[fill=black] (2.5,1.5) rectangle ++(2\squaresize,2\squaresize);
\end{tikzpicture}
indicates that interpretation showed less slowdown than \Perm.
Over all executable queries, row-level interpretation levies
a provenance tax of factor~$5.1$ while \Perm{} imposes a factor of~$18.9$ (geometric means).

\begin{changed}
\cref{fig:perm-row-performance} shows that the advantage of interpretation over \Perm{} increases with
query complexity.  \Perm's log-free approach pays off in
category~\circled{A} of scans or simple joins and (grouped) aggregation.
The price of writing a large log and correlation in Phase~$\ph{1}$
(see~\cref{sec:performance-phase1}) makes
\tpchQ{21} the only complex \TPCH{} query for which \Perm{} outperforms
interpretation.  As discussed above, tupling for~\tpchQ{21}
could tip the scales in favor of interpretation, though. The queries
in~\circled{B} are characterized by increasing predicate complexity and
join width, with the latter contributing to the discussed space
overhead: \Perm{} generates queries that emit wide rows (of~$62$ instead
of the original $2$~columns for
\TPCH{} query~\tpchQ{8}, for example).  Some queries in~\circled{B} and
\emph{all} queries in~\circled{C} feature nested subqueries, which amplify
\Perm's provenance representation size problem:\footnote{We thus omitted
these queries from~\cref{fig:provenance-rep-size}---for~\tpchQ{22} and
its two scalar and existentially quantified subqueries, \Perm{} incurs
a representation size overhead of factor~$25\,000$.} a row of the outer query is replicated~$n$ times
if the subquery emits $n$~rows---even if the subquery is existentially
quantified~\cite{provenance-subqueries}.
Three queries in category~\circled{C} that \Perm{} failed to process
within 4~hours are marked
\tikz[baseline=1pt]\draw[draw=none,pattern=stripes,pattern color=bodycolor!30] (0,0) rectangle (0.5,0.25);
in~\cref{fig:perm-row-performance}.
\end{changed}

\begin{changed}
\smallskip\noindent
\textbf{\GProM-style provenance-aware query optimization.}
After rewriting for provenance, \Perm{} has been found to generate query
shapes that significantly deviate from the original subject query. Plans
take on a form that challenges existing query processors or may lead to
the duplication of work (\emph{e.g.}, see~\Perm{}'s \sql{GROUP\;BY} translation
rule~\textbf{R5} in~\cite{perm-rewriting}). These observations led to follow-up work on
successor project \GProM{} that identifies specific algebraic
optimizations tuned to cope with challenging query
structure~\cite{gprom,gprom-overview,gprom-opt}. These provenance-aware
optimizations primarily target grouping and aggregation and, for some
queries, can offer a speed-up of up to factor~$3$ (personal
communication with the author and~\cite{gprom-opt}). With
these---partially heuristic, partially cost-based---algebraic rewrites,
\GProM{} reaches even deeper into the underlying RDBMS than
\Perm{}.

\smallskip\noindent
However, provenance-specific optimizations also apply to the
interpretation of \SQL{}.  The principle can be adapted to
\begin{compactitem}
\item match our provenance model (dependency sets),
\item be non-invasive, \emph{i.e.}, not reach inside the RDBMS kernel,
\item be easily expressible on the \SQL{} language level, \emph{i.e.}, in terms of
  a shape-preserving source-level transformation.
\end{compactitem}
One particular transformation relates to the occurrence of a closed
(non-correlated) subquery~$q_1$ under an aggregrate.  In Phase~$\ph{2}$ we have:
$$
\setagg_{\mathclap{r \in \inphase{2}{\sql{r}}}}\,\bigl(q_1 \cup q_2(r)\bigr)
\quad\equiv\quad
q_1\, \cup\, \setagg_{\mathclap{r \in \inphase{2}{\sql{r}}}}\,q_2(r)
\quad\text{, $r$ not free in $q_1$.}
$$
Note that this rewrite is specific for set aggregation and would be
incorrect in a subject query that uses~\sql{SUM}/\sql{+}, for example.
In \TPCH{}, such constellations arise for~\tpchQ{11},
\tpchQ{16}, \tpchQ{18}, \tpchQ{20}, and \tpchQ{22}, where the transformation
reduces interpretation time in Phase~$\ph{2}$ by factors between~$2$
and~$160$ (the latter for~\tpchQ{22}). The experiments of this section
have been performed with the transformation enabled.
\end{changed}

\section{More Related Work}
\label{sec:related-work}

The traced evaluation of subject queries is a defining feature of the
present work.  Phase~$\ph{1}$ identifies rows that \emph{actually
participated} in query evaluation;  Phase~$\ph{2}$ adds cell-level
dependencies and aggregates the Phase~$\ph{1}$ findings as needed. This
places the approach in the landscape of established provenance notions.

To form \where-provenance, we collect those input cells that were
\emph{used to compute the value} of an output cell---this includes those
input cells that were copied verbatim and thus generalizes the notion of
\where-provenance as defined by Buneman in~\cite{why-and-where}.
\Perm{}~\cite{perm-rewriting}
argues for and implements the same generalization.  If we combine the
\where- and \why-provenance derived by interpretation, for each output
cell~$o$ we obtain \emph{one} set of input cells that witness~$o$. This
provides a cell-level analog to \emph{lineage}~\cite{lineage-tracing} or
\emph{influence contribution}~\cite{glavic-phd,perm-rewriting}, concepts orignally established at
the coarser row level. In deviation from Buneman's definition of
\why-provenance, we do not derive all possible witnesses but the
particular set of input cells that were indeed used by the system to
produce~$o$.
% A re-evaluation of the subject query on the reduced
% database instance that only contains the rows whose cells are found in
% the witness, will yield~$o$ (among other rows).
This is invaluable in
declarative query debugging where such database size reductions can help
to prevent users from ``drowning in a sea of observations''~\cite{best-bang-for-bug}.
Let us note that non-standard interpretation is a member of the \emph{annotation propagation}
family of approaches~\cite{provenance-color} which fail to derive provenance in the presence of
empty intermediate results~\cite{why-not}.

The shift from values to computation over dependency sets~$\ptype$ relates to the
\emph{provenance semiring} that derives lineage for the positive bag
algebra by Green~\emph{et al.}~\cite{semirings}.  In a nutshell, Phase~$\ph{2}$ realizes a
\SQL{} semantics interpreted in the particular semiring
$(\ptype,\bot,\varnothing,\cup_L,\cup_S)$,\footnote{See~\cite[Sections~1.3~and~5.1]{provenance-in-databases}
for the definitions of $\cup_L$, $\cup_S$ and their interaction
with~$\bot$.} in which rows are annotated with dependency sets.  To illustrate, in the
treatment of~${\text{\large$\sigma$}}_{\textbf{P}}(R)$
in~\cite{semirings}, rows~$t$ that fail to satisfy
predicate~$\textbf{P}$ are mapped to~$\bot$ which effectively
discards~$t$'s provenance contribution (case~\textbf{selection} of
Definition~3.2 in~\cite{semirings}). In the present work, this role
of~$\textbf{P}$ is assumed by the
\sql{LATERAL}~join with function~$\logread{JOIN$\langle m\rangle$}$ which discards~$t$ if
$t\sql{.}\row$ cannot be found in the associated log (see the
redefinition of interpreter~$\inphase{2}{i}$ of~\cref{rule:join}
in~\cref{sec:from-cell-to-row}, set $m=1$ to obtain a direct
correspondence with~\cite{semirings}).

We understand provenance derivation as dynamic data dependency analysis
and share this view with Cheney~\emph{et
al.}~\cite{provenance-as-dependency,queries-explain-work}.  The
interpreters defined in the rules of
\ifarXiv
\cref{fig:rules,fig:rules-extended}
\else
\cref{fig:rules}
\fi
propagate and
accumulate dependency sets much like the \emph{provenance tracking
semantics} defined in Figures~5~and~6
of~\cite{provenance-as-dependency}.  The authors state that
``\emph{[d]ynamic provenance may be [expensive to compute and]
non-trivial to implement in a standard relational database system}.''
Our present effort addresses just this challenge.

\smallskip\noindent
Given a piece~$o$ of the output, \emph{backward
slicing}~\cite{slicing,slicing-survey,slicing-and-provenance} finds
those parts---or:~slices---of a program that are involved in
producing~$o$.  In~\cite{prov-sql-abstract-int,how-provenance}, we demonstrated the
derivation of provenance through the application of slicing to
imperative programs that simulate the semantics of \SQL{} queries.  In
the present work, instead, we directly realize a dynamic variant of slicing for
\SQL{} but are only interested in input data slices on
which~$o$'s value depends. If, however, we associate identifiers with
\SQL{} subexpressions (instead of cells), interpretation could instead
identify the subject query slices relevant to the computation of~$o$.
This paves the way for a notion of
\emph{how}-provenance~\cite{provenance-in-databases} whose findings
directly relate to \SQL{}'s surface syntax (instead of algebraic plans,
say).

\smallskip\noindent
C.\ Barry Jay has explored the decomposition of data structures into
their \emph{shape} and contained
values~\cite{shape-in-computing,shapely-types}. We have deliberately
designed a two-phase approach that preserves data shape (original input
and output tables share row width and cardinality with those of
Phases~$\ph{1}$/$\ph{2}$, respectively) and query shape (recall the discussion of
parametricity of~\cref{sec:parametricity}). We reap the benefits in
terms of a straightforward, extensible formulation of inference rules
and plans that do not swamp the DBMS's optimizer and executor.  This
focus on shape preservation tells this work apart from related efforts
where data and its provenance are tightly bundled and then threaded
jointly through the
computation~\cite{links-provenance,provenance-as-dependency,queries-explain-work,perm-rewriting,polygen,provenance-color}.
This reshapes input, intermediate, and output data as well as the
computation process itself---sometimes dramatically so---and ultimately
leads to restrictions on what data and query sizes are considered
tractable~\cite{queries-explain-work,gprom}.  Bundling has the advantage
that queries may post-process data and its provenance together, however.
We can offer this integrated view through a join of the Phase~$\ph{1}$
and~$\ph{2}$ outputs: consider $\inphase{1}{\sql{output}} \bowtie_\row
\inphase{2}{\sql{output}}$ in~\cref{fig:prov-row}, for example.

\section{Wrap-Up}
\label{sec:conclusions}

The desire to move complex computation---like tasks in machine learning
or graph processing---close to their data sources led to a steep growth in
query complexity.  As this trend will only continue, this work is an attempt
to develop provenance derivation for \SQL{} that catches up and helps to
explain the resulting intricate queries.  We shift from value-
to dependency-based computation through a non-standard interpretation of
the \SQL{} semantics that can derive provenance at either the cell level
or the coarser row granularity.  The approach embraces a rich dialect of
\SQL{} constructs---including recursion, windowed aggregates, or
table-valued and user-defined functions---and relies on a two-phase
evaluation process designed to not
overwhelm the underlying database system.

\smallskip\noindent
This work is extensible in several dimensions.  We believe that the idea
of non-standard interpretation does not break if further \SQL{}
constructs are added to the dialect.  Currently, we explore the treatment
of \SQL{} DML statements (\sql{INSERT}, \sql{UPDATE}, \sql{DELETE}) and
functions defined in PL/\SQL{}---this is also related to recent work on the
re-enactment of transactions~\cite{reenactment}.  Further, the provenance model
realized by the approach is subject to tuning.  Phase~$\ph{1}$, for example, may
employ ``lazy'' or ``greedy'' variants of~\sql{EXISTS} to decide whether the
provenance of a subquery includes one particular row or all rows that satisfied
the quantifier (see~\cite{perm-subqueries} for a discussion of possible semantics).

We pursue optimizations that can help to boost Phase~$\ph{1}$.  Data
flow analysis can reveal inclusion relationships between log files and
thus render $\logwrite{$\Box$}$ at some call sites obsolete.  Likewise,
we can statically infer particular \emph{write once} safeguards
(\cref{sec:log-writing-reading}) to be superfluous.

Lastly, the ``onion-style'' normalization of \SQL{} has helped to keep the
inference rule set of
\ifarXiv
\cref{fig:rules,fig:rules-extended}
\else
\cref{fig:rules}
\fi
orthogonal and compact.  We
conjecture that this syntactic normal form can generally benefit efforts
that rely on a source-level analysis and transformation of \SQL{}.  We will
follow up in an independent thread of work.

\ifarXiv
\else
\smallskip\noindent
\textbf{Acknowledgments.}
We thank Boris Glavic for insightful dicussions about provenance derivation
for \SQL{} and his help in configuring and operating \Perm.  Noah Doersing
wrote a \Pg{} extension that integrates \emph{roaring bitmaps} into the system.  This
research is supported by the German Research Council (DFG) under grant no.\ GR\,2036/4-1.
We thank the anonymous reviewers for their observations and suggestions which
greatly helped to improve this paper.
\fi

%% ....................................................................
%% bibliography
\clearpage
\bibliographystyle{abbrv}
\bibliography{provenance-pure-SQL}

%% ....................................................................
%% appendix

\color{changed}
\appendix
\color{black}

\ifarXiv

\section{Extended Inference Rule Set}
\label{app:rules}

We extend the set of inference rules for mapping $q \comp
\twophase{\inphase{1}{i}}{\inphase{2}{i}}$ that derives the
instrumented variant~$\inphase{1}{i}$ and interpreter~$\inphase{2}{i}$
for \SQL{} query~$q$.  Jointly, the rules of
\cref{fig:rules,fig:rules-extended} handle a \SQL{} dialect that covers
the example query of~\cref{fig:visibility-query}, all 22~queries of the
\TPCH{} benchmark~\cite{tpch}, as well as recursive queries (expressed
via~\sql{WITH RECURSIVE}).  This rule set extension retains the
compositionality property of the rules of~\cref{fig:rules}.  None of the
core rule principles discussed in~\cref{sec:abs-int} need to be reworked.

\smallskip\noindent
\cref{rule:tbl-fun} derives instrumentation~$\inphase{1}{i}$
and interpreter~$\inphase{2}{i}$ for invocations of table-valued
functions~$f$ (like the built-in \sql{generate\_series}).  Instrumentation wraps
calls~\sql{$f$($\inphase{1}{e_1}$,$\dots$,$\inphase{1}{e_n}$)} such
that~\sql{$\logwrite{TBLF}$} can attach a unique identifier~$\row$ to
every row emitted by~$f$.  Interpretation does not re-run~$f$ but
instead uses~\sql{$\logread{TBLF}$} to reproduce these identifiers from
the log.  In the present variant of \cref{rule:tbl-fun}, we assume that
all rows output by~$f$ depend on all function
arguments---$\inphase{2}{i}$ thus assigns the union of the dependency
sets $\abs{\var{args}\sql{(D)}} \equiv \abs{\inphase{2}{e_{1}}
\cup \cdots \cup \inphase{2}{e_{n}}}$ as the rows' \where-provenance.
More specific provenance derivation rules for particular table
functions~$f$ can easily be added to the rule set.

\cref{rule:map} offers a simplified treatment for the specific case of a single-table
\sql{FROM}~clause that is not accompanied by a~\sql{WHERE} predicate.
Since no value-based decisions are made, instrumentation and logging are
not required. (The general case of an $m$-table \sql{FROM}~clause
with~\sql{WHERE} predicate is handled by~\cref{rule:join}
of~\cref{fig:rules}.)

\begin{changed}
\smallskip\noindent
\cref{rule:case} transforms an $(n+1)$-way \sql{CASE}--\sql{WHEN}
conditional. (We use the notation~$\sql{CASE}^v$ here to indicate that
the guard and branch expressions $e_{w,i}$, $e_{t,i}$ depend on a single
row variable~$v$---normalization can always ensure this syntactic
restriction.) Phase~$\ph{1}$ identifies the \sql{WHEN}~branch taken
($\var{branch} \in \{\sql{0},\sql{1},\ndots,n\}$) and
uses~\logwrite{CASE} to save this decision for the interpreter.
Note that the rule duplicates the~\sql{CASE} expression but each
guard~$\inphase{1}{e_{w,i}}$ and branch
expression~$\inphase{1}{e_{t,i}}$ is evaluated at most once.  Phase~$\ph{2}$
interprets the selected branch---say~$i$---and adds \why-provenance for
\emph{all preceding} guard expressions~$\inphase{2}{e_{w,1}},\ndots,\inphase{2}{e_{w,i}}$
(this reflects top-down guard evaluation as specified by the \SQL{}
standard~\cite[\S\,6.12]{sql-2016}).

% notes on rule:case

% normalized input:
% * CASE WHEN ELSE - style only
% * one variable (no more, no less)
%   * if > 1: merge columns into one variable (not visible, employ the JOIN rule)
%   * if 0: add constant/dummy variable with tuid=1
% translated into:
% * Phase 1:
%   * CASE e WHEN ELSE style (outer)
%   * CASE WHEN ELSE style (inner)
% * Phase 2: CASE e WHEN ELSE style

% please take note:
% * one input CASE becomes two CASEs in Phase 1
%   however, the subexpressions are *not* duplicated:
%   * all WHEN expressions (only) go into the inner CASE
%   * all THEN expressions (only) go into the outer CASE
%   advantages:
%   * logging happens at a single, dedicated spot
%   * data dependencies are right (can't get optimized away)
%   * just one logging call per CASE evaluation (instead of one for each THEN)

\smallskip\noindent
A \SQL{} subquery may refer to---and thus depend on---a \emph{free} row variable~$v$
that has been bound outside the subquery (correlated subquery,
see~\cite[\S\,7.6]{sql-2016}).  To make the dependence on this free row
variable explicit for mapping~$\comp$, query normalization introduces a
local binding that shadows the outer binding for~$v$ but otherwise
does not alter query semantics (assume that the rows bound to~$v$ have
columns~$c_1,\dots,c_n$):
$$
\begin{array}{@{}l@{~}l@{}}
  \sql{\textbf{SELECT}} & e(v,t)\\
  \sql{\textbf{FROM}}   & q~\sql{\textbf{AS}}~t
\end{array}
\underset{\text{normalize}}{\longrightarrow}
\begin{array}{@{}l@{~}l@{}}
  \sql{\textbf{SELECT}} & e(v,t) \\
  \sql{\textbf{FROM}}   & \sql{(BIND}~c_1\sql{,}\dots\sql{,}c_n\sql{)}~\sql{\textbf{AS}}~v\sql{,} \\
                        & q~\sql{\textbf{AS}}~t
\end{array}\enskip.
$$
Here, \sql{BIND} is a pseudo \SQL{} construct that is subsequently traded for a
\SQL{} \sql{SELECT} by~\cref{rule:bind}. We are left with a
regular join that~\cref{rule:join} will transform as described earlier
in~\cref{sec:abs-int}.  In particular, the dependence of expression~$e$ on
row variables~$v$ (formerly free) and~$t$ will be handled correctly.
We exploit the semantic similarities of \SQL{}'s correlation and
join: both form Cartesian products of row variable bindings. \cref{rule:bind}
transforms the former into the latter, once more rendering the rule set
more compact.

\smallskip\noindent
\cref{rule:nested-subquery}: A parenthesized subquery, evaluating to
a---possibly singleton---set of values and optional set membership
(\sql{$e$\;$[$NOT$]$\;IN}) or emptiness tests (\sql{$[$NOT$]$\;EXISTS})
are retained in Phase~$\ph{1}$.  Phase~$\ph{2}$ collects the
dependencies of~$e$ and all set members to form the subquery's overall
provenance.  Note that a single rule suffices to process both correlated
and non-correlated subqueries, thanks to the preparatory work performed
by normalization as well as~\cref{rule:bind,rule:join} (see above).

The design of this rule implies that the derivation of provenance for
subqueries with existential semantics directly follows that of the
underlying RDBMS.  In particular, interpretation is as lazy as
value-based evaluation: if the system shortcuts a membership (\sql{IN})
or existence (\sql{EXISTS}) test as soon as one element~$v$ is found,
the \where-dependencies of the subquery will be the dependencies of~$v$,
disregarding all other elements (if any).  This deviates from the
greedy semantics of \Perm's sublink query provenance~\cite{provenance-subqueries}
which collects all values returned by the subquery.  \Perm's design
has significant consequences for provenance representation size as discussed in~\cref{sec:perm}.

\smallskip\noindent
In~\cref{rule:left-join}, $\inphase{1}{i}$ logs the join pairs generated
by a left outer join
via~$\logwrite{LEFT}(\location{\ell},t_1\sql{.}\row,t_2\sql{.}\row)$.
Here, row identifier~$t_2\sql{.}\row$ will be~\sql{NULL} should~$t_1$
fail to find a join partner. These~\sql{NULL}s are retrieved
when~$\logread{LEFT}(\location{\ell},t_1\sql{.}\row)$ reads the row
identifiers~\sql{\var{join}.right} of $t_1$'s join partners from the log
in Phase~$\ph{2}$. If~\sql{\var{join}.right} indeed is~\sql{NULL}, the
left outer join with~$\inphase{2}{q_2}$ will fail and generate
a~\sql{NULL} binding for row variable~$t_2$  (just like in
Phase~$\ph{1}$).  Dependency set operations like~$\cup$ and
$\smash{\setagg}$ interpret~\sql{NULL} as~$\varnothing$ such that left
join-generated \sql{NULL}s effectively make no provenance contribution,
as desired. Inference rules for right and full outer joins are not shown
here but operate likewise.
\end{changed}

\begin{figure*}
  \begin{narrow}{-5mm}{-5mm}
  \centering\small
  \renewcommand{\arraystretch}{0.9}
  \begin{tabular}{@{}c@{}}

  \inferrule{%
    \fn{f} \coloncolon \tau_1 \times \cdots \times \tau_n \shortrightarrow \sql{TABLE($c_1$\,$\tau'_1$,$\ndots$,$c_m$\,$\tau'_m$)}
    \and
    \bigl\lvert e_i \comp \twophase{\inphase{1}{e_i}}{\inphase{2}{e_i}}\bigr\rvert_{i=1\ndots n} \and
    \location{\ell} = \fresh \and\and\and
    \inphase{1}{i} =
      \begin{array}{@{}l@{}l@{~}l@{}}
        \sql{(} & \sql{\textbf{SELECT}}   & \logwrite{TBLF}\sql{(}\location{\ell}\sql{)}~\sql{\textbf{AS}}~\row\sql{,}~
                                            t\sql{.}c_1~\sql{\textbf{AS}}~c_1\sql{,} \ndots\sql{,} t\sql{.}c_m~\sql{\textbf{AS}}~c_m \\
                & \sql{\textbf{FROM}}     & \fn{f}\sql{(}\inphase{1}{e_{1}}\sql{,}\ndots\sql{,}\inphase{1}{e_{n}}\sql{)}~\sql{\textbf{AS}}~t\sql{(}c_1\sql{,}\ndots\sql{,}c_m\sql{)}
        \sql{)}
      \end{array}
    \and
    \inphase{2}{i} =
      \begin{array}{@{}l@{}l@{~}l@{}}
        \sql{(} & \sql{\textbf{SELECT}}   & \var{set}\sql{.}\row~\sql{\textbf{AS}}~\row\sql{,}~
                                            \abs{\var{args}\sql{.D}}~\sql{\textbf{AS}}~c_1\sql{,} \ndots\sql{,} \abs{\var{args}\sql{.D}}~\sql{\textbf{AS}}~c_m \\
                & \sql{\textbf{FROM}}     & \sql{(}\sql{\textbf{VALUES}}~\sql{(}\abs{\inphase{2}{e_1} \cup\cdots\cup \inphase{2}{e_n}}\sql{))}~\sql{\textbf{AS}}~\var{args}\sql{(D)},~\\
                &                         & \logread{TBLF}\sql{(}\location{\ell}\sql{)}~\sql{\textbf{AS}}~\var{set}\sql{(}\row\sql{)}
        \sql{)}
      \end{array}
    }{%
    \fn{f}\sql{(}e_{1}\sql{,}\ndots\sql{,}e_{n}\sql{)}
    \comp
    \twophase{\inphase{1}{i}}{\inphase{2}{i}}}
    \thisisrule{TblFun}\label{rule:tbl-fun}\\
    % table function / set-returning function / collection derived table
  \\[6mm]

  \inferrule{%
    \bigl\lvert e_i \comp \twophase{\inphase{1}{e_i}}{\inphase{2}{e_i}}\bigr\rvert_{i=1\ndots n} \qquad
    q \comp \twophase{\inphase{1}{q}}{\inphase{2}{q}} \qquad
    \inphase{1}{i} =
      \begin{array}{@{}l@{~}l@{}}
        \sql{\textbf{SELECT}}   & t\sql{.}\row~\sql{\textbf{AS}}~\row\sql{,}~
                                  \inphase{1}{e_1}~\sql{\textbf{AS}}~c_1\sql{,} \ndots\sql{,} \inphase{1}{e_n}~\sql{\textbf{AS}}~c_n \\
        \sql{\textbf{FROM}}     & \inphase{1}{q}~\sql{\textbf{AS}}~t \\
      \end{array}
    \qquad
    \inphase{2}{i} =
      \begin{array}{@{}l@{~}l@{}}
        \sql{\textbf{SELECT}}   & t\sql{.}\row~\sql{\textbf{AS}}~\row\sql{,}~
                                  \abs{\inphase{2}{e_1}}~\sql{\textbf{AS}}~c_1\sql{,} \ndots\sql{,} \abs{\inphase{2}{e_n}}~\sql{\textbf{AS}}~c_n \\
        \sql{\textbf{FROM}}     & \inphase{2}{q}~\sql{\textbf{AS}}~t \\
      \end{array}
    }{%
    \begin{array}{l@{~}l}
      \sql{\textbf{SELECT}}   & e_1~\sql{\textbf{AS}}~c_{1}\sql{,} \ndots\sql{,} e_n~\sql{\textbf{AS}}~c_n \\
      \sql{\textbf{FROM}}     & q~\sql{\textbf{AS}}~t \\
    \end{array}
    \comp
    \twophase{\inphase{1}{i}}{\inphase{2}{i}}}
    \thisisrule{Map}\label{rule:map}\\
  \\[5mm]

  \inferrule{
    \bigl\lvert e_{w,i} \comp \twophase{\inphase{1}{e_{w,i}}}{\inphase{2}{e_{w,i}}}\bigr\rvert_{i=1\ndots n} \and
    \bigl\lvert e_{t,i} \comp \twophase{\inphase{1}{e_{t,i}}}{\inphase{2}{e_{t,i}}}\bigr\rvert_{i=1\ndots n} \and
    \location{\ell} = \fresh \and\and\and
    \var{branch} =
      \begin{array}{@{}l@{~}l@{~}l@{}}
        \sql{CASE} & \sql{WHEN}~\inphase{1}{e_{w,1}} & \sql{THEN 1} \\[-1ex]
                   & \multicolumn{1}{l}{\vdots} \\
                   & \sql{WHEN}~\inphase{1}{e_{w,n}} & \sql{THEN $n$} \\
                   & \sql{ELSE 0} \\
        \sql{END}  & \\
      \end{array}
    \and
    \inphase{1}{i} =
      \begin{array}{@{}l@{~}l@{~}l@{}}
        \sql{CASE} & \multicolumn{2}{@{}l}{\logwrite{CASE}\sql{(}\location{\ell}\sql{,}v\sql{.}\row\sql{,}\var{branch}\sql{)}} \\
                   & \sql{WHEN 1} & \sql{THEN}~\inphase{1}{e_{t,1}} \\[-1ex]
                   & \multicolumn{1}{l}{\vdots} \\
                   & \sql{WHEN $n$} & \sql{THEN}~\inphase{1}{e_{t,n}} \\
                   & \multicolumn{2}{@{}l}{\sql{ELSE}~\inphase{1}{e_{t,0}}} \\
        \sql{END}  & \\
      \end{array}
    \and
    \inphase{2}{i} =
      \begin{array}{@{}l@{~}l@{~}l@{}}
        \sql{CASE} & \multicolumn{2}{@{}l}{\logread{CASE}\sql{(}\location{\ell}\sql{,}v\sql{.}\row\sql{)}} \\
                   & \sql{WHEN 1} & \sql{THEN}~\inphase{2}{e_{t,1}} \Y{\cup \toY\sql{(}\inphase{2}{e_{w,1}}\sql{)}} \\[-1ex]
                   & \multicolumn{1}{l}{\vdots} \\
                   & \sql{WHEN $n$} & \sql{THEN}~\inphase{2}{e_{t,n}} \Y{\cup \toY\sql{(}\inphase{2}{e_{w,1}} \cup \cdots \cup \inphase{2}{e_{w,n}}\sql{)}} \\
                   & \multicolumn{2}{@{}l}{\sql{ELSE}~\inphase{2}{e_{t,0}} \Y{\cup \toY\sql{(}\inphase{2}{e_{w,1}} \cup \cdots\cup \inphase{2}{e_{w,n}}\sql{)}}} \\
        \sql{END}  & \\
      \end{array}
    }{%
    \begin{array}{@{}l@{~}l@{~}l@{}}
      \sql{CASE}^{v} & \sql{WHEN}~e_{w,1} & \sql{THEN}~e_{t,1} \\[-1ex]
                     & \multicolumn{1}{l}{\vdots} \\
                     & \sql{WHEN}~e_{w,n} & \sql{THEN}~e_{t,n} \\
                     & \sql{ELSE}~e_{t,0} \\
      \sql{END}      & \\
    \end{array}
    \comp
    \twophase{\inphase{1}{i}}{\inphase{2}{i}}}
    \thisisrule{Case}\label{rule:case} \\
  \\[6mm]

  \inferrule{%
    i =
    \sql{\textbf{(SELECT}}~v\sql{.}\row~\sql{\textbf{AS}}~\row\sql{,}~v\sql{.}c_1~\sql{\textbf{AS}}~c_1\sql{,}\ndots\sql{,}v\sql{.}c_n~\sql{\textbf{AS}}~c_n\sql{)}~\sql{\textbf{AS}}~v}{%
    \sql{(BIND}~c_1\sql{,}\ndots\sql{,}c_n\sql{)}~\sql{\textbf{AS}}~v
    \comp
    \twophase{i}{i}}
    \thisisrule{Bind}\label{rule:bind} \\
  \\[6mm]

  \inferrule{% SCALAR / IN / NOT IN / EXISTS / NOT EXISTS, correlated / uncorrelated
     q \comp \twophase{\inphase{1}{q}}{\inphase{2}{q}} \and
     e \comp \twophase{\inphase{1}{e}}{\inphase{2}{e}} \and
    \inphase{1}{i} =
    \begin{array}{@{}l@{}l@{~}l}
      [\,\inphase{1}{e}~[\sql{NOT}]~\sql{IN}\mid[\sql{NOT}]~\sql{EXISTS}\,]~\sql{(}
        & \sql{\textbf{SELECT}}   & t\sql{.}c~\sql{\textbf{AS}}~c \\
        & \sql{\textbf{FROM}}     & \inphase{1}{q}~\sql{\textbf{AS}}~t \sql{)}
    \end{array}
    \and
    \inphase{2}{i} =
    \begin{array}{@{}l@{}l@{~}l}
      \sql{(}
        & \sql{\textbf{SELECT}}   & [\,\inphase{2}{e}\,\cup\,]~\raisebox{3pt}{\smash[t]{$\displaystyle\setagg$}} \{t\sql{.}c\} \\
        & \sql{\textbf{FROM}}     & q^{\ph{2}}~\sql{\textbf{AS}}~t \sql{)}
    \end{array}
    }{%
    \begin{array}{@{}l@{}l@{~}l}
      [\,e~[\sql{NOT}]~\sql{IN}\mid[\sql{NOT}]~\sql{EXISTS}\,]~\sql{(}
        & \sql{\textbf{SELECT}}   & t\sql{.}c~\sql{\textbf{AS}}~c \\
        & \sql{\textbf{FROM}}     & q~\sql{\textbf{AS}}~t \sql{)}
    \end{array}
    \comp
    \twophase{\inphase{1}{i}}{\inphase{2}{i}}}
    \thisisrule{NestedSubquery}\label{rule:nested-subquery} \\
  \\[6mm]

  \inferrule{% LEFT OUTER JOIN
    \bigl\lvert e_i \comp \twophase{\inphase{1}{e_i}}{\inphase{2}{e_i}}\bigr\rvert_{i=1\ndots n} \and
    \bigl\lvert q_i \comp \twophase{\inphase{1}{q_i}}{\inphase{2}{q_i}}\bigr\rvert_{i=1,2} \and
    p \comp \twophase{\inphase{1}{p}}{\inphase{2}{p}} \and
    \location{\ell} = \fresh \and\and\and
    \inphase{1}{i} =
      \begin{array}{l@{~}l}
        \sql{\textbf{SELECT}}   & \logwrite{LEFT}\sql{(}\location{\ell}\sql{,} t_1\sql{.}\row\sql{,} t_2\sql{.}\row\sql{)}~\sql{\textbf{AS}}~\row\sql{,} \\
                                & \inphase{1}{e_1}~\sql{\textbf{AS}}~c_1\sql{,} \ndots\sql{,} \inphase{1}{e_n}~\sql{\textbf{AS}}~c_n \\
        \sql{\textbf{FROM}}     & \inphase{1}{q_1}~\sql{\textbf{AS}}~t_1
                                ~ \sql{\textbf{LEFT}}~\sql{\textbf{OUTER}}~\sql{\textbf{JOIN}}~\inphase{1}{q_2}~\sql{\textbf{AS}}~t_2~\sql{\textbf{ON}}~\sql{\textbf{(}}\inphase{1}{p}\sql{\textbf{)}} \\
      \end{array}
    \and
    \inphase{2}{i} =
      \begin{array}{l@{~}l}
        \sql{\textbf{SELECT}}   & \var{join}\sql{.}\row~\sql{\textbf{AS}}~\row\sql{,}\
                                  \abs{\inphase{2}{e_1} \Y{\cup \varY{join}}}~\sql{\textbf{AS}}~c_1\sql{,} \ndots\sql{,} \abs{\inphase{2}{e_n} \Y{\cup \varY{join}}}~\sql{\textbf{AS}}~c_n \\
        \sql{\textbf{FROM}}     & \inphase{2}{q_1}~\sql{\textbf{AS}}~t_1
                                ~ \sql{\textbf{CROSS}}~\sql{\textbf{JOIN}}~\sql{\textbf{LATERAL}}~\logread{LEFT}\sql{(}\location{\ell}\sql{,} t_1\sql{.}\row\sql{)}~\sql{\textbf{AS}}~\var{join}\sql{(}\row\sql{,right}\sql{)} \\
                                & \sql{\textbf{LEFT}}~\sql{\textbf{OUTER}}~\sql{\textbf{JOIN}}~\inphase{2}{q_2}~\sql{\textbf{AS}}~t_2~\sql{\textbf{ON}}~\sql{\textbf{(}}\var{join}\sql{.right}~\sql{=}~t_2\sql{.}\row\sql{\textbf{)}}\sql{,} \\
                                & \Y{\sql{\textbf{LATERAL}}~\abs{\toY\sql{(}\inphase{2}{p}\sql{)}}~\sql{\textbf{AS}}~\varY{join}} \\
      \end{array}
    }{%
    \begin{array}{l@{~}l}
      \sql{\textbf{SELECT}}   & e_1~\sql{\textbf{AS}}~c_1\sql{,} \ndots\sql{,} e_n~\sql{\textbf{AS}}~c_n \\
      \sql{\textbf{FROM}}     & q_1~\sql{\textbf{AS}}~t_1
                              ~ \sql{\textbf{LEFT}}~\sql{\textbf{OUTER}}~\sql{\textbf{JOIN}}~q_2~\sql{\textbf{AS}}~t_2~\sql{\textbf{ON}}~\sql{\textbf{(}}p\sql{\textbf{)}} \\
    \end{array}
    \comp
    \twophase{\inphase{1}{i}}{\inphase{2}{i}}}
    \thisisrule{LeftJoin}\label{rule:left-join} \\
  \\[4mm]

  \end{tabular}
  \end{narrow}
  \caption{%
    \color{changed}
    Mapping $q \comp \twophase{\inphase{1}{i}}{\inphase{2}{i}}$:
    additions to the inference rule set of~\cref{fig:rules}. (Continued on next page.)}
  \label{fig:rules-extended}
\end{figure*}

\begin{figure*}
  \ContinuedFloat
  \begin{narrow}{-5mm}{-5mm}
  \centering\small
  \renewcommand{\arraystretch}{0.9}
  \begin{tabular}{@{}c@{}}
  \inferrule{% DISTINCT ON, namely normalized version of COUNT(DISTINCT ...), do we need ORDER BY here?
     q \comp \twophase{\inphase{1}{q}}{\inphase{2}{q}} \and
    \bigl\lvert e_i \comp \twophase{\inphase{1}{e_i}}{\inphase{2}{e_i}}\bigr\rvert_{i=1\ndots n} \and
    \bigl\lvert o_i \comp \twophase{\inphase{1}{o_i}}{\inphase{2}{o_i}}\bigr\rvert_{i=1\ndots m+l} \and
    \location{\ell} = \fresh \and\and\and
    \inphase{1}{i} =
      \begin{array}{l@{~}l@{}}
        \sql{\textbf{SELECT}}   & \logwrite{FILTER}\sql{(}\location{\ell}\sql{,}~\var{t}\sql{.}\row\sql{)}~\sql{\textbf{AS}}~\row\sql{,}~
                                            \var{t}\sql{.}c_1~\sql{\textbf{AS}}~c_1\sql{,} \ndots\sql{,} \var{t}\sql{.}c_n~\sql{\textbf{AS}}~c_n \\
        \sql{\textbf{FROM}}     & \sql{(}
                                  \begin{array}[t]{@{}l@{~}l}
                                    \sql{\textbf{SELECT}}   & \sql{\textbf{DISTINCT}}~\sql{\textbf{ON}}~\sql{\textbf{(}}\inphase{1}{o_1}\sql{,} \ndots\sql{,} \inphase{1}{o_m}\sql{\textbf{)}}
                                                            ~t\sql{.}\row~\sql{\textbf{AS}}~\row\sql{,}\\
                                                            & \inphase{1}{e_1}~\sql{\textbf{AS}}~c_1\sql{,} \ndots\sql{,} \inphase{1}{e_n}~\sql{\textbf{AS}}~c_n \\
                                    \sql{\textbf{FROM}}     & \inphase{1}{q}~\sql{\textbf{AS}}~t \\
                                    \sql{\textbf{ORDER BY}} & \inphase{1}{o_1}\sql{,} \ndots\sql{,} \inphase{1}{o_m}\sql{,} \ndots\sql{,} \inphase{1}{o_{m+l}}
                                  \sql{)}~\sql{\textbf{AS}}~\var{t}
                                  \end{array} \\
      \end{array}
    \and
    \inphase{2}{i} =
      \begin{array}{l@{~}l}
        \sql{\textbf{SELECT}}   & \var{filter}\sql{.}\row~\sql{\textbf{AS}}~\row\sql{,}~\abs{\inphase{2}{e_1} \Y{\cup \varY{filter}}}~\sql{\textbf{AS}}~c_1\sql{,} \ndots\sql{,} \abs{\inphase{2}{e_n} \Y{\cup \varY{filter}}}~\sql{\textbf{AS}}~c_n \\
        \sql{\textbf{FROM}}     & \inphase{2}{q}~\sql{\textbf{AS}}~t\sql{,} \\
                                & \sql{LATERAL}~\logread{FILTER}\sql{(}\location{\ell}\sql{,} t\sql{.}\row\sql{)}~\sql{\textbf{AS}}~\var{filter}\sql{(}\row\sql{)}, \\
                                & \Y{\sql{LATERAL}~\abs{\toY\sql{(}\inphase{2}{o_1} \cup \cdots \cup \inphase{2}{o_m}\sql{)}}~\sql{\textbf{AS}}~\varY{filter}} \\
      \end{array}
    }{%
    \begin{array}{l@{~}l}
      \sql{\textbf{SELECT}}   & \sql{\textbf{DISTINCT}}~\sql{\textbf{ON}}~\sql{\textbf{(}}o_1\sql{,} \ndots\sql{,} o_m\sql{\textbf{)}}
                              ~ e_1~\sql{\textbf{AS}}~c_1\sql{,} \ndots\sql{,} e_n~\sql{\textbf{AS}}~c_n \\
      \sql{\textbf{FROM}}     & q~\sql{\textbf{AS}}~t \\
      \sql{\textbf{ORDER BY}} & o_1\sql{,} \ndots\sql{,} o_m\sql{,} \ndots\sql{,} o_{m+l} \\
    \end{array}
    \comp
    \twophase{\inphase{1}{i}}{\inphase{2}{i}}}
    \thisisrule{Distinct}\label{rule:distinct} \\
  \\[6mm]

  \inferrule{% optimized version of ORDER BY without LIMIT does not need logging (alternatively writes sequence -> shape preservation?)
     q \comp \twophase{\inphase{1}{q}}{\inphase{2}{q}} \and
    \bigl\lvert e_i \comp \twophase{\inphase{1}{e_i}}{\inphase{2}{e_i}}\bigr\rvert_{i=1\ndots n} \and
    \bigl\lvert o_i \comp \twophase{\inphase{1}{o_i}}{\inphase{2}{o_i}}\bigr\rvert_{i=1\ndots m} \and
    \location{\ell} = \fresh \and\and\and
    \inphase{1}{i} =
      \begin{array}{l@{~}l@{}}
        \sql{\textbf{SELECT}}   & \logwrite{FILTER}\sql{(}\location{\ell}\sql{,}~\var{t}\sql{.}\row\sql{)}~\sql{\textbf{AS}}~\row\sql{,}~
                                            \var{t}\sql{.}c_1~\sql{\textbf{AS}}~c_1\sql{,} \ndots\sql{,} \var{t}\sql{.}c_n~\sql{\textbf{AS}}~c_n \\
        \sql{\textbf{FROM}}     & \sql{(}
                                  \begin{array}[t]{@{}l@{~}l}
                                    \sql{\textbf{SELECT}}   & t\sql{.}\row~\sql{\textbf{AS}}~\row\sql{,}
                                                            ~ \inphase{1}{e_1}~\sql{\textbf{AS}}~c_1\sql{,} \ndots\sql{,} \inphase{1}{e_n}~\sql{\textbf{AS}}~c_n \\
                                    \sql{\textbf{FROM}}     & \inphase{1}{q}~\sql{\textbf{AS}}~t \\
                                    \sql{\textbf{ORDER BY}} & \inphase{1}{o_1}\sql{,} \ndots\sql{,} \inphase{1}{o_m} \\
                                    \sql{\textbf{OFFSET}}   & k \\
                                    \sql{\textbf{LIMIT}}    & l
                                  \sql{)}~\sql{\textbf{AS}}~\var{t}
                                  \end{array} \\
      \end{array}
    \and
    \inphase{2}{i} =
      \begin{array}{l@{~}l}
        \sql{\textbf{SELECT}}   & \var{filter}\sql{.}\row~\sql{\textbf{AS}}~\row\sql{,}~\abs{\inphase{2}{e_1} \Y{\cup \varY{filter}}}~\sql{\textbf{AS}}~c_1\sql{,} \ndots\sql{,} \abs{\inphase{2}{e_n} \Y{\cup \varY{filter}}}~\sql{\textbf{AS}}~c_n \\
        \sql{\textbf{FROM}}     & \inphase{2}{q}~\sql{\textbf{AS}}~t\sql{,} \\
                                & \sql{LATERAL}~\logread{FILTER}\sql{(}\location{\ell}\sql{,} t\sql{.}\row\sql{)}~\sql{\textbf{AS}}~\var{filter}\sql{(}\row\sql{)}, \\
                                & \Y{\sql{LATERAL}~\abs{\toY\sql{(}\inphase{2}{o_1} \cup \cdots \cup \inphase{2}{o_m}\sql{)}}~\sql{\textbf{AS}}~\varY{filter}} \\
      \end{array}
    }{%
    \begin{array}{l@{~}l}
      \sql{\textbf{SELECT}}   & e_1~\sql{\textbf{AS}}~c_1\sql{,} \ndots\sql{,} e_n~\sql{\textbf{AS}}~c_n \\
      \sql{\textbf{FROM}}     & q~\sql{\textbf{AS}}~t \\
      \sql{\textbf{ORDER BY}} & o_1\sql{,} \ndots\sql{,} o_m \\
      \sql{\textbf{OFFSET}}   & k \\
      \sql{\textbf{LIMIT}}    & l \\
    \end{array}
    \comp
    \twophase{\inphase{1}{i}}{\inphase{2}{i}}}
    \thisisrule{OrderBy}\label{rule:order-by} \\
  \end{tabular}
  \end{narrow}
  \caption{%
    \color{changed}
    Mapping $q \comp \twophase{\inphase{1}{i}}{\inphase{2}{i}}$:
    additions to the inference rule set.}
\end{figure*}

\begin{changed}
\smallskip\noindent
From the viewpoint of~\cref{rule:distinct}, \SQL{}'s
\sql{DISTINCT\;ON}\footnote{Normalization rewrites occurrences
of~\sql{DISTINCT} into equivalent~\sql{DISTINCT\;ON} clauses.  This also
helps to avoid that the prepended row identifier columns~$\row$ impact
duplicate removal.} acts like a filter: in Phase~$\ph{1}$, the rule
uses~\logwrite{FILTER} to log the row identifiers~$t\sql{.}\row$ of
those rows~$t$ that survive duplicate elimination. A row~$t$ remains if
the evaluation of the expressions $o_1,\dots,o_m$ yields a unique
combination of values.  For these remaining rows, Phases~$\ph{2}$ thus
assembles \why-provenance from the associated dependency
sets~$\inphase{2}{o_1},\dots,\inphase{2}{o_m}$.

Similar observations apply to~\cref{rule:order-by}: once rows have been
orderded by~\sql{ORDER\;BY~$o_1$,$\dots$,$o_m$}, rows are eliminated by
clauses~\sql{OFFSET~$k$} and~\sql{LIMIT~$l$} \cite[\S\,4.15.3]{sql-2016}. The
instrumented query~$\inphase{1}{i}$ calls~\logwrite{FILTER} to log the
identifiers~$t\sql{.}\row$ of all rows that remain.  The
interpreter~$\inphase{2}{i}$ uses~\logread{FILTER} to re-enact this
filtering behavior in Phase~$\ph{2}$. While the filtering semantics
of~\sql{ORDER\;BY}--\sql{OFFSET}--\sql{LIMIT} is relevant for dependency
set computation, row ordering itself is not: note that~$\inphase{2}{i}$
does not feature an~\sql{ORDER\;BY} clause.  Since row inclusion is
decided by the ordering criteria~$o_1,\dots,o_m$, the rule augments all result
cells with \why-provenance~$\inphase{2}{o_{1}} \cup \cdots \cup
\inphase{2}{o_{m}}$.

% * describe semantics of y-provenance
%   * DISTINCT: all columns in DISTINCT ON(...) contribute
%   * inter-row provenance is not considered
% * add reference to earlier description of log_filter (explaining _LATERAL readFilter(.)_ et al.)

% open question: does this case matter? (part of tpc-h)
% count(distinct e) -> count(e) FROM DISTINCT
\end{changed}

\begin{changed}
\section{Log File Contents and \\ Relational Implementation}
\label{app:log-files}

Log files enable the interpreters of Phase~$\ph{2}$ to re-enact
particular decisions and computation performed by the value-based
Phase~$\ph{1}$. The inference rules
of~\cref{fig:rules,fig:rules-extended} define how queries are augmented
to write to and read from these logs. \cref{sec:log-writing-reading}
describes the behavior of the
associated~\logwrite{$\Box$}/\logread{$\Box$} routines and how these
call on the lower-level routines $\var{put}_{\Box}$ and
$\var{get}_{\Box}$ to interact with the log files. Here, we show the
actual log contents produced by  representative
\SQL{} queries in Phase~$\ph{1}$.  Logging and log files may be
implemented in a variety of ways. Below
in~\cref{app:log-files-relational}, we discuss one possible in-RDBMS
implementation of logging.

\smallskip\noindent
All log files exhibit the following general properties:
\begin{compactitem}
\item Log files \emph{never} hold actual input table cell values but exclusively
  (see a comment on~$\var{log}_{\sql{WIN}}$ below) store compact call
  site and row identifiers: $\location{\ell}$ and $\row_i$ .  Log file
  sizes thus are not impacted by large cell contents like text, arrays,
  or multimedia data.  This is reminiscient of the use of row identifiers
  in the \emph{rid-indexes} of~\cite{smoke-lineage}.
\item A call to $\var{put}_{\Box}(\location{\ell}, k, e)$ adds a log
  entry $\langle\sql{site} = \location{\ell}, \sql{key} = k, \sql{entry}
  = e\rangle$ to log file~$\var{log}_{\Box}$.  Only unique $(\sql{site},\sql{key})$ combinations are
  entered (\emph{write once} safeguard,
  see~\cref{sec:log-writing-reading}): log files \emph{never} need to
  hold duplicate entries.
\item A lookup~$\var{get}_{\Box}(\location{\ell}, k)$ in Phase~$\ph{2}$ retrieves~$e$ from
  the unqiue entry $\langle\sql{site} = \location{\ell}, \sql{key} = k, \sql{entry} = e\rangle$
  in~$\var{log}_{\Box}$.
  Lookups \emph{never} rely on entry order.
\end{compactitem}

\smallskip\noindent
To illustrate log file contents, we evaluate three deliberately simple
\SQL{} queries over two input tables~\sql{r} and~\sql{s}.  All queries contain
a single logging call site~$\location{1}$.
\cref{fig:log-files-input} shows the Phase~$\ph{1}$ variants of both
tables in which the row identifiers~$\row_i$ are exposed. The
instrumented queries---generated following the rules
of~\cref{fig:rules}---and the log file contents they create are
displayed in~\cref{fig:log-files-queries-contents}:
\begin{compactdesc}
\item Query~$Q_j$: Entry $\langle\sql{site} = \location{1}, \sql{key} =
  \langle\row_1,\row_6\rangle, \sql{entry} = \row_{10}\rangle$ in
  $\var{log}_{\sql{JOIN}\langle 2\rangle}$ indicates that the query has
  joined rows~$\row_1$ of~\sql{r} and~$\row_6$ of~\sql{s} to form a
  new joined row~$\row_{10}$.  Subsequently, \emph{e.g.}, in queries
  embracing subquery~$Q_j$, only $\row_{10}$ is used to represent the
  joined pair, but $\var{log}_{\sql{JOIN}\langle 2\rangle}$ allows to
  trace back the pair's provenance to input rows~$\row_{1,6}$.

  Note that the number of log file entries depends on the selectivity of
  join predicate~$t_1\sql{.}\sql{b}\,\sql{=}\,t_2\sql{.}\sql{c}$: only
  successfully joined pairs find their way into the log. \cref{sec:performance-phase1}
  discusses how this helps to reduce the provenance tax.

\item Query~$Q_g$: Entry $\langle\sql{site} = \location{1}, \sql{key} =
  \{\row_1,\row_3,\row_5\}, \sql{entry} = \row_{13}\rangle$ in
  $\var{log}_{\sql{GRP}}$ records that rows $\row_{1,3,5}$ of~\sql{r}
  are members of group~$\row_{13}$.  Each group formed is represented by
  one log entry.  Since keys are of variable size and $\logread{GRP}$ performs
  lookup by group member (see~\cref{sec:log-writing-reading}), it may pay off
  to normalize the representation of~$\sql{key}$ (see a relational
  implementation below).

\item Query~$Q_w$: in~$\var{log}_{\sql{WIN}}$, the entry
  $\langle\sql{site} = \location{1}, \sql{key} = \row_3, \sql{entry} =
  \langle\row_{1},{\scriptstyle 2}\rangle\rangle$ shows
  that row~$\row_3$ of~\sql{r} has been placed at rank~$2$ in the
  same window as~$\row_1$ (the first row in that window).  \cref{sec:abs-int} and
  \cref{fig:window-row} explain how this suffices to re-enact the behavior
  of \SQL{}'s window functions.
\end{compactdesc}
\end{changed}

\begin{figure*}
  \newcommand{\incolumn}[1]{\text{\scriptsize (#1)}} %% describes what's in a log column
  \begin{narrow}{-3mm}{-3mm}
  \centering\small
  \subcaptionbox{Tables~\sql{r} and \sql{s} in Phase~$\ph{1}$.\label{fig:log-files-input}}{%
    \begin{littbl}
      \setlength\minrowclearance{1pt}%
      \begin{tabular}[b]{@{}>{\color{black!60}}c@{\,}>{\,}c<{\,}|>{\,}c<{\,}|}
                 & \tabname{2}{$\inphase{1}{\sql{r}}$\strut} \\
        $\row$   & \colhd{a} & \colhd{b} \\
        $\row_1$ & \multicolumn{1}{|>{\,}c<{\,}|}{1} & 1 \\
        $\row_2$ & \multicolumn{1}{|>{\,}c<{\,}|}{2} & 0 \\
        $\row_3$ & \multicolumn{1}{|>{\,}c<{\,}|}{3} & 1 \\
        $\row_4$ & \multicolumn{1}{|>{\,}c<{\,}|}{4} & 0 \\
        $\row_5$ & \multicolumn{1}{|>{\,}c<{\,}|}{5} & 1 \\
        \cline{2-3}
      \end{tabular}
    \end{littbl}
    \begin{littbl}
      \setlength\minrowclearance{1pt}%
      \begin{tabular}[b]{@{}>{\color{black!60}}c@{\,}>{\,}c<{\,}|}
                 & \tabname{1}{$\inphase{1}{\sql{s}}$\strut} \\
        $\row$   & \colhd{c} \\
        $\row_6$ & \multicolumn{1}{|>{\,}c<{\,}|}{1} \\
        $\row_7$ & \multicolumn{1}{|>{\,}c<{\,}|}{2} \\
        \cline{2-2}
      \end{tabular}%
    \end{littbl}%
    \vspace*{5mm}
  }
  \hfill\vrule\hfill
  \subcaptionbox{Sample instrumented join, grouping, and window queries, instrumented for Phase~$\ph{1}$.  When these queries
    are evaluated over input tables~\sql{r} and \sql{s} (left), the log files
    $\var{log}_{\Box}$ shown here are created as a side effect.\label{fig:log-files-queries-contents}}{%
    \begin{tabular}{@{}c!{~\color{black!40}\vrule~}c!{~\color{black!40}\vrule~}c@{}}
      $
      \begin{array}[t]{@{}l@{~}l}
        \mskip-5mu Q_j{:} \\
        \sql{\textbf{SELECT}} & \logwrite{JOIN$\langle 2\rangle$}\sql{(}\location{1}\sql{,} t_1\sql{.}\row\sql{,} t_2\sql{.}\row\sql{)}\sql{,} \\
                              & e(t_1, t_2) \\
        \sql{\textbf{FROM}}   & \inphase{1}{\sql{r}}~\sql{\textbf{AS}}~t_1\sql{,} \inphase{1}{\sql{s}}~\sql{\textbf{AS}}~t_2 \\
        \sql{\textbf{WHERE}}  & t_1\sql{.}\sql{b}~\sql{=}~t_2\sql{.}\sql{c}
      \end{array}
      $
      &
      $
      \begin{array}[t]{@{}l@{~}l}
        \mskip-5mu Q_g{:} \\
        \sql{\textbf{SELECT}} & \logwrite{GRP}\sql{(}\location{1}\sql{,} \smash{\setagg}\,\{t\sql{.}\row\}\sql{)}\sql{,} \\
                              & \sql{\var{AGG}(}e(t)\sql{)} \\
        \sql{\textbf{FROM}}   & \inphase{1}{\sql{r}}~\sql{\textbf{AS}}~t \\
        \sql{\textbf{GROUP}}  & \sql{BY}~t\sql{.}{\sql{b}}
      \end{array}
      $
      &
      $
      \begin{array}[t]{@{}l@{~}l}
        \mskip-5mu  Q_w{:} \\
        \sql{\textbf{SELECT}} & \logwrite{WIN}\sql{(}
                                \begin{array}[t]{@{}l@{}}
                                  \location{1}\sql{,} t\sql{.}\row\sql{,} \\
                                  \sql{\textbf{FIRST\_VALUE}}\sql{(}t\sql{.}\row\sql{)}~\sql{\textbf{OVER}}~\sql{(}\var{w}\sql{)}\sql{,} \\
                                  \sql{\textbf{RANK()}}~\sql{\textbf{OVER}}~\sql{(}\var{w}\sql{)}\sql{)}
                                \end{array} \\
                              & \sql{\var{AGG}(}e(t)\sql{)}~\sql{\textbf{OVER}}~\sql{(}\var{w}\sql{)}\\
        \sql{\textbf{FROM}}   & \inphase{1}{\sql{r}}~\sql{\textbf{AS}}~t \\
        \sql{\textbf{WINDOW}} & \var{w}~\sql{\textbf{AS}}~\sql{(}\sql{\textbf{PARTITION BY}}~t\sql{.}\sql{b}~\sql{\textbf{ORDER BY}}~t\sql{.}\sql{a}\sql{)}
      \end{array}
      $
      \\ && \\
      $
      \begin{array}[t]{ccc}
        \multicolumn{3}{@{\bigstrut}c}{\var{log}_{\sql{JOIN}\langle 2\rangle}} \\
        \toprule
        \sql{site} & \sql{key}                & \sql{entry}            \\
                   & \incolumn{left/right partner} & \incolumn{pair} \\
        \cmidrule(lr){1-2}\cmidrule(lr){3-3}
        \location{1} & \langle \row_1, \row_6\rangle & \row_{10} \\
        \location{1} & \langle \row_3, \row_6\rangle & \row_{11} \\
        \location{1} & \langle \row_5, \row_6\rangle & \row_{12} \\
        \bottomrule
      \end{array}
      $
      &
      $
      \begin{array}[t]{ccc}
        \multicolumn{3}{@{\bigstrut}c}{\var{log}_{\sql{GRP}}} \\
        \toprule
        \sql{site} & \sql{key}          & \sql{entry}    \\
                   & \incolumn{members} & \incolumn{group} \\
        \cmidrule(lr){1-2}\cmidrule(lr){3-3}
        \location{1} & \{\row_1, \row_3, \row_5\} & \row_{13} \\
        \location{1} & \{\row_2, \row_4\}         & \row_{14} \\
        \bottomrule
      \end{array}
      $
      &
      $
      \begin{array}[t]{ccc}
        \multicolumn{3}{@{\bigstrut}c}{\var{log}_{\sql{WIN}}} \\
        \toprule
        \sql{site} & \sql{key}      & \sql{entry} \\
                   & \incolumn{row} & \incolumn{placement in window} \\
        \cmidrule(lr){1-2}\cmidrule(lr){3-3}
        \location{1} & \row_1 & \langle \row_1, {\scriptstyle 1}\rangle \\
        \location{1} & \row_3 & \langle \row_1, {\scriptstyle 2}\rangle \\
        \location{1} & \row_5 & \langle \row_1, {\scriptstyle 3}\rangle \\
        \location{1} & \row_2 & \langle \row_2, {\scriptstyle 1}\rangle \\
        \location{1} & \row_4 & \langle \row_2, {\scriptstyle 2}\rangle \\
        \bottomrule
      \end{array}
      $
    \end{tabular}}
  \end{narrow}
  \vskip-2mm
  \caption{%
    \color{changed}
    Sample queries and the log file contents they create in Phase~$\ph{1}.$  Later on, Phase~$\ph{2}$
    uses $(\sql{site},\sql{key})$ to lookup~$\sql{entry}$.}
  \label{fig:log-files}
\end{figure*}

\begin{changed}
\subsection{Relational Logging}
\label{app:log-files-relational}

Hosting the logs inside the database itself opens the opportunity to
devise a purely relational implementation of data provenance for \SQL{}.
The simple log file structures lead to straightforward relational
implementations.  \cref{fig:log-files-relational} shows the resulting
tables for the log files~$\var{log}_{\sql{JOIN}\langle 2\rangle}$
and~$\var{log}_{\sql{GRP}}$ of~\cref{fig:log-files-queries-contents}.
The experiments of~\cref{fig:performance} have used exactly the following
tabular log file encodings, indexes, and \SQL{} functions.

\smallskip\noindent
It is apparent from~\cref{fig:log-file-relational-JOIN} that the tabular
$\var{log}_{\sql{JOIN}\langle 2\rangle}$ takes the form of a
join index between tables~\sql{r} and~\sql{s}~\cite{join-indices}.
The creation of an index on~$(\sql{site},\sql{left},\sql{right})$ benefits
Phase~$\ph{2}$ which then saves the actual join effort,
resulting in interpretation that may beat regular evaluation by an order of
magnitude (\cref{sec:performance}).  An additional index with flipped
key~$(\sql{site},\sql{pair})$ admits the efficient lookup of the
provenance relationship between a joined pair and its two source rows.
This is reminiscent of the use of access support
relations~\cite{access-support-rels} in the work on the provenance
language~ProQL by Karvounarakis, Ives, and Tannen~\cite{proql}.

\cref{fig:log-file-relational-GRP} displays the tabular encoding of log
file~$\var{log}_{\sql{GRP}}$.  The table uses a normalized
representation in which each group member occupies one row.  An index on
$(\sql{site},\sql{member})$ constitutes a purely index-based
implementation of  function~$\logread{GRP}(\location{\ell}, \row_i)$
that retrieves the identifier of the group containing member row~$\row_i$.  Similar
to the join case, the flipped index~$(\sql{site}, \sql{group})$ returns
all member rows that form the provenance for the given group.  This
flipped index directly relates to the \emph{1-to-N
rid-indexes} used by~\cite[Section~3.1]{smoke-lineage} to represent the
backward lineage of~\sql{GROUP\,BY}.

\smallskip\noindent
If we employ B$^+$-trees, note that the leading~\sql{site} column in the
above index keys effectively builds \sql{site}-partitioned indexes, an
efficient single-table encoding of the log entries for multiple call
sites~\cite{partitioned-btrees}.
\end{changed}

\begin{figure}
  \centering\small
  \subcaptionbox{Table~$\var{log}_{\sql{JOIN}\langle 2\rangle}$.\label{fig:log-file-relational-JOIN}}{%
    \begin{littbl}
      \begin{tabular}[b]{|c|c|c|c|}
        \tabname{3}{\strut$\var{log}_{\sql{JOIN}\langle 2\rangle}$} \\
        \keyhd{site} & \keyhd{left} & \keyhd{right} & \colhd{pair} \\
        \strut
        $\location{1}$ & $\row_1$ & $\row_6$ & $\row_{10}$ \\
        $\location{1}$ & $\row_3$ & $\row_6$ & $\row_{11}$ \\
        $\location{1}$ & $\row_5$ & $\row_6$ & $\row_{12}$ \\
        \cline{1-4}
      \end{tabular}
    \end{littbl}}
  \hfill\vrule\hfill%
  \subcaptionbox{Table~$\var{log}_{\sql{GRP}}$ (normalized).\label{fig:log-file-relational-GRP}}[4cm]{%
    \begin{littbl}
      \begin{tabular}[b]{|c|c|c|}
        \tabname{3}{\strut$\var{log}_{\sql{GRP}}$} \\
        \keyhd{site} & \keyhd{member} & \colhd{group} \\
        \strut
        $\location{1}$ & $\row_1$ & $\row_{13}$ \\
        $\location{1}$ & $\row_3$ & $\row_{13}$ \\
        $\location{1}$ & $\row_5$ & $\row_{13}$ \\
        $\location{1}$ & $\row_2$ & $\row_{14}$ \\
        $\location{1}$ & $\row_4$ & $\row_{14}$ \\
        \cline{1-3}
      \end{tabular}
    \end{littbl}}
  \caption{%
    \color{changed}
    Tabular encodings of log files~$\var{log}_{\sql{JOIN}\langle 2\rangle}$ and~$\var{log}_{\sql{GRP}}$.}
  \label{fig:log-files-relational}
\end{figure}

\begin{figure}
  \centering\small
\begin{lstlisting}[language=sql]
CREATE FUNCTION writeJOIN2(%$\location{\ell}$% int,%$\langle\row_1$% rid,%$\row_2$% rid%$\rangle$%)
  RETURNS rid AS
$$
  DECLARE %$\row_{\var{join}}$% rid;
  BEGIN
    INSERT INTO %\smash[b]{$\var{log}_{\sql{JOIN}\langle 2\rangle}$}%(site,left,right,pair)
      VALUES (%$\location{\ell}$%,%$\row_1$%,%$\row_2$%,DEFAULT)
      RETURNING pair INTO %$\row_{\var{join}}$%;
    RETURN %$\row_{\var{join}}$%; -- return row ID of joined pair
  EXCEPTION
    WHEN UNIQUE_VIOLATION THEN
      RETURN readJOIN2(%$\location{\ell}$%, %$\langle\row_1$%,%$\row_2\rangle$%);
  END;
$$ LANGUAGE PLPGSQL VOLATILE
\end{lstlisting}

\begin{lstlisting}[language=sql]
CREATE FUNCTION readJOIN2(%$\location{\ell}$% int,%$\langle\row_1$% rid,%$\row_2$% rid%$\rangle$%)
  RETURNS TABLE(%$\row$% rid) AS
$$
  SELECT log.pair
  FROM   %\smash[b]{$\var{log}_{\sql{JOIN}\langle 2\rangle}$}% AS log
  WHERE  log.site = %$\location{\ell}$% AND (log.left,log.right) = (%$\row_1$%,%$\row_2$%)
$$ LANGUAGE SQL STABLE
\end{lstlisting}
  \vskip-3mm
  \caption{%
    \color{changed}
    Companion SQL UDFs~\sql{writeJOIN2} and \sql{readJOIN2}.
    Type~\sql{rid} represents row identifiers.}
  \label{fig:log-files-write-read-UDFs}
\end{figure}

\begin{changed}
\smallskip\noindent
\textbf{\SQL{} implementations
of~$\logwrite{$\Box$}$/$\logread{$\Box$}$}. The tabular log files are
accompanied by user-defined \SQL{} functions that implement the actual
log writing and reading. For~$\var{log}_{\sql{JOIN}\langle 2\rangle}$,
\cref{fig:log-files-write-read-UDFs} shows the \Pg{} variants of these
UDFs.  These functions have been used in the experiments
of~\cref{sec:performance}.

The code for \sql{writeJOIN2} rolls the semantics
of~$\logwrite{JOIN$\langle 2\rangle$}$ and~$\var{put}_{\sql{JOIN}\langle
2\rangle}$ (see~\cref{sec:log-writing-reading}) into a single
\SQL{} UDF for efficiency reasons.  The \emph{write once} safeguard
 is implemented in terms of a~\sql{UNIQUE\_VIOLATION} that is triggered
whenever a duplicate $(\sql{site},\sql{left},\sql{right})$ insertion
would violate the primary key of table~$\var{log}_{\sql{JOIN}\langle
2\rangle}$.\footnote{\SQL's~\sql{ON CONFLICT DO NOTHING} or~\sql{MERGE}
clauses~\cite[\S\,14.12]{sql-2016} provide alternative implementations
of the \emph{write once} safeguard.} The assignment of new join pair
identifiers in column~$\sql{pair}$---\emph{i.e.}, $\row_{10,11,12}$
in~\cref{fig:log-files-queries-contents}--- is realized
using an auto-incrementing \SQL{} sequence. Note that~\sql{writeJOIN2}
has been marked as~\sql{VOLATILE} to announce its side effect on the
logging table~\cite[\S\,37.6]{pgsql}.
This constrains the optimizer's rewriting choices for instrumented
queries---we quantify the effects in~\cref{sec:performance}.

Companion UDF~\sql{readJOIN2}, instead, acts like a pure (\SQL{}:
\sql{STABLE}) function.  The routine performs a simple index-supported lookup for
key $(\sql{site},\sql{left},\sql{right})$ in
table~$\var{log}_{\sql{JOIN}\langle 2\rangle}$ and returns a table of
zero or one join pair identifiers as required by~\cref{rule:join} (\cref{fig:rules}).
\end{changed}

\begin{changed}
\section{Scalability Experiment \\
  (\TPCH{} Scale Factor~10)}
\label{app:tpch-ten-gb}

To assess the scalability of cell-level provenance derivation through
the non-standard interpretation of \SQL{}, we repeated the \TPCH{}-based
experiments reported in~\cref{fig:performance} on a larger database
instance.  Here, we set the benchmark's scale factor $\mathit{sf} = 10$ such that
central table~\sql{lineitem} holds $60\,000\,000$~rows.  Otherwise, the
experimental setup exactly matches that of~\cref{sec:performance}.  In
particular, we did not modify the instrumented queries
(Phase~$\ph{1}$) or interpreters (Phase~$\ph{2}$) that were fed into
\Pg.
\end{changed}

\begin{figure*}
  \centering\small
  %% the plot
  \begin{tikzpicture}[x=0.72mm,y=8mm]
    \begin{scope}[thick]
      %% x axis (TPC-H queries)
      \draw (-22,0) -- (220,0);
      %% y axes
      \draw[->] (-1, 0.05) -- (-1, 4.4);
      \node[anchor=south,align=center] at (-1,4.4) {slowdown};
      \draw[->] (-1,-0.05) -- (-1,-1.7);
      \node[anchor=east,align=right] at (-3,-1.7) {log size \\\relax [kBytes]};
    \end{scope}
    %% queries (TPC-H query Q#/norm ratio/p1 ratio/p1+p2 ratio/p1+p2y ratio/log cells/write sites/show measurement?)
    \foreach \tpch/\n/\p/\pp/\ppy/\logsize/\sites/\measure [count=\x from 0] in {
      %% 10GB experiments  (21.05.2018)
      1  /1.00  /  4.99  /-999    /-999    /2557520  /1 /0,
      2  /1.00  /  1.54  /  1.54  /   1.54 /664      /3 /1,
      3  /0.95  /  1.57  /  2.17  /   2.87 /28480    /3 /0,
      4  /0.99  /  3.69  /  4.48  /   9.23 /45504    /2 /1,
      5  /0.94  /  1.12  /  1.23  /   1.43 /7456     /2 /0,
      6  /0.99  /  2.48  /  4.56  /   7.73 /40328    /1 /0,
      7  /1.00  /  1.14  /  1.27  /   1.76 /5968     /2 /0,
      8  /1.00  /  1.20  /  1.41  /   1.76 /3736     /3 /0,
      9  /0.78  /  2.06  /  2.93  /   6.13 /332912   /2 /1,
      10 /0.95  /  3.12  /  6.41  /  15.97 /108072   /3 /0,
      11 /1.01  /  6.27  /  6.47  /2843.78 /33552    /3 /1,
      12 /1.00  /  2.10  /  2.72  /   3.82 /53784    /4 /0,
      13 /1.01  / 54.26  / 55.73  /  59.87 /1754680  /3 /0,
      14 /0.99  /  3.82  /  6.62  /   7.40 /64800    /2 /0,
      15 /1.00  /  1.73  /  2.90  /   6.58 /97992    /2 /0,
      16 /1.07  /  5.03  /  7.53  /  15.34 /102632   /3 /0,
      17 /0.99  /127.76  /127.81  / 128.91 /2896     /2 /1,
      18 /0.95  /  1.14  /  1.14  /   1.43 /608      /3 /1,
      19 /0.99  /  1.02  /  1.04  /   1.05 /56       /1 /1,
      20 /0.99  /  1.04  /  1.05  / 126.13 /7168     /4 /0,
      21 /1.01  / 33.89  / 33.91  /  33.99 /3079672  /4 /0,
      22 /0.95  /  5.17  /  5.47  /  10.22 /21776    /3 /0
    } {
      %% measurements: query runtime ratios (+ log size)
      \begin{scope}[yshift=5mm]
        %% normalized query
        \pgfmathsetmacro\y{log10(\n)}
        \draw[black,fill=white] (2+10*\x,\y)+(-\squaresize,-\squaresize) rectangle ++(\squaresize,\squaresize);
        \ifnum\x=2
          \node[legend] at (2+10*\x,\y) {normalized};
        \fi
        %% p1
        \pgfmathsetmacro\y{log10(\p)}
        \draw[fill=black] (4+10*\x,\y)+(-\squaresize,-\squaresize) rectangle ++(\squaresize,\squaresize);
        \ifnum\x=2
          \node[legend] at (4+10*\x,\y) {$\ph{1}$};
        \fi
        \ifnum-999=\pp
          \typeout{10GB DNF}
        \else
          %% p1+p2
          \pgfmathsetmacro\y{log10(\pp)}
          \draw[black,fill=white] (6+10*\x,\y) circle (\circlesize);
          \ifnum\x=2
            \node[legend] at (6+10*\x,\y) {$\ph{1}\mathord{+}\ph{2}$};
          \fi
          %% p1+p2y
          \pgfmathsetmacro\y{log10(\ppy)}
          \draw[fill=black] (8+10*\x,\y) circle (1.5pt);
          \ifnum\x=2
            \node[legend] at (8+10*\x,\y) {$\ph{1}\mathord{+}\ph{2}$\toY};
          \fi
        \fi
        %% selected measurements
        \ifnum\measure=1
          \pgfkeys{/pgf/number format/.cd,std,precision=0}
          \node[legend,anchor=east,black,yshift=-1pt] at (8+10*\x,\y) {\pgfmathprintnumber{\ppy}};
        \fi
      \end{scope}
      %% measurements: p1 log size
      \begin{scope}
        \pgfkeys{/pgf/fpu=true,/pgf/fpu/output format=fixed} %% activate high-precision arithmetics in TikZ
        \pgfmathsetmacro\y{-(log10(\logsize) - 1) / 5}
        \pgfkeys{/pgf/fpu=false}
        \draw[black,fill=black] (5+10*\x,-0.02)+(-\barsize,0) rectangle ++(\barsize,\y);
        \node[legend,black,left=-2pt] at (3+10*\x,0) {$\pgfmathprintnumber{\logsize}$};
        %% # of write sites
        \node[legend,black] at (5.2+10*\x,\y-0.45) {\circled{\sites}};
      \end{scope}
      %% query columns
      \begin{pgfonlayer}{background}
        \pgfmathsetmacro\even{15*(1-mod(int(\tpch),2)}
        \ifnum-999=\pp
          \draw[draw=none,pattern=stripes,pattern color=black!10] (0+10*\x,-1.5) rectangle (10+10*\x,4.2);
        \else
          \draw[draw=none,fill=black!\even] (0+10*\x,-1.5) rectangle (10+10*\x,4.2);
        \fi
      \end{pgfonlayer}
      %% TPC-H query number
      \node[anchor=north,font=\scriptsize] at (5+10*\x, -1.5) {\tpchQ{\tpch}};
    };
    %% bands (slowdown)
    \begin{pgfonlayer}{background}
      \begin{scope}[yshift=5mm,band]
        \foreach \slowdown in {1,10,100,1000} {
          \pgfmathsetmacro\y{log10(\slowdown)}
          \draw (-3,\y) -- (220,\y);
          \node[anchor=east] at (-3,\y) {$\mathop{\times} \pgfmathprintnumber{\slowdown}$};
        };
      \end{scope}
      %% bands (log size)
      \begin{scope}[band]
        \foreach \logsize in {1000,1000000} {
          \pgfkeys{/pgf/fpu=true,/pgf/fpu/output format=fixed} %% activate high-precision arithmetics in TikZ
          \pgfmathsetmacro\y{-(log10(\logsize) - 1) / 5}
          \pgfkeys{/pgf/fpu=false}
          \draw (-3,\y) -- (220,\y);
          \node[anchor=east] at (-3,\y) {$\pgfmathprintnumber{\logsize}$};
        };
      \end{scope}
    \end{pgfonlayer}
  \end{tikzpicture}
  \caption{%
    \color{changed}
    Repetition of experiments reported in~\cref{fig:performance} at \TPCH{} scale factor~10.
    Slowdown of normalization
    (\protect\tikz\protect\draw[fill=white] (2.5,2.5) rectangle ++(2\squaresize,2\squaresize);),
    Phase~$\ph{1}$ (\protect\tikz\protect\draw[fill=black] (2.5,2.5) rectangle ++(2\squaresize,2\squaresize);),
    Phases~$\ph{1}$+$\ph{2}$
    \mbox{(\protect\tikz\protect\draw[fill=white] (0,0) circle (\circlesize);
    without/\protect\tikz\protect\draw[fill=black] (0,0) circle (\circlesize);} with \why-provenance)
    relative to value-based \TPCH{}.
    The dependency set cardinality for query~\tpchQ{1} exceeded the \Pg{} size limit for array
    type~\sql{int[]}.}
  \label{fig:performance-ten-gb}
\end{figure*}

\begin{figure}
  \begin{narrow}{-8mm}{-6mm}
  %\makebox[\linewidth][r]{%  let figure stick out into the margin
  \centering\small
   %% the plot
  \begin{tikzpicture}[x=0.36mm,y=7mm]
    \begin{scope}[thick]
      %% x axis (TPC-H queries)
      \draw (0,0) -- (220,0);
      %% y axes
      \draw[->] (-1, 0.05) -- (-1, 4.65);
      \node[anchor=south,align=center] at (-1,4.7) {slowdown};
    \end{scope}
    %% queries (TPC-H query Q#/p1+p2y ratio 1GB/p1+p2y ratio 10GB
    \foreach \tpch/\ppy/\ppytengb/\measure [count=\x from 0] in {
      %% experiment with relational logs (1GB: 2018/05/13, 10GB: 2018/05/13 and 2018/05/14)
      %% geometric mean slowdown: 1.18
      1  /  53.88/-999   /0,  % max array size exceeded?
      2  /   1.38/   1.54/1,
      3  /   4.27/   2.87/0,
      4  /   8.48/   9.23/0,
      5  /   1.50/   1.43/0,
      6  /   3.66/   7.73/1,
      7  /   1.66/   1.76/0,
      8  /   2.51/   1.76/1,
      9  /   8.25/   6.13/0,
      10 /  17.77/  15.97/0,
      11 /1039.12/2843.78/1,
      12 /   3.54/   3.82/0,
      13 /  22.63/  59.87/1,
      14 /   5.68/   7.40/0,
      15 /   6.01/   6.58/0,
      16 /  17.06/  15.34/0,
      17 /  99.86/ 128.91/0,
      18 /   1.22/   1.43/0,
      19 /   1.24/   1.05/0,
      20 /  42.49/ 125.73/1,
      21 /  28.42/  33.99/0,
      22 /  11.46/  10.22/0
    } {
      % %% measurements: query runtime ratios
      % \begin{scope}[yshift=12mm]
      %   %% p1+p2y (1GB)
      %   \pgfmathsetmacro\y{log10(\ppy)}
      %   \draw[fill=black] (2+10*\x,\y) circle (1.5pt);
      %   \ifnum\x=3
      %     \node[legend] at (2+10*\x,\y) {$\ph{2}$\toY~(1~GB)};
      %   \fi
      %   %% p1+p2y (10GB)
      %   \ifnum-999=\ppytengb
      %     \typeout{10GB DNF}
      %   \else
      %     \pgfmathsetmacro\z{log10(\ppytengb)}
      %     \draw[draw=black,fill=white] (8+10*\x,\z) circle (1.5pt);
      %     \ifnum\x=3
      %       \node[legend] at (8+10*\x,\z) {$\ph{2}$\toY~(10~GB)};
      %     \fi

      %% measurements: query runtime ratios
      \begin{scope}[yshift=5mm]
        %% p1+p2y (1GB)
        \pgfmathsetmacro\y{log10(\ppy)}
        \draw[fill=black] (2+10*\x,\y) circle (1.5pt);
        \ifnum\x=3
          \node[legend] at (2+10*\x,\y) {$\ph{1}\mathord{+}\ph{2}$\toY~(1GB)};
        \fi
        %% p1+p2y (10GB)
        \ifnum-999=\ppytengb
          \typeout{10GB DNF}
        \else
          \pgfmathsetmacro\z{log10(\ppytengb)}
          \draw[draw=black,fill=white] (8+10*\x,\z) circle (1.5pt);
          \ifnum\x=3
            \node[legend] at (8+10*\x,\z) {$\ph{1}\mathord{+}\ph{2}$\toY~(10GB)};
          \fi
          %% trend line
          \begin{pgfonlayer}{annotation}
            \ifnum\measure=1
              \pgfmathsetmacro\ratio{\ppytengb > \ppy ? \ppytengb / \ppy : \ppy / \ppytengb}
              \pgfkeys{/pgf/number format/.cd,std,precision=1}
              \draw[trend] (2+10*\x,\y) to node[black,ratio,sloped,above=-1pt] {$\pgfmathprintnumber{\ratio}\scalebox{0.8}{$\times$}$} (8+10*\x,\z);
            \else
              \draw[trend] (2+10*\x,\y) -- (8+10*\x,\z);
            \fi
          \end{pgfonlayer}
        \fi
      \end{scope}
      %% query columns
      \begin{pgfonlayer}{background}
        \pgfmathsetmacro\odd{10*isodd(\x)}
        \ifnum-999=\ppytengb
          \draw[draw=none,pattern=stripes,pattern color=black!10] (0+10*\x,0) rectangle (10+10*\x,4.4);
        \else
          \draw[draw=none,fill=black!\odd] (0+10*\x,0) rectangle (10+10*\x,4.4);
        \fi
      \end{pgfonlayer}
      %% TPC-H query number
      \node[anchor=east,rotate=90,font=\scriptsize] at (5+10*\x, 0) {\tpchQ{\tpch}};
    };
    %% bands (slowdown)
    \begin{pgfonlayer}{background}
      \begin{scope}[yshift=5mm,band]
        \foreach \slowdown in {1,10,100,1000} {
          \pgfmathsetmacro\y{log10(\slowdown)}
          \draw (-3,\y) -- (220,\y);
          \node[anchor=east] at (-3,\y) {$\mathop{\times} \pgfmathprintnumber{\slowdown}$};
        };
      \end{scope}
    \end{pgfonlayer}
  \end{tikzpicture}
  \end{narrow}
  \caption{%
    \color{changed}
    Phases~$\ph{1}+\ph{2}$ (with \why-provenance):
    slowdown remains stable for \TPCH{} scale
    factors~1~(\protect\tikz\protect\draw[fill=black] (0,0) circle (\circlesize);)
    and~10~(\protect\tikz\protect\draw[fill=white] (0,0) circle (\circlesize);).}
  \label{fig:tpch-ten-gb}
\end{figure}

\begin{changed}
\smallskip\noindent
\cref{fig:performance-ten-gb} summarizes our observations regarding
logging effort and provenance tax.  Compared to~\cref{fig:performance},
we see that log sizes consistently grow by a factor of~$10$.  This is
expected due to the nature of \TPCH{} in which the sizes of intermediate
results grow linearly and selectivities remain
stable~\cite{tpch,tpch-pain-points}. The scale-up does not, however,
affect the number of~\logwrite{$\Box$} call sites (shown
in~\circled{\phantom{2}}) as these exclusively depend on the syntactic
structure of the subject queries.

\smallskip\noindent
Most importantly, we find that the provenance tax remains stable: regarding
slowdown, \cref{fig:performance-ten-gb} largely presents a mirror image of
\cref{fig:performance}.  Indeed, we find a mean ratio
$\nicefrac{\text{slowdown($\mathit{sf}\!=\!10$)}}{\text{slowdown($\mathit{sf}\!=\!1$)}}$ of $1.18$
across the benchmark.  \cref{fig:tpch-ten-gb} aims to visualize that
the slowdown essentially remains stable.  Partly, deviations
from perfect scalability, \emph{i.e.}, a horizontal trend
\begin{tikzpicture}[sketch]
  \draw[trend] (0,1.0) -- (2.5,1.0);
  \draw[fill] (0,1.0) circle (\circlesize);
  \draw[fill=white] (2.5,1.0) circle (\circlesize);
\end{tikzpicture},
are to be attributed to \Pg's plan choices.  If we force \Pg{} to
stick to an indexed nested loops join to interpret~\tpchQ{6}, for example,
the slowdown at scale factors~$1$ and~$10$ is identical.\footnote{It is important
to note that, here, we
enforce plan choices solely to find causes of query slowdown/speed-up.  We
pursue query shape preservation precisely to support the RDBMS in identifying
efficient plans on its own.}
For other queries, the figure documents that a growth in data volume amplifies the
effects we have already discussed in~\cref{sec:performance-phase1,sec:performance-phase2}:
\begin{compactitem}
\item correlation in~\tpchQ{11} now requires Phase~$\ph{1}$ to avoid duplicate
  entries in log files of ten-fold the size,
\item the comparably large log tables written by~\tpchQ{13} stress the
  database buffer (an enlarged buffer restores almost perfect scalability, though),
\item the occurrence of subqueries nested to depth two in the
  \sql{WHERE} clause of~\tpchQ{20} leads to a disproportionate growth of
  \why-dependencies (by a factor of about~$200$ instead of~$10$).
  Note that this constitutes a property of the query and the
  dependencies it induces, not a defect of the approach.
\end{compactitem}

\smallskip\noindent
High provenance sizes in~\tpchQ{1} exceeded the 1\,GB size
limit for \Pg{}'s array type~\sql{int[]} which we have used throughout this
paper to represent dependency sets (also see~\cref{app:dep-set-rep} below).
Specifically, the grouped aggregate over table~\sql{lineitem}
$$
\sql{SUM(l\_extendedprice * (1-l\_discount) * (1+l\_tax))}
$$%
collects about~$3~\text{(columns)} \times 60\,000\,000~\text{(rows)}
\div 4~\text{(groups)} = 45\,000\,000$
\where-dependencies for each individual output cell.  With
\why-dependencies enabled, this number doubles due to the dependencies incurred by three additional
columns (the two grouping critera~\sql{l\_returnflag},
\sql{l\_linestatus} and column~\sql{l\_shipdate} that is referenced in
the query's~\sql{WHERE} predicate, see~\cref{rule:group,rule:join}).  A
deployment of the approach on a specific RDBMS should take such
high-volume dependencies---that are inherent to the source data and
query workload---into account when an implementation type for~$\ptype$
is chosen.
\end{changed}

\section{\!Dependency Set Representation}
\label{app:dep-set-rep}

The shift from values to dependency sets in Phase~$\ph{2}$ brings with
it a significant increase in data volume that has to be processed. The
original query~\tpchQ{1} returns $40 = 4 \times 10$ (rows $\times$
columns) cells of scalar values.  Given the same query, \where- and
\why-provenance derivation accumulates $40$~dependency sets which
together contain $236\,663\,640$~cell identifiers. For the \TPCH{}
benchmark, the mean size of a dependency set is about $10\,000$ elements. The
bottom of~\cref{fig:full-pset} gives an impression of the overall
provenance cardinalities we obtain for the 22~queries.    It is expected
that the interpreters spend significant time in building these large
dependency sets and that the choice of set representation will have a
measureable impact on Phase~$\ph{2}$.

In the experiments of~\cref{fig:performance,fig:performance-ten-gb}, we represent the
dependency set type~$\ptype$ in terms of \Pg{}'s array type~\sql{int[\,]}.
The set operations $\cup$ and $\smash\bigcup$ map to array concatenation~\sql{\!\textbar\!\textbar\!}
and aggregation~\sql{array\_agg}, respectively.  Set semantics is enforced
via explicit duplicate elimination at the query top-level.  If we trade
this native (but naive) implementation for a bit set representation---based on the
dynamically growing \emph{roaring bitmaps}~\cite{roaring}---we
significantly reduce the running time of Phase~$\ph{2}$, up to an
order of magnitude (see the slowdown drops
\begin{tikzpicture}[sketch]
  \draw[trend] (0,2) -- (2,-0.5);
  \draw[fill=black] (0,2)    circle (\circlesize);
  \draw[fill=white] (2,-0.5) circle (\circlesize);
\end{tikzpicture}
in~\cref{fig:full-pset}).
Across the benchmark, we see a geo\-metric mean improvement of factor~$2.1$.
Interpreters for queries like~\tpchQ{1}, \tpchQ{11}, and \tpchQ{20}
which spend $75$--$90\%$ of their execution time building huge dependency sets
particularly benefit.  Unlike these, the evaluation
of~\tpchQ{17} and~\tpchQ{19} collects comparably tiny sets and thus see no improvement.
Indeed, when query interpretation requires few dependency set operations (as for
\tpchQ{6}/\tpchQ{14}: one/two set aggregrations~$\setagg$ only) or is dominated by
non-set operations (as for~\tpchQ{5} whose evaluation is dominated by a large
six-way join), a change of set representation has a lesser impact.

\begin{figure}
  \begin{narrow}{-10mm}{-4mm}
  %\makebox[\linewidth][r]{%  let figure stick out into the margin
  \centering\small
   %% the plot
  \begin{tikzpicture}[x=0.36mm,y=7mm]
    \begin{scope}[thick]
      %% x axis (TPC-H queries)
      \draw (-42,0) -- (220,0);
      %% y axes
      \draw[->] (-1, 0.05) -- (-1, 5.6);
      \node[anchor=south,align=center] at (-1,5.6) {slowdown/speed-up};
      \draw[->] (-1,-0.05) -- (-1,-2.1);
      \node[anchor=east,align=right] at (-3,-2.1) {provenance\\\relax cardinality};
    \end{scope}
    %% queries (TPC-H query Q#/p2y ratio/p2y-pset-rb ratio/provenance size
    \foreach \tpch/\ppy/\ppypset\provsize/\measure [count=\x from 0] in {
      %% experiment with relational logs and roaring-transient (15.05.2018)
      1 / 49.86  / 13.32  / 236663640/1,
      2 / 0.11   / 0.01   / 14392    /0,
      3 / 2.31   / 1.15   / 1280     /0,
      4 / 5.03   / 2.83   / 630276   /0,
      5 / 0.34   / 0.27   / 128038   /0,
      6 / 1.21   / 1.23   / 456640   /0,
      7 / 0.55   / 0.27   / 176248   /0,
      8 / 1.31   / 0.51   / 62112    /1,
      9 / 5.78   / 3.40   / 8621066  /0,
      10/ 14.47  / 3.43   / 12080    /0,
      11/ 1022.43/ 73.08  / 200870896/1,
      12/ 1.49   / 1.10   / 614815   /0,
      13/ 9.24   / 5.83   / 9203508  /0,
      14/ 2.23   / 2.23   / 430156   /0,  %% ↓
      15/ 4.44   / 3.24   / 4519088  /0,
      16/ 12.13  / 3.23   / 2005344  /0,
      17/ 0.74   / 0.74   / 12914    /0,  %% ↓
      18/ 0.03   / 0.008  / 274968   /0,
      19/ 0.13   / 0.11   / 1138     /0,
      20/ 40.42  / 4.35   / 22556034 /1,
      21/ 0.22   / 0.13   / 70366    /0,
      22/ 6.02   / 1.74   / 1601040  /1
    } {
      %% measurements: query runtime ratios
      \begin{scope}[yshift=16mm]
        %% p1+p2y
        \pgfmathsetmacro\y{log10(\ppy)}
        \draw[fill=black] (2+10*\x,\y) circle (1.5pt);
        \ifnum\x=3
          \node[legend] at (2+10*\x,\y) {$\ph{2}$\toY~(array)};
        \fi
        %% p1+p2y-tuple
        \pgfmathsetmacro\z{log10(\ppypset)}
        \draw[draw=black,fill=white] (8+10*\x,\z) circle (1.5pt);
        \ifnum\x=3
          \node[legend] at (8+10*\x,\z) {$\ph{2}$\toY~(bit~set)};
        \fi
        %% selected measurements
        \ifnum\measure=1
          \ifdim\ppy pt<1pt\pgfkeys{/pgf/number format/precision=2}
            \else\pgfkeys{/pgf/number format/precision=1}\fi
          \node[legend,anchor=east,black] at (2+10*\x,\y) {\pgfmathprintnumber{\ppy}};
          \ifdim\ppypset pt<1pt\pgfkeys{/pgf/number format/precision=2}
             \else\pgfkeys{/pgf/number format/precision=1}\fi
          \node[legend,anchor=east,black] at (8+10*\x,\z) {\pgfmathprintnumber{\ppypset}};
        \fi
        %% trend line
        \begin{pgfonlayer}{annotation}
          \draw[trend] (2+10*\x,\y) -- (8+10*\x,\z);
        \end{pgfonlayer}
      \end{scope}
      %% measurements: provenance cardinality
      \begin{scope}
        \pgfkeys{/pgf/fpu=true,/pgf/fpu/output format=fixed} %% activate high-precision arithmetics in TikZ
        \pgfmathsetmacro\y{-(log10(\provsize) - 1) / 5}
        \pgfkeys{/pgf/fpu=false}
        \draw[black,fill=black] (7+10*\x,0)+(-\barsize,0) rectangle ++(\barsize,\y);
        \node[legend,black,left=-2pt] at (4+10*\x,0) {$\pgfmathprintnumber{\provsize}$};
      \end{scope}
      %% query columns
      \begin{pgfonlayer}{background}
        \pgfmathsetmacro\even{15*(1-mod(int(\tpch),2))}
        \draw[draw=none,fill=black!\even] (0+10*\x,-1.9) rectangle (10+10*\x,5.35);
      \end{pgfonlayer}
      %% TPC-H query number
      \node[anchor=east,rotate=90,font=\scriptsize] at (5+10*\x, -1.9) {\tpchQ{\tpch}};
    };
    %% bands (seconds)
    \begin{pgfonlayer}{background}
      \begin{scope}[yshift=16mm,band]
        \pgfkeys{/pgf/number format/precision=3}
        \foreach \slowdown in {0.01,0.1,1,10,100,1000} {
          \pgfmathsetmacro\y{log10(\slowdown)}
          \draw (-3,\y) -- (220,\y);
          \node[anchor=east] at (-3,\y) {$\mathop{\times} \pgfmathprintnumber{\slowdown}$};
        };
      \end{scope}
      %% bands (provenance cardinality)
      \begin{scope}[band]
        \foreach \provsize in {1000,1000000} {
          \pgfkeys{/pgf/fpu=true,/pgf/fpu/output format=fixed} %% activate high-precision arithmetics in TikZ
          \pgfmathsetmacro\y{-(log10(\provsize) - 1) / 5}
          \pgfkeys{/pgf/fpu=false}
          \draw (-3,\y) -- (220,\y);
          \node[anchor=east] at (-3,\y) {$\pgfmathprintnumber{\provsize}$};
        };
      \end{scope}
    \end{pgfonlayer}
  \end{tikzpicture}
  \end{narrow}
  \caption{Efficient dependency set representations reduce interpretation
    overhead in Phase~$\ph{2}$: native
    (\protect\tikz\protect\draw[fill=black] (0,0) circle (\circlesize);) vs.\ bit set
    (\protect\tikz\protect\draw[fill=white] (0,0) circle (\circlesize);).}
  \label{fig:full-pset}
\end{figure}
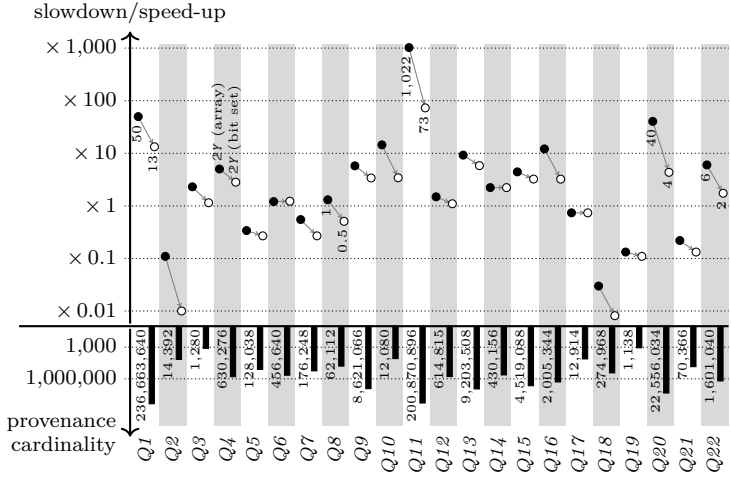

\smallskip\noindent
\begin{changed}
We see that, as query complexity and dependency set cardinalities grow,
it pays off to invest in efficient representations for type~$\ptype$.
The bitmap-based sets discussed above come packaged as a \Pg{} extension
and thus do reach inside the RDBMS---unlike the rest of the approach.
With its \SQL:2003
revision~\cite[\S\,4.10.3]{sql-2003}, however, the \SQL{} standard has
introduced the data type~\sql{$a$~MULTISET} that represents bags
(and sets) of elements of type~$a$.  Type~\sql{int~MULTISET} thus
provides another native implementation for the type~$\ptype$ of
dependency sets.  In this alternative, $\cup$ and $\smash\bigcup$ map
directly to the standard operations
\sql{$\cdot$\;MULTISET UNION DISTINCT\;$\cdot$} and
\sql{SET(FUSION($\cdot$))}, respectively.  The bitmap-based set representation
effectively adds support for type~\sql{int~MULTISET} and its
operations to \Pg.
\end{changed}

\else

An extended version of this paper has been posted on \emph{arXiv}
and is available at
\begin{center}
\url{https://arxiv.org/abs/1805.11517} .
\end{center}
The appendix of this version contain extensions to the inference
rule set of~\cref{fig:rules}, an in-depth discussion of log file
contents and implementation, \TPCH{} scalability experiments, and
an exploration of an alternative representation of dependency sets.

\fi

\end{document}